\documentclass[a4paper,fleqn,usenatbib]{mnras}

\usepackage[T1]{fontenc}
\usepackage{ae,aecompl}

\usepackage{graphicx}	
\usepackage{amsmath}	
\usepackage{amssymb}	

\usepackage{txfonts}
\usepackage{color, colortbl}
\usepackage{enumitem}
\usepackage{yfonts}

\newcommand{\be}{\begin{equation}}
\newcommand{\ee}{\end{equation}}

\def\Mvir{M_{\rm vir}}
\def\Rvir{R_{\rm vir}}

\def\rp{r^\prime}

\def\rmd{{\rm d}}
\def\vir{{\rm vir}}

\defcitealias{Dekel2017}{D17}
\defcitealias{Einasto1965}{Einasto}

\defcitealias{DiCintio2014b}{\mbox{Di Cintio+}}
\defcitealias{Zhao1996}{\mbox{Z96}}
\defcitealias{An2013}{\mbox{AZ13}}
\defcitealias{Freundlich2020}{\mbox{F20}}

\title[The Dekel-Zhao profile]{The Dekel-Zhao profile: A mass-dependent dark-matter density profile with flexible inner slope and analytic potential, velocity dispersion, and lensing properties}

\author[J. Freundlich et al.]{Jonathan Freundlich,$^{1,2}$\thanks{E-mail: jonathan.freundlich@mail.huji.ac.il}
Fangzhou Jiang,$^{1}$
Avishai Dekel,$^{1,3}$ 
Nicolas Cornuault,$^{1}$
Omry 
\newauthor
Ginzburg,$^{1}$
R\'emy Koskas,$^{1,4}$
Sharon Lapiner,$^{1}$
Aaron Dutton,$^{5}$
and Andrea V. Macci\`o$^{5,6}$
\\
$^1$Centre for Astrophysics and Planetary Science, Racah Institute of Physics, The Hebrew University, Jerusalem 91904, Israel\\
$^2$School of Physics and Astronomy, Tel Aviv University, Tel Aviv 69978, Israel\\
$^3$Santa Cruz Institute for Particle Physics, University of California, Santa Cruz, CA 95064, USA\\
$^4$\'Ecole Nationale des Ponts et Chauss\'ees, 77420 Champs-sur-Marne, France\\
$^5$New York University Abu Dhabi, PO Box 129188, Saadiyat Island, Abu Dhabi, United Arab Emirates\\
$^6$Max Planck Institute f\"ur Astronomie, K\"onigstuhl 17, 69117 Heidelberg, Germany\\
}

\date{Accepted 2020 September 9. Received 2020 September 9; in original form 2020 April 17.}

\pubyear{2020}

\begin{document}
\label{firstpage}
\pagerange{\pageref{firstpage}--\pageref{lastpage}}
\maketitle

\begin{abstract}
	
We explore a function with two shape parameters for the dark-matter halo density profile subject to baryonic effects, which is a special case of the general Zhao family of models applied to simulated dark matter haloes by  Dekel et al. This profile has variable inner slope and concentration parameter, and analytic expressions for the gravitational potential, velocity dispersion, and lensing properties. Using the NIHAO cosmological simulations, we find that it
provides better fits than the Einasto profile and the generalized NFW profile with variable inner slope, in particular towards the halo centers. We show that the profile parameters are correlated with the stellar-to-halo mass ratio $M_{\rm star}/M_{\rm vir}$. This defines a mass-dependent density profile describing the average dark matter profiles in all galaxies, which can be directly applied to observed rotation curves of galaxies, gravitational lenses, and semi-analytic models of galaxy formation or satellite-galaxy evolution. The effect of baryons manifests itself by a significant flattening of the inner density slope and a 20\% decrease of the concentration parameter for $M_{\rm star}/M_{\rm vir} \!=\! 10^{-3.5}$ to $10^{-2}$, corresponding to $M_{\rm star} \!\sim\! 10^{7-10}\, M_\odot$. The accuracy by which this profile fits simulated galaxies is similar to certain multi-parameter, mass-dependent profiles, but its fewer parameters and analytic nature make it most desirable for many purposes.
\end{abstract}




\section{Introduction}
\label{section:intro}

Dark matter (DM) halo density profiles in DM-only cosmological simulations are 
well described by the `NFW' parametrization \citep{NFW1996,NFW1997,Springel2008,Navarro2010} from dwarf halos to large clusters,  although with some systematic deviations \citep[e.g.,][]{Navarro2004,Navarro2010, Maccio2008, Gao2008,Springel2008}. This density profile scales with radius as
\begin{equation}
\label{eq:NFW}
\rho_{\rm NFW}(r) = \frac{\rho_c}{x (1+x)^2},
\end{equation}
with $x=r/r_s$, $r_s$ being a characteristic scale radius at which the density logarithmic slope equals $2$ in absolute value. This radius defines a concentration 
$c_{\rm DMO}=R_\vir/r_s$, which depends on the halo virial mass $M_\vir$ and redshift \citep[e.g.,][]{Bullock2001,Wechsler2002,Dutton2014} -- both $M_\vir$ and the virial radius $R_\vir$ being set by cosmology. 
The inner $\rho \propto r^{-1}$ `cusp' of the NFW parametrization is at odds with observations of DM dominated dwarf, low-surface-brightness and dwarf satellite galaxies as well as clusters, which infer shallower `cores' \citep[e.g., ][]{Flores1994,Moore1994,McGaugh1998a,vandenBosch2001, deBlok2008,deBlok2010, Kuzio2011,Oh2011,Oh2015,Newman2013a,Newman2013b,Adams2014}.
The introduction of baryonic processes such as cooling, star formation and feedback resulting from star formation or active galactic nuclei (AGN) in the simulations can alleviate this `cusp-core discrepancy' by transforming cusps into cores \citep[e.g.,][]{Governato2010, Governato2012,Maccio2012b,Maccio2020, Zolotov2012, Martizzi2013,Teyssier2013, DiCintio2014,Chan2015, Tollet2016, Peirani2017}.

Baryonic processes can affect DM haloes in different ways. 
When baryons cool slowly and accumulate at the center of a DM halo, they steepen the potential well, leading to an adiabatic contraction of the DM distribution and even more severe cusps \citep{Blumenthal1986, Gnedin2004,Onorbe2007}.
When a clump of gas or a satellite galaxy moves within the halo, it can transfer part of its orbital energy and angular momentum to the DM background through dynamical friction \citep{Chandrasekhar1943,Tremaine1984}. This latter process dynamically 'heats' the DM halo and has been shown to contribute to core formation \citep{El-Zant2001,El-Zant2004,Tonini2006,RomanoDiaz2008,DelPopolo2009, Goerdt2010,Cole2011,Nipoti2015}. 
When stellar winds, supernova explosions or AGNs generate outflows, they induce mass and potential fluctuations that can also dynamically heat the DM and form cores \citep{Dekel1986, Dekel2003a, Dekel2003b, Read2005, Mashchenko2006, Mashchenko2008, Penarrubia2012, Pontzen2012, Pontzen2014, Governato2012, Zolotov2012, Martizzi2013, Teyssier2013, Madau2014, Dutton2016, El-Zant2016, Peirani2017,Freundlich2020}.
Other processes such as galactic bars \citep{Weinberg2002} or tidal effects at the halo outskirts \citep{More2015} may also affect the DM distribution.

These different processes are reflected in hydrodynamical simulations, which display a variety of DM halo responses to the introduction of baryons, notably depending on stellar and halo masses.
In particular, \cite{DiCintio2014}, \cite{Chan2015}, \cite{Tollet2016} and \cite{Dutton2016} show that 
the inner slope of simulated DM haloes displays a minimum for stellar masses between $10^7$ and $10^{10} ~\rm M_\odot$ while it rises above the NFW slope when the stellar mass exceeds $10^{10}~\rm M_\odot$. 
This behaviour can be interpreted in terms of a competition between outflows induced by feedback and the confinement imposed by halo gravity \citep[e.g.,][]{Dekel1986, Penarrubia2012}: for very low stellar masses, the inner slope follows that of DM-only NFW haloes; between $10^7$ and $10^{10} ~\rm M_\odot$, outflows overcome halo gravity, leading to the expansion of the halo; above $10^{10}~\rm M_\odot$, the accumulation of baryons leads to adiabatic contraction,   
although the introduction of AGN feedback in simulations can partially counteract adiabatic contraction at high halo mass \citep{Maccio2020}. 
Hydrodynamical simulations of dwarf galaxies by \cite{Mashchenko2008}, \cite{Madau2014}, \cite{Verbeke2015}, \cite{Read2016}, and \cite{Dutton2016} further suggest that the main parameter driving the halo response is the stellar-to-halo mass ratio rather than  the stellar or halo mass itself. 
The different responses of the DM halo as well as the potentially smooth transition between cusps and cores motivates a parametrization of DM halo density profiles that would reflect the different halo shapes induced by baryonic physics or environment. 
In particular, a parametrization with free inner slope in addition to a free concentration parameter would enable to follow the transition between cusps and cores.

Different parametrizations allowing some inner slope flexibility have been proposed \citep{Einasto1965, Jaffe1983, Hernquist1990, Dehnen1993, Evans1994, Tremaine1994, Burkert1995, Zhao1996, Jing2000, Navarro2004, Stoehr2006, Merritt2006, An2013, DiCintio2014, Schaller2015, Oldham2016, Dekel2017}.
Amongst them, the \citetalias{Einasto1965} profile \citep{Einasto1965, Navarro2004,Mamon2010, Retana-Montenegro2012, An2013} with two free shape parameters provides excellent fits to DM cusps and analytic expressions for the mass and the gravitational potential (involving incomplete gamma functions for the potential) as well as for the surface density, the deflection angle and the deflection potential relevant for lensing studies \citep[involving Fox $H$ functions, cf. Eq.~(\protect\ref{eq:Fox}) below for their definition and ][]{Retana-Montenegro2012},
but does not seem to fully recover the innermost part of shallower density profiles \citep[][and Section \protect\ref{section:fitting}]{Dekel2017}. 
Modified NFW and Einasto profiles allowing constant-density cores have been proposed by \cite{Read2016} and \cite{Lazar2020}, but at the expense of analyticity (in particular, the analycity of the concentration).
The profile proposed by \cite{Dehnen1993} and \cite{Tremaine1994} has the particularity to have analytic expressions for the mass, the gravitational potential, and the velocity dispersion (in terms of elementary functions) and, in certain cases, for the distribution function and the surface density (in terms of elementary functions for some of the cases), but its unique shape parameter does not allow to recover the diversity of DM haloes. 
More generally, \citet[][hereafter \citetalias{Zhao1996}]{Zhao1996} shows that double power-law density profiles of the form 
\begin{equation}
\label{eq:rho_abc}
\rho(r) = \frac{\rho_c}{x^a (1+x^{1/b})^{b(g-a)}}
\end{equation}
where $x=r/r_c$, $r_c$ a characteristic radius, and $\rho_c$ a characteristic density, have analytic expressions for the gravitational potential, the enclosed mass, and the velocity dispersion (in terms of elementary functions) provided that $b=n$ and $g=3+k/n$, where $n$ and $k$ can be any natural numbers. 
Within this general Zhao family of profiles with four shape parameters ($a$, $b$, $g$, and the concentration $c=R_{\rm vir}/r_c$ associated to the characteristic radius), \citet[][hereafter \citetalias{Dekel2017}]{Dekel2017} show that the specific profile with $n=2$ and $k=1$, i.e., $b=2$ and $g=3.5$ in Eq.~(\ref{eq:rho_abc}), provides excellent fits for DM haloes in simulations with and without baryons, ranging from steep cusps to flat cores. This specific profile with two remaining shape parameters ($a$ and $c$), hereafter referred to as the Dekel-Zhao (DZ) profile, notably captures cores better than the \citetalias{Einasto1965} profile. 
In \citet[][hereafter \citetalias{Freundlich2020}]{Freundlich2020}, we accordingly used it to model the cusp-core transformation by outflow episodes induced by feedback, and further derived analytic expressions for the velocity dispersion in such DM halos with additional fiducial baryonic mass distributions (in terms of incomplete beta functions). 
We note that \cite{Zhao1997} provides analytic approximations for the distribution function and the projected line-of-sight velocity dispersion of this profile, while \citet[][hereafter \citetalias{An2013}]{An2013} offers a general parametrization of density profiles\footnote{The \citetalias{An2013} parametrisation is characterized by a logarithmic density slope
\be
\frac{d\ln \rho}{d\ln r} = -\frac{a+x^{1/b}}{1+s x^{1/b}},
\ee
which leads to Eq.~(\ref{eq:rho_abc}) with $g=s^{-1}$ for the density profile when $s>0$ and to the \citetalias{Einasto1965} density profile when $s=0$ (cf. their equations (5a) and (6a)). 
}
that includes both double power-law profiles (including the NFW and other profiles) and the \citetalias{Einasto1965} profile, with general analytic expressions for the gravitational potential, the enclosed mass, the velocity dispersion (in terms of incomplete beta and gamma functions), and the surface density (in terms of Fox $H$ functions).

Without being concerned by the non-analyticity of the potential and kinetic energy associated with most density profiles given by Eq.~(\ref{eq:rho_abc}), 
\cite{DiCintio2014} analyse a suite of hydrodynamical simulations to obtain functional forms for the shape parameters $a, b, g$ and the concentration parameter associated to $r_c$ as a function of the stellar-to-halo mass ratio $M_{\rm star}/M_{\rm vir}$ at redshift $z=0$. 
This enables them to define a mass-dependent density profile (hereafter \citetalias{DiCintio2014b}) for DM haloes, whose parameters are entirely set by the stellar and halo masses and which reflects the halo response to baryonic processes, since $M_{\rm star}/M_{\rm vir}$ represents an integrated star formation efficiency including the effects of feedback. 
The \citetalias{DiCintio2014b} profile not only enables to fit simulated DM distributions, but it is widely used to model observed rotation curves \citep[e.g.][]{Allaert2017, VanDokkum2019, Wasserman2019, Cautun2019} and at times to parametrize semi-analytical models of satellite evolution \citep[e.g.][]{Carleton2019}.
It however lacks analytic expressions for the gravitational potential, the velocity dispersion, and lensing properties such as the projected surface density and mass, the deflection angle and the magnification.

In the present article, we review the analytic properties of the DZ parametrization of DM density profiles, as established in \citetalias{Zhao1996}, \citetalias{An2013}, \citetalias{Dekel2017}, and \citetalias{Freundlich2020}, and further derive expressions for its lensing properties in terms of Fox $H$ functions and series expansions. 
We systematically test this parametrization in a large suite of cosmological hydrodynamical zoom-in simulations, compare it both to the \citetalias{Einasto1965} model and the generalized NFW model with variable inner slope, and obtain the dependences of its two shape parameters on stellar and halo mass. This enables us to establish it as a mass-dependent profile including the influence of baryons, whose accuracy is comparable to the  \citetalias{DiCintio2014b} profile but with the advantage of having analytic expressions for the gravitational potential and the velocity dispersion. We further give an integral expression for its associated isotropic distribution function.  
This model can be directly applied to model rotation curves for assessing halo masses, and also to gravitational lenses and semi-analytical models.

This article unfolds as follows: 
in Section~\ref{section:dekel}, we recall the analytic properties of the spherically-symmetric DZ profile, in particular its associated gravitational potential and velocity dispersion, and derive analytic expressions for its lensing properties; 
in Section~\ref{section:results}, we systematically test the profile in the NIHAO suite of hydrodynamical cosmological simulations \citep{Wang2015} and quantify the mass-dependence of its two free parameters, the inner logarithmic slope $s_1$ and the concentration $c_2$; 
in Section~\ref{section:mass}, we provide prescriptions to describe DM haloes given their stellar and halo masses and to model rotation curves with the DZ profile.

\section{Analytics}
\label{section:dekel}

\subsection{General case}
\label{section:dekel_general}

\subsubsection{Mean density profile}

To describe the transition from cusps to cores and alterations of the DM distribution due to environmental effects while enabling straightforward analytic expressions of the density, mass and circular velocity profiles of DM haloes, \citetalias{Dekel2017} proposed a functional form similar to Eq.~(\ref{eq:rho_abc}) for the \emph{mean} density profile within a sphere of radius $r$, 
\begin{equation}
\label{eq:rhob}
\overline{\rho}(r) = \frac{\overline{\rho_c}}{x^a (1+x^{1/b})^{b(\overline{g}-a)}},
\end{equation}
where $\overline{\rho_c}$ is a characteristic density, $x=r/r_c$ with $r_c=R_{\rm vir}/c$ an intermediate characteristic radius, $a$ and $\overline{g}$ the inner and outer asymptotic slopes, $b$ a middle shape parameter and $c$ a concentration parameter. 
The normalisation factor $\overline{\rho_c}$ can be expressed as 
$\overline{\rho_c}= c^3 \mu \overline{\rho_\vir}$, 
with 
$\mu = c^{a-3} (1+c^{1/b})^{b(\overline{g}-a)}$, 
and $\overline{\rho_{\rm vir}}=3\Mvir/4\pi \Rvir^3$ the mean mass density within $R_{\vir}$.
As the virial radius $R_{\vir}$ is set by cosmology for a given halo mass through $\overline{\rho_\vir}= \Delta \rho_{\rm crit}$ with $\Delta$ the overdensity, this functional form effectively depends on four shape parameters: $a$, $b$, $\overline{g}$ and $c$.

\subsubsection{Mass, velocity, force and density profiles}

The enclosed mass, circular velocity, and force profiles stemming from Eq.~(\ref{eq:rhob}) can be expressed as  
\be
\label{eq:M(r)}
M(r) = \frac{4\pi r^3}{3}\overline{\rho}(r) = \mu M_{\rm vir} x^3 \overline{\rho}(r)/\overline{\rho_c},
\ee
\be
\label{eq:V2(r)}
V^2(r) = \frac{GM(r)}{r} = c \mu V_{\rm vir}^2 x^2 \overline{\rho}(r)/\overline{\rho_c}
\ee
and 
\be
\label{eq:F(r)}
F(r) = -\frac{GM(r)}{r^2} = - c^2 \mu F_{\rm vir} x \overline{\rho}(r)/\overline{\rho_c}
\ee
where 
\mbox{$V_{\rm vir}^2 = G M_{\rm vir}/R_{\rm vir}$} and \mbox{$F_{\rm vir} = - G M_{\rm vir}/R_{\rm vir}^2$}. 
In turn, the density profile is obtained by derivating the expression of the enclosed mass:
\begin{equation}
\label{eq:rho}
\rho(r) = \frac{1}{4\pi r^2} \frac{dM}{dr} = \frac{3-a}{3}\left( 1+\frac{3-\overline{g}}{3-a} x^{1/b} \right) \frac{1}{1+x^{1/b}} \overline{\rho}(r).
\end{equation}
This expression reduces to Eq.~(\ref{eq:rho_abc}) when $\overline{g}=3$, with $g=3+1/b$ and $\rho_c=(1-a/3) \overline{\rho_c}$. 
More generally, each term of Eq.~(\ref{eq:rho}) is analogous to Eq.~(\ref{eq:rho_abc}) with $g=\overline{g}+1/b$ so the results of \citetalias{Zhao1996} apply: this density profile allows analytic expressions for the gravitational potential and the velocity dispersion provided that $b=n$ and $\overline{g}=3+k/n$, where $n$ is a natural number and $k$ a positive or null integer.

\subsubsection{Inner slope and concentration}

In the density profile derived from Eq.~(\ref{eq:rhob}), the shape parameter $a$ may not be the slope at the resolution limit \citep[$0.01 R_\vir$ in the case of the NIHAO simulations, cf.][]{Wang2015} and $c$ does not necessarily reflect the actual concentration of the halo as for an NFW profile. 
The logarithmic slope of the density profile expressed in Eq.~(\ref{eq:rho}) is 
\be
\label{eq:s}
s(r)=-\frac{d\ln \rho}{d \ln r}=\frac{a+(\overline{g}+b^{-1})x^{1/b}}{1+x^{1/b}}
-\frac{3-\overline{g}}{3-a}\frac{b^{-1}x^{1/b}}{1+\frac{3-\overline{g}}{3-a}x^{1/b}}, 
\ee
so $s_1=s(0.01 R_{\rm vir})$ measures the inner logarithmic slope at the resolution limit in the NIHAO simulations. 
This Eq.~(\ref{eq:s}) further enables to define a concentration parameter $c_{2}$ similar to the NFW parameter, corresponding to the radius $r_{2}$ at which the logarithmic slope $s$ of the density profile equals $2$. This radius is such that
\be
\label{eq:c-2}
c_{2} \equiv \frac{R_\vir}{r_{2}} = c \left( \frac{\overline{g}+b^{-1}-2}{2-a} \right)^b,
\ee
which coincides with $c$ when $a+\overline{g}+b^{-1}=4$. 
Another concentration parameter, $c_{\rm max}$, can be defined from the radius $r_{\rm max}$ at which the circular velocity peaks (cf. Appendix~\ref{appendix:cmax}). 
The logarithmic slope at the resolution limit ($s_1$ for the NIHAO simulations) and $c_{2}$ (or $c_{\rm max}$)  can be used as effective inner slope and concentration when describing the density profile.

\subsection{The Dekel-Zhao profile}
\label{section:dekel+}

\begin{figure*}
	\includegraphics[width=0.33\textwidth,trim={0.cm 0.cm 0.cm 1.6cm},clip]{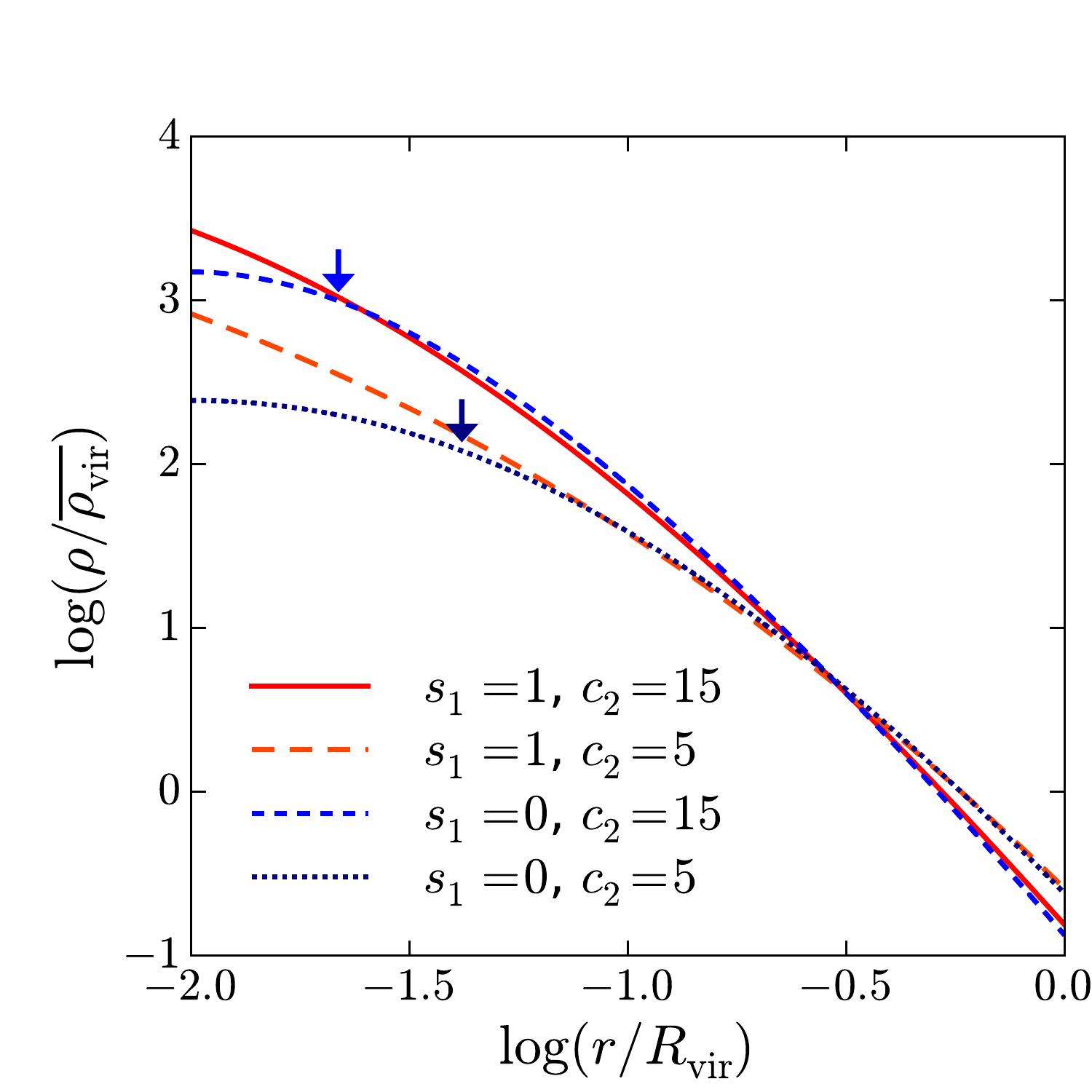}
	\includegraphics[width=0.33\textwidth,trim={0.cm 0.cm 0.cm 1.6cm},clip]{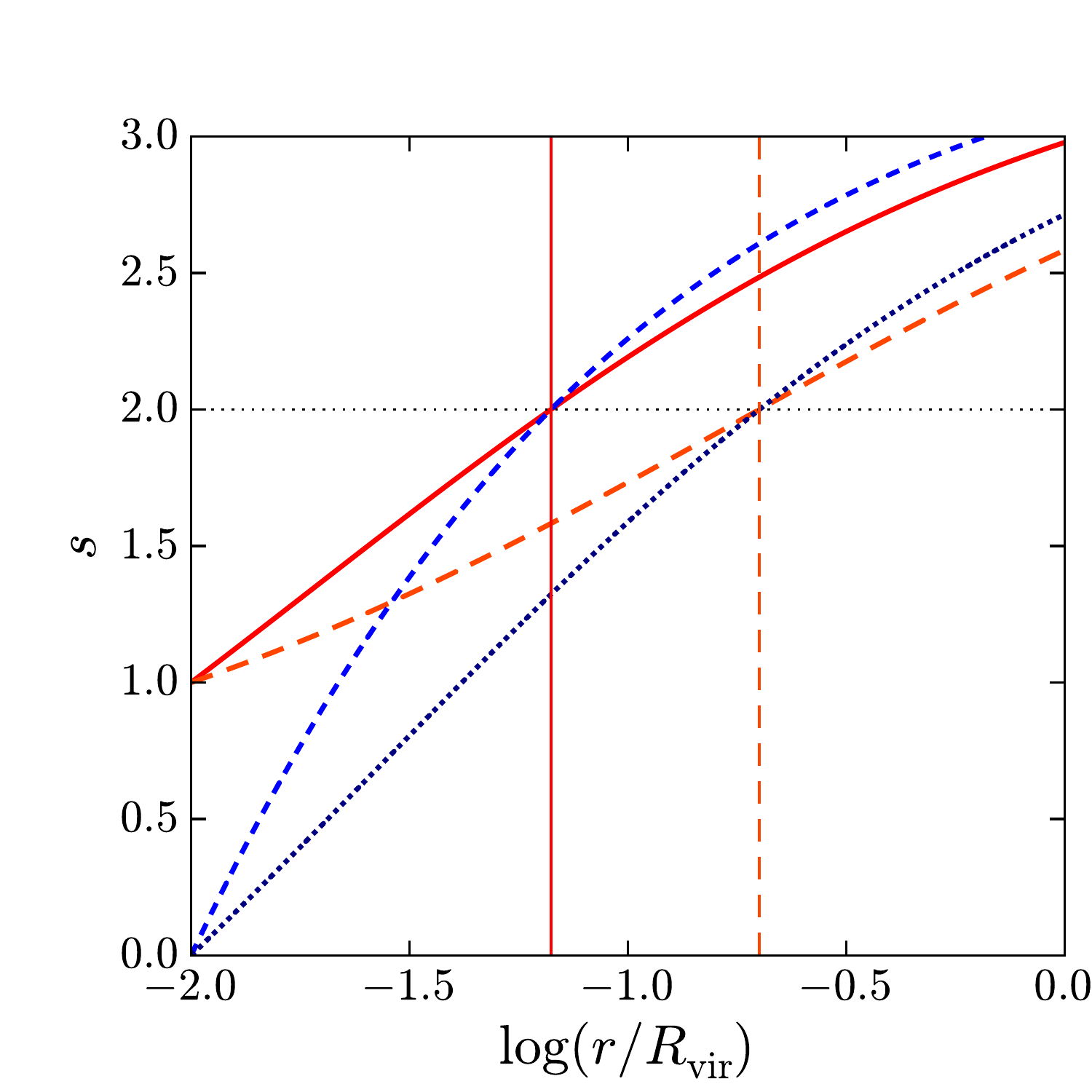}
	\includegraphics[width=0.33\textwidth,trim={0.cm 0.cm 0.cm 1.6cm},clip]{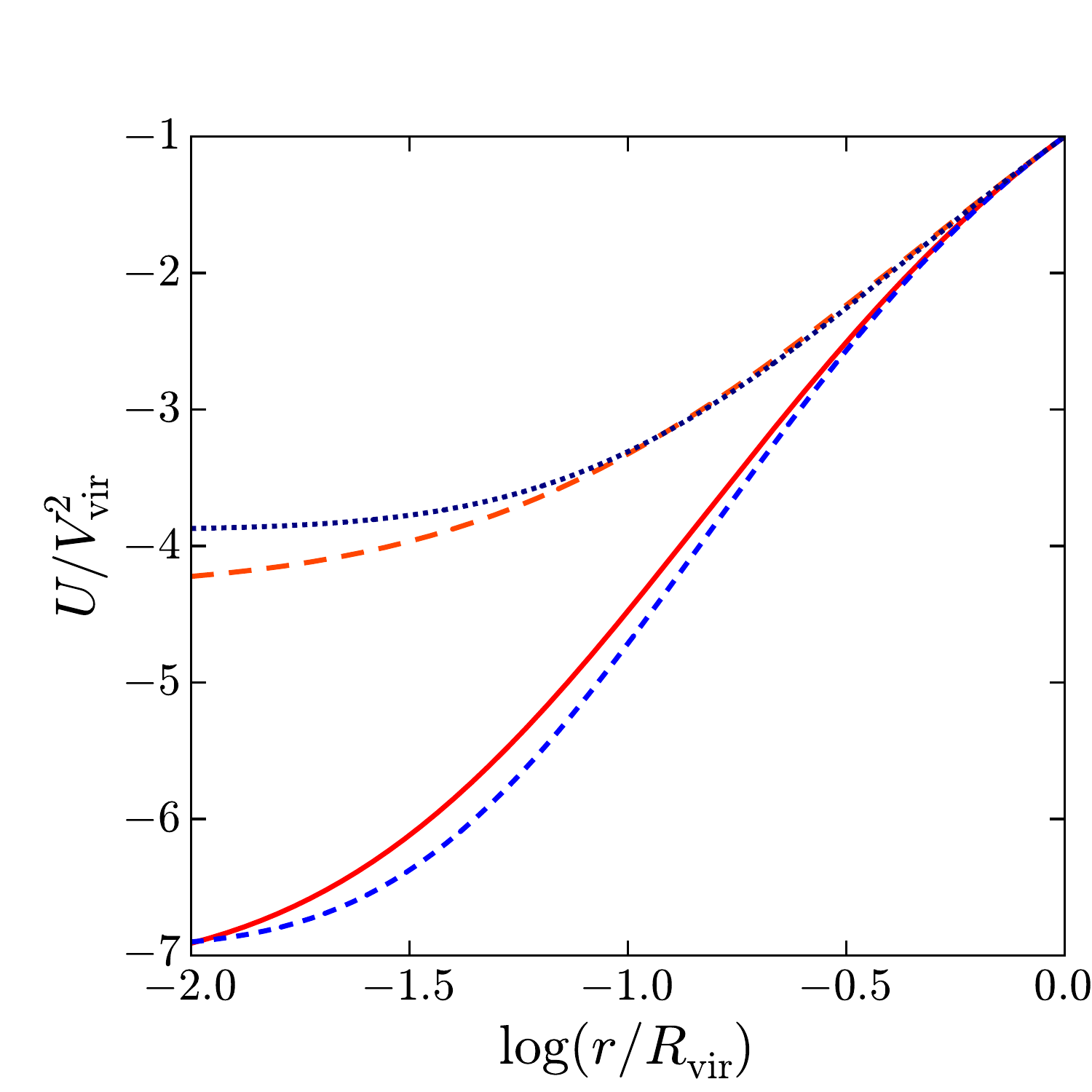}
	\\
	\includegraphics[width=0.33\textwidth,trim={0.cm 0.cm 0.cm 1.6cm},clip]{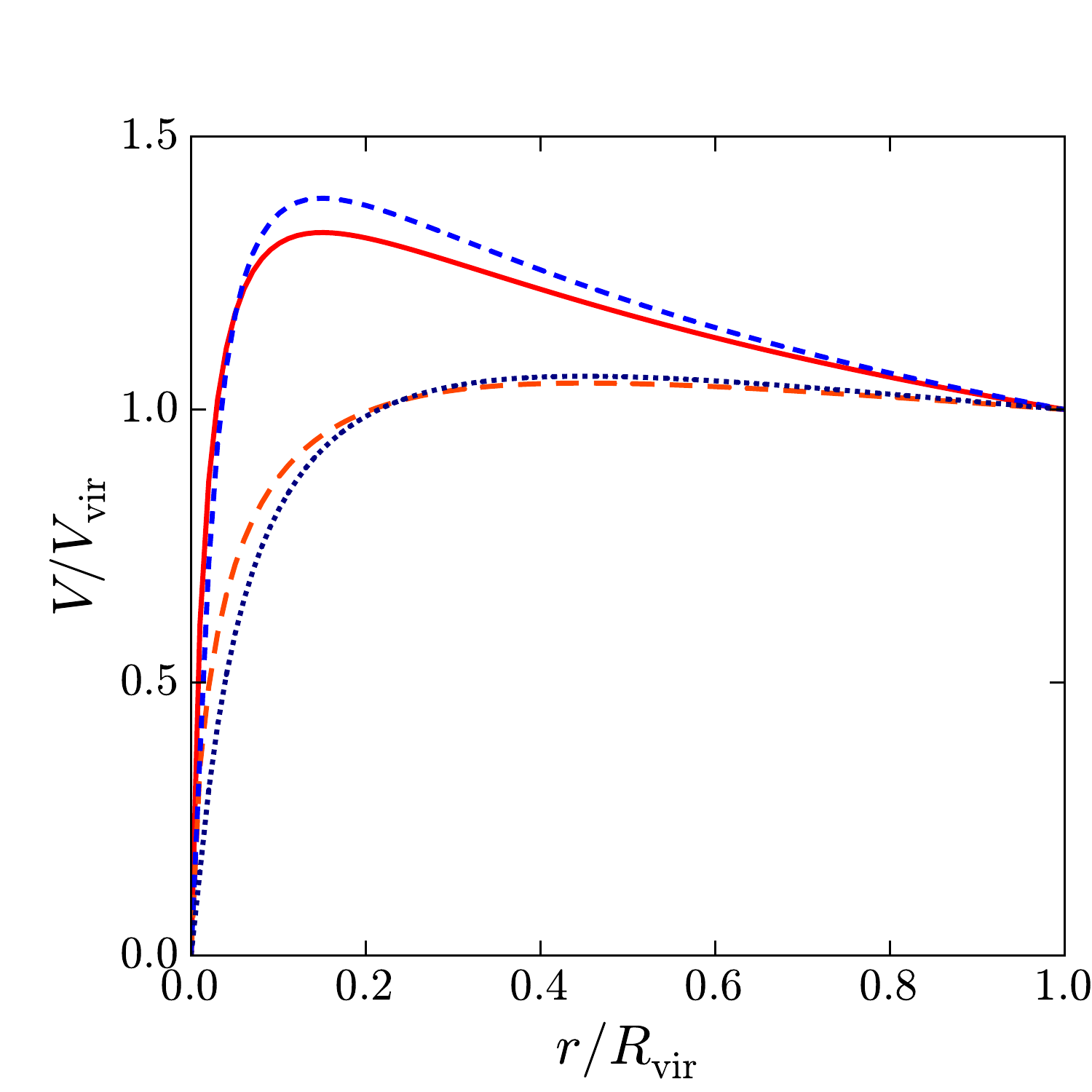}
	\includegraphics[width=0.33\textwidth,trim={0.cm 0.cm 0.cm 1.6cm},clip]{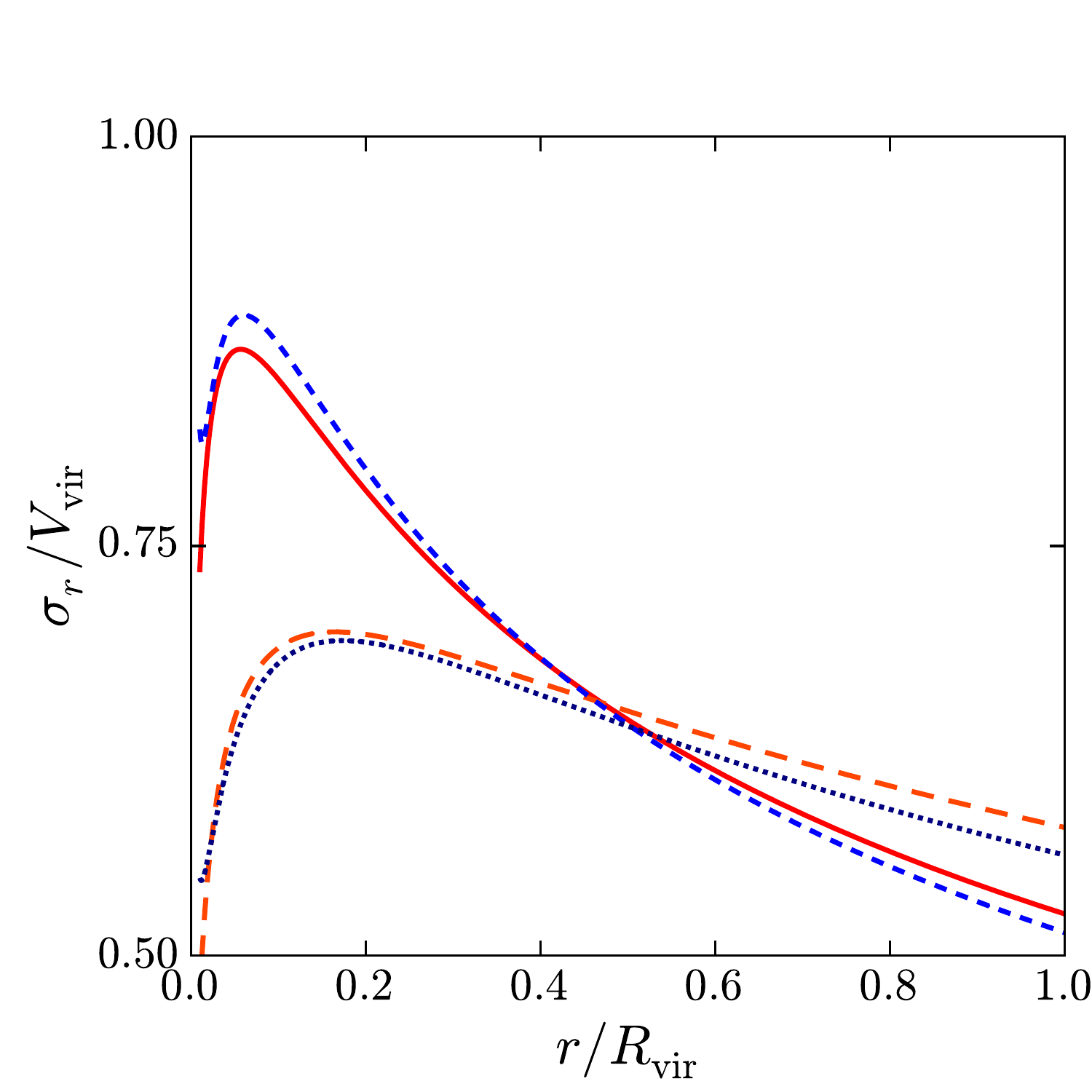}
	\includegraphics[width=0.33\textwidth,trim={0.cm 0.cm 0.cm 1.6cm},clip]{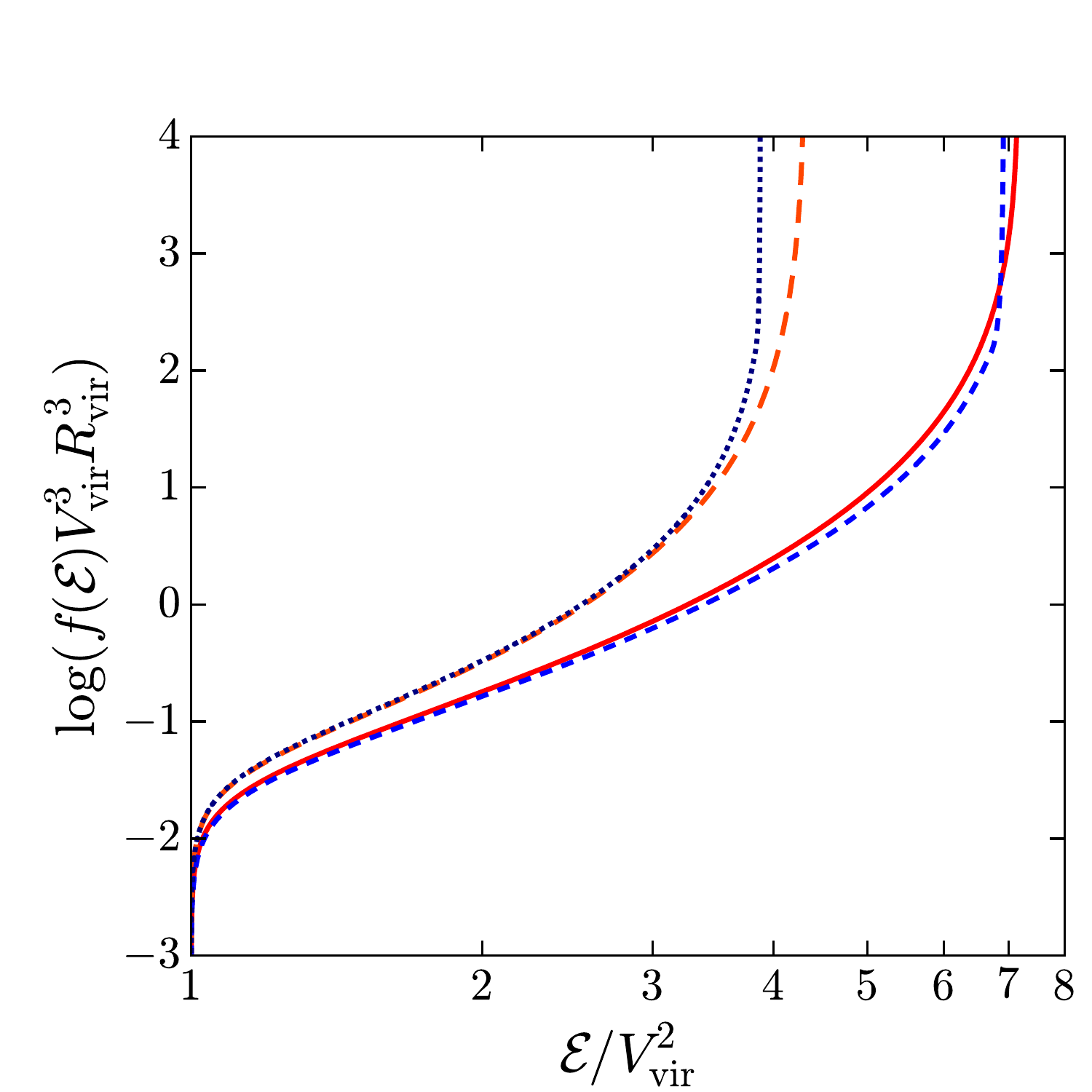}
	\vspace{-0.6cm}
	\caption{Fiducial DZ profiles: density ($\rho$), logarithmic slope ($s$), circular velocity ($V$), radial velocity dispersion ($\sigma_r$), and gravitational potential per unit mass ($U$) as a function of radius as well as the distribution function $f(\mathcal{E})$ associated to four DZ haloes truncated at the virial radius with different inner slope ($s_1=0$ or $1$) and concentration ($c_2=5$ or $15$). Eqs.~(\ref{eq:rho32}), (\ref{eq:s}), (\ref{eq:V2(r)}), (\ref{eq:sigma_dekel}), (\ref{eq:U32_1}) respectively provide analytic expressions for the radial profiles, while the distribution function is obtained by numerically integrating Eq.~(\ref{eq:fE_int2}). 
	Arrows in the upper left panel indicate for the two cored profiles the core radii defined by Eq.~(\ref{eq:rcore}) with $s_{\rm core}=1$.
	The vertical lines in the upper middle panel highlight the radius $r_2 = \Rvir/c_2$ where $s=2$. 
	Dimensional quantities are in virial units, with $\overline{\rho_{\rm vir}}=3\Mvir/4\pi \Rvir^3$ and $V_{\rm vir} = \sqrt{G\Mvir/\Rvir}$. 
	The DZ profile enables to capture a variety of DM density profiles from cusps to cores with different concentrations. 
	The concentration $c_2$ sets the depth of the potential well and hence the behaviours of the circular velocity, velocity dispersion, and distribution function. 
	}
	\label{fig:examples_dekel}
\end{figure*}

Using three pairs of simulated haloes at different masses with and without baryons at $z=0$ from the NIHAO suite of simulations \citep{Wang2015}, \citetalias{Dekel2017} show that the functional form of Eq.~(\ref{eq:rho}) with $b=2$ and $\overline{g}=3$ yields excellent fits for haloes ranging from steep cusps to flat cores. They notably show that this parametrization, here referred to as the Dekel-Zhao (DZ) profile, matches simulated profiles better than the NFW and Einasto profiles, capturing cores better, in addition to providing fully analytic expressions for the density, the mass, the gravitational potential, and the velocity dispersion. 
We further show in \citetalias{Freundlich2020} that density profile fits using this parametrization enable to recover the simulated gravitational potentials and the velocity dispersions of simulated haloes. 
The upper left panel of Fig.~\ref{fig:examples_dekel} highlights the variety of density profiles from cusps to cores that can be described by the DZ profile, with four examples of different inner slope ($s_1=0$ and $1$) and concentration ($c_2=5$ and $15$). These fiducial examples correspond to different rotation curves, velocity dispersions, gravitational potentials and distribution functions.

In the following subsections, we recall the analytic expressions of the gravitational potential and velocity dispersion.  In Section~\ref{section:lensing}, we obtain analytic expressions for quantities relevant to gravitational lensing. 
In Appendix~\ref{appendix:cmax}, we further express the DZ profile in terms of $r_{\rm max}$ and $V_{\rm max}$, which can notably be useful to describe satellite haloes \citep[e.g.,][]{Jiang2020}. 
In Appendices~\ref{appendix:sigmar_sum} and \ref{appendix:additional}, we recall sum expressions for the velocity dispersion obtained by \citetalias{Zhao1996} and \citetalias{Freundlich2020}, which enable to express this quantity in terms of elementary functions, as well as expressions for the velocity dispersion in haloes with fiducial baryonic components from \citetalias{Freundlich2020}. 
In Appendix~\ref{appendix:DF}, we give an integral expression of the distribution function.
Finally, in the next Sections~\ref{section:results} and \ref{section:mass}, we  test the DZ profile over the whole NIHAO suite of simulations and establish it as a mass-dependant profile whose shape parameters $s_1$ and $c_2$ only depend on the stellar-to-halo mass ratio.

\subsubsection{Shape parameters}
\label{section:shape}

Introducing $\overline{g}=3$ and $b=2$ in Eq.~(\ref{eq:rho}), the DZ density profile  is
\begin{equation}
\label{eq:rho32}
\rho(r)  =  \frac{\rho_c}{x^a (1+x^{1/2})^{2(3.5-a)}} 
\end{equation}
with 
$x=r/r_c$, 
$\rho_c=(1-a/3)\overline{\rho_c}$ while
$\overline{\rho_c}= c^3 \mu \overline{\rho_\vir}$, $\mu = c^{a-3} (1+c^{1/2})^{2(3-a)}$ and $\overline{\rho_{\rm vir}}=3\Mvir/4\pi \Rvir^3$, 
and two shape parameters $a$ and $c=\Rvir/r_c$. 
The inner logarithmic slope $s_1$ at the resolution $r_1$ from Eq.~(\ref{eq:s}) is 
\be
\label{eq:s1_23}
s_1=\frac{a+3.5 c^{1/2}(r_1/\Rvir)^{1/2}}{1+c^{1/2}(r_1/\Rvir)^{1/2}}, 
\ee
while the concentration parameters is 
\be
\label{eq:c2_23}
c_{2}
=c\left(\frac{1.5}{2-a}\right)^2. 
\ee
A positive density imposes $a\leq 3$, a positive inner logarithmic slope $a+3.5c^{1/2}(r_1/\Rvir)^{1/2}\geq 0$: 
negative values of $a$ can be compatible with a positive logarithmic slope at the resolution limit, in particular for large values of $c$. 
Since the logarithmic slope tends to $a$ when the radius goes to zero, $c_2$ is only defined when $a\leq 2$.

There are bijections between the couples $(a,c)$ and $(s_{\rm 1}, c_{\rm 2})$ (and $(s_{\rm 1}, c_{\rm max})$, cf. Appendix~\ref{appendix:cmax}) so these couples are equivalent in describing the density profile. 
Indeed, $a$ and $c$ can be expressed  as functions of $s_1$ and $c_{2}$, 
\be
\label{eq:a(s1,c2)}
a = \frac{1.5 s_1 -2 \left( 3.5-s_1\right)\left(r_1/R_\vir\right)^{1/2}c_{2}^{1/2} }{1.5 - \left(3.5-s_1\right)\left(r_1/R_\vir\right)^{1/2}c_{2}^{1/2} }
\ee
and
\be
\label{eq:c(s1,c2)}
c= \left( \frac{s_1-2}{\left(3.5-s_1\right)\left(r_1/R_\vir \right)^{1/2} -1.5 c_{2}^{-1/2}} \right)^2. 
\ee
In the following, analytic expressions are expressed in terms of ($a$, $c$) while numerical tests focus on ($s_1$, $c_2$). Eqs.~(\ref{eq:s1_23}), (\ref{eq:c2_23}), (\ref{eq:a(s1,c2)}), and (\ref{eq:c(s1,c2)}) enable to switch from the two couples of parameters at will. 

It is further possible to define a core radius $r_{\rm core}$ corresponding to a given value of the logarithmic slope, namely
	\be
	\label{eq:rcore}
	r_{\rm core} = \frac{R_\vir}{c} \left(\frac{s_{\rm core}-a}{3.5-s_{\rm core}}\right)^2
	\ee 
with $s_{\rm core}=s(r_{\rm core})$. Since the logarithmic slope $s$ is an increasing function of radius with $s(r=0)=a$, this equation is only valid when $s_{\rm core}\geq a$. 
We find that $s_{\rm core}=1$ enables to retrieve a radius close to what one's eye identifies as a core (cf. Fig.~\ref{fig:examples_dekel}). This value also corresponds to the slope at the core radius of a pseudo-isothermal halo. Moreover, we note from Fig.~\ref{fig:mass-dependence-rho} below that $s_1=1$ lies right below the $1\sigma$ scatter of the inner slope $s_1$ at low mass and hence marks the threshold below which core formation occurs. By analogy with the \cite{Burkert1995} and ``Lucky13'' \citep{Li2020} cored profiles, one could also choose $s_{\rm core}=1.5$. We point out that the slopes at the core radii of the ``core-NFW'' \citep{Read2016} and ``core-Einasto'' \citep{Lazar2020} profiles are not fixed to a specific value. 
At given $a$ and $c$, the core radii from Eq.~(\ref{eq:rcore}) defined at different $s_{\rm core}$ can be related to one another through constant factors depending only on $a$.

Eq.~(\ref{eq:M(r)}) also enables to express the half-mass radius, or more generally the radius 
\be
\label{eq:rf}
r_f = \frac{R_\vir}{c} \left(\left(\frac{\mu}{f}\right)^{1/(6-2a)} -1\right)^{-2}
\ee
enclosing a DM mass $M(r_f)=f M_{\rm vir}$. The half-mass radius of a DZ halo truncated at the virial radius corresponds to  $f=0.5$ in this equation. 
We stress that neither the cuspy NFW profile, nor the cored pseudo-isothermal, \cite{Burkert1995}, and ``Lucky13'' \citep{Li2020} profiles, nor the \citetalias{Einasto1965}, ``core-Einasto'' \citep{Lazar2020}, ``core-NFW'' \citep{Read2016}, and generalized NFW profiles with flexible inner slope have analytic expressions for the half-mass radius and therefore $r_f$ (cf. also the table of Fig.~\ref{fig:profiles_compare}).

\subsubsection{Gravitational potential}
\label{section:potential_velocity}

The mass, circular velocity, force and logarithmic slope profiles of the DZ profile can be expressed analytically from Eqs.~(\ref{eq:M(r)}), (\ref{eq:V2(r)}), (\ref{eq:F(r)}), and (\ref{eq:s}) with $b=2$ and $\overline{g}=3$. 
Its density (Eq.~(\ref{eq:rho32})) follows the form of Eq.~(\ref{eq:rho_abc}) with $b=2$ and $g=3+1/2$ so the DZ profile also allows analytic expressions for the gravitational potential and the velocity dispersion (\citetalias{Zhao1996}).

Assuming that the gravitational potential vanishes at infinity and that the halo density profile is truncated at the virial radius yields the gravitational potential per unit mass 
\footnote{We use the variable change $\zeta = z^{1/2}/(1+z^{1/2})$ with $z=y/r_c$, which is such that $z^{1/2}=\zeta/(1-\zeta)$, $1+z^{1/2}=1/(1-\zeta)$, and $dz=2\zeta (1-\zeta)^{-3}d\zeta$.}
\begin{align} 
\label{eq:U32}
\displaystyle U(r)  
&=\!-\frac{G\Mvir}{\Rvir}\!-\!\!\!\int_r^{\Rvir}\! \frac{GM(y)}{y^2}dy
=\! -\!V_\vir^2\!~\Bigg( 1\!+\! 2c\mu \!\!\int_\chi^{\chi_c} \zeta^{3-2a}(1\!-\!\zeta) d\zeta \Bigg)
\end{align} 
within the virial radius, with $V_\vir^2 = GM_\vir/R_\vir$, $x=r/r_c$, \mbox{$\chi=x^{1/2}/(1+x^{1/2})$}, and $\chi_c = c^{1/2}/(1+c^{1/2})$. 
When $a\neq 2$ and $a\neq5/2$, this yields
\footnote{If $a=2$, it instead yields 
$U(r) =-V_\vir^2~(1 + 2c\mu [ \ln(\chi_c/\chi)+\chi-\chi_c])$
and if $a=5/2$, 
$U(r) =-V_\vir^2~(1 + 2c\mu [ 1/\chi-1/\chi_c-\ln(\chi_c/\chi) ])$
but such specific rational values of $a$ are unlikely to arise from fits. } 
\begin{equation}
\label{eq:U32_1}
\medmuskip=0.5mu
\thinmuskip=0.5mu
\thickmuskip=0.5mu
\displaystyle  U(r) =-V_\vir^2~\Bigg(1 + 2c\mu \left[ \frac{\chi_c^{2(2-a)}-\chi^{2(2-a)}}{2(2-a)} - \frac{\chi_c^{2(2-a)+1}-\chi^{2(2-a)+1}}{2(2-a)+1} \right]\Bigg). 
\end{equation}
As noted in \cite{Zhao1997} and \citetalias{An2013}, Eq.~(\ref{eq:U32}) and hence Eq.~(\ref{eq:U32_1}) can be rewritten in terms of incomplete beta functions (cf. also Appendix \ref{appendix:power-law}).

\subsubsection{Velocity dispersion}

The equilibrium of a spherical collisionless system can be described by the 
spherical Jeans equation stemming from the Boltzmann equation \citep[][Eq.~(4.215)]{BinneyTremaine2008}, which yields the radial velocity dispersion 
\be
\label{eq:sigmar2-0}
\sigma_r^2 (r)
= \frac{G}{\rho(r)}\int_{r}^{\Rvir}\rho(\rp)M(\rp)r^{-2}\rmd\rp
\ee 
for a halo truncated at the virial radius when the anisotropy parameter $\beta\equiv 1-\sigma_t^2/2\sigma_r^2$, where $\sigma_t$ is the tangential velocity dispersion, is null (isotropic case) and the boundary condition is $\lim_{r\rightarrow +\infty} \sigma_r^2 = 0$.  
For a DZ density profile as in Eq.~(\ref{eq:rho32}), this leads to 
\be
\label{eq:sigma_dekel_int}
\sigma_r^2(r)
= 2 c\mu \frac{GM_{\vir}}{R_{\vir}} \frac{{\rho_c}}{\rho(r)}
\int_\chi^{\chi_c}\zeta^{3-4a} (1-\zeta)^8 \rmd \zeta, 
\ee
or
\be
\label{eq:sigma_dekel}
\sigma_r^2(r)
= 2 c\mu \frac{GM_{\vir}}{R_{\vir}} \frac{{\rho_c}}{\rho(r)}
\Big[ \mathcal{B}(4-4a,9,\zeta)\Big]_\chi^{\chi_c}
\ee
where $\mathcal{B}(a,b,x) = \int_0^x t^{a-1}(1-t)^{b-1} dt$ is the incomplete beta function and the brackets denote the difference of the enclosed function between 1 and $\chi$, i.e., $\left[ f(\zeta)\right]_\chi^{\chi_c} \equiv f(\chi_c)-f(\chi)$.
We extend here the definition of the incomplete beta function appearing inside the brackets to negative parameters since the integral of Eq.~(\ref{eq:sigma_dekel_int}) is well-defined as long as $\chi>0$ such that the bracketted term is also well-defined. 
This equation is a specific case of Eq.~(B6) of \citetalias{An2013}, and it can further be expressed in terms of finite sums (\citetalias{Zhao1996}, \citetalias{Freundlich2020}), as recalled in the present Appendix~\ref{appendix:sigmar_sum}. The sum expressions enable to express the velocity dispersion in terms of elementary functions.

In Appendix \ref{appendix:additional}, we further recall expressions from Appendix B of \citetalias{Freundlich2020} for the velocity dispersion in haloes with baryons (i) where the ratio between the DM and the total masses follows a power-law, (ii) where the baryons are concentrated to a central point mass, (iii) where they  constitute a uniform sphere, (iv) where they constitute a singular isothermal sphere, and (v) where they themselves follow the DZ profile.

When the anisotropy parameter $\beta$ is constant but not necessarily equal to zero, the Jeans equation corresponds to a differential equation in $\rho \sigma_r^2$ whose solution is 
\be
\label{eq:vrsqr}
\sigma_r^2 (r) = \frac{G}{r^{2\beta}\rho(r)}\int_{r}^{\Rvir}\rho(\rp)M(\rp)\rp{}^{2\beta-2}\rmd\rp
\ee
assuming that $\lim_{r\rightarrow +\infty} \sigma_r^2 = 0$ \citep[][Eq.~(4.216)]{BinneyTremaine2008}. 
Following similar steps as for Eq.~(\ref{eq:sigma_dekel}), this leads to 
\be
\label{eq:sigma_dekel_beta}
\sigma_r^2(r) = 2 c\mu \frac{GM_{\vir}}{R_{\vir}} \frac{{\rho_c}}{\rho(r)}\frac{1}{x^{2\beta}}
\Big[ \mathcal{B}(4-4a+4\beta,9+4\beta,\zeta)\Big]_\chi^{\chi_c}.
\ee

\subsection{Lensing properties}
\label{section:lensing}

\subsubsection{Surface density}

The mass surface density of a spherically-symmetric lens is obtained by integrating the three-dimensional density profile along the line of sight, 
\be
\label{eq:Sigma}
\Sigma(R) = \int_{-\infty}^{+\infty} \rho(r)~dz
\ee
where $R$ is the projected radius measured from the center of the lens and $r=\sqrt{R^2+z^2}$ is the three-dimensional radius. This expression can be written as the Abel transform 
\be
\label{eq:Sigma_Abel}
\Sigma(R) = 2 \int_R^{+\infty} \frac{\rho(r) r dr}{\sqrt{r^2-R^2}},
\ee
which yields 
\be
\label{eq:Sigma_dekel}
\Sigma(X)=2\rho_c r_c \int_{X}^{c} \frac{xdx}{x^a(1+x^{1/2})^{2(3.5-a)} \sqrt{x^2-X^2}}
\ee
with $X=R/r_c$ and $c=\Rvir/r_c$ for a DZ density profile truncated at the virial radius. 
This integral can be broken into two terms such that $\Sigma(X)=\widetilde{\Sigma}(X)-\widetilde{\Sigma}(c)$ with 
\be
\label{eq:Sigma_dekel_f}
\widetilde{\Sigma}(X)=2\rho_c r_c \int_{X}^{\infty} \frac{xdx}{x^a(1+x^{1/2})^{2(3.5-a)} \sqrt{x^2-X^2}}
\ee
the surface density associated with an untrucated DZ profile.
When $a<1$, this expression yields at the center
\be
\label{eq:Sigma_0}
\widetilde{\Sigma}(0) = 4\rho_c r_c \mathcal{B}(2-2a,5) 
\ee
with the variable change used to obtain Eqs.~(\ref{eq:U32_1}) and (\ref{eq:sigma_dekel}).
However, the integral can not be easily expressed in terms of elementary functions for all values of $a$ when $X \neq 0$. 
Following \cite{Mazure2002}, \cite{Baes2011a}, \cite{Baes2011b} and \cite{Retana-Montenegro2012}, who expressed similar integrals involving S\'ersic and Einasto profiles in terms of the Meijer $G$ and Fox $H$ functions, we use the Mellin transform method \citep{Marichev1985, Adamchick1996, Fikioris2007} to evaluate it 
as the Mellin-Barnes integral
\be
\label{eq:Sigma_int}
\widetilde{\Sigma}(R)  = 4 \sqrt{\pi} \rho_c r_c \frac{X}{2\pi i} \int_{\mathcal{L}} \frac{\Gamma(4y\!-\!2a) \Gamma(7\!-\!4y)}{\Gamma(7\!-\!2a)} \frac{\Gamma(y\!-\!\frac{1}{2})}{\Gamma(y)} \left[ X^2\right]^{-y} dy
\ee
where $\mathcal{L}$ is a vertical line in the complex plane (cf. Appendix \ref{appendix:mellin}). 
This integral can be recognized as a Fox $H$ function \citep[e.g., ][]{Fox1961, Mathai1978, Srivastava1982, Kilbas1999, Kilbas2004, Mathai2009}, which is generally defined as the inverse Mellin transform of a product of gamma functions, 
\be
\label{eq:Fox}
H_{p,q}^{m,n} 
\left[
\left.
\!\!\!\!\!
\begin{array}{c}
	(\mathbf{a}, \!\mathbf{A})\\
	(\mathbf{b}, \!\mathbf{B})
\end{array}
\!\!\!
\right|
z
\right]
\!
= 
\!
\frac{1}{2\pi i} \int_{\mathcal{L}}
\!\!
\frac{\Pi_{j=1}^m \Gamma(b_j\!+\!B_j y) \Pi_{j=1}^n \Gamma(1\!-\!a_j\!-\!A_j y)}{\Pi_{j=m+1}^q \Gamma(1\!-\!b_j\!-\!B_j y) \Pi_{j=n+1}^p\Gamma(a_j\!+\!A_j y)} z^{-y} dy
\ee 
where the couples $(\mathbf{a}, \!\mathbf{A})$ and $(\mathbf{b}, \!\mathbf{B})$ indicate the coefficients in the gamma functions with $A_j, B_j>0$ and $a_j,b_j$ complex numbers while $0\leq m\leq q$ and $0\leq n \leq p$ are integers. With this definition, the surface density associated with the untruncated DZ profile can be compactly written as
\be
\label{eq:Sigma_H}
\widetilde{\Sigma}(X) = \frac{4 \sqrt{\pi} \rho_c r_c}{\Gamma(7-2a)} X~
H_{2,2}^{2,1} 
\left[
\left.
\!\!\!\!
\begin{array}{c}
	(-6,4), (0,1)\\
	(-\frac{1}{2},1), (-2a,4)
\end{array}
\!\!
\right|
X^2
\right]. 
\ee 
This expression has explicit series expansions depending on the nature of the poles of the gamma functions at the denominator of the integrand of the Mellin-Barnes integral \citep[e.g.,][]{Kilbas1999, Baes2011b}, which are given in Appendix~\ref{appendix:lensing}. 
We note that Eq.~(\ref{eq:Sigma_H}) is a specific case of Eq.~(C1) of \citetalias{An2013}, which includes both other double power-law profiles and the \citetalias{Einasto1965} profile, and that \citetalias{An2013} also provide analytic expressions for the limiting behaviours of this surface density when $X \rightarrow 0$ and $X \rightarrow \infty$ in terms of elementary functions.

The cumulative mass contained within an infinite cylinder of radius $R$ is 
\be
\label{eq:Mcum}
\widetilde{\mathcal{M}}(R) = 2\pi \int_0^R \widetilde{\Sigma}(R^\prime) R^\prime dR^\prime
\ee
for an untruncated DZ profile and $\mathcal{M}(R)=\widetilde{\mathcal{M}}(R)-\pi R^2 \widetilde{\Sigma}(c)$ for a DZ profile truncated at the virial radius. 
Injecting Eq.~(\ref{eq:Sigma_int}) and inverting the two integrals involved yields 
\begin{align}
\label{eq:M2D_H}
\widetilde{\mathcal{M}}(X) & = 
\frac{4 \pi^{3/2} \rho_c r_c^3}{\Gamma(7\!-\!2a)} X^3~
H_{3,3}^{2,2} 
\left[
\left.
\!\!\!\!
\begin{array}{c}
(-6,4), (-\frac{1}{2},1), (0,1)\\
(-\frac{1}{2},1), (-2a,4), (-\frac{3}{2}, 1)
\end{array}
\!\!
\right|
X^2
\right], 
\end{align}
which also has an explicit series expansion (Appendix~\ref{appendix:lensing}).

\subsubsection{Deflection angle}

A gravitational lens deflects light from background sources depending on their projected distance $R$ in the lens plane. The deflection angle $\hat{\alpha}(R)$ of a thin axially-symmetric lens where the distances between the source, the lens, and the observer are much larger than the size of the lens is directly related to its cumulative mass $\mathcal{M}(R)$ through
\be
\label{eq:lens_alpha}
\hat{\alpha}(R) = \frac{4 G\mathcal{M}(R)}{c^2 R}
\ee
\citep[][Eq. (8.5)]{Schneider1992}, $c$ being here the speed of light. 
Introducing $D_{\rm L}$, $D_{\rm S}$, and $D_{\rm LS}$ the angular distances respectively between the observer and the lens, between the observer and the source, and between the lens and the source, one can express the scaled deflection angle
\be
\alpha(R)\equiv\frac{D_{\rm L}D_{\rm LS}}{r_c D_{\rm S}} \hat{\alpha}(R)
\ee
and the convergence
\be
\label{eq:lensing_kappa}
\kappa(R) \equiv \frac{{\Sigma}(R)}{\Sigma_{\rm crit}}
\ee
where distances in the lens plane are scaled in units of $r_c$ and $\Sigma_{\rm crit}= c^2 D_{\rm S}/4\pi G D_{\rm L} D_{\rm LS}$ is the lensing critical surface density. 
Introducing
$\widetilde{\kappa_0} \equiv \widetilde{\Sigma}(0)/\Sigma_{\rm crit}$ 
with Eq.~(\ref{eq:Sigma_0}) and $\mathcal{B}(a,b)=\Gamma(a)\Gamma(b)/\Gamma(a+b)$, 
the scaled deflection angle for an untruncated DZ profile yields
\be
\label{eq:lensing_alpha_H}
\widetilde{\alpha}(X) = \frac{\sqrt{\pi} ~\widetilde{\kappa_0} }{\Gamma(2-2a)\Gamma(5)} X^2~
H_{3,3}^{2,2} 
\left[
\left.
\!\!\!\!
\begin{array}{c}
	(-6,4), (-\frac{1}{2},1), (0,1)\\
	(-\frac{1}{2},1), (-2a,4), (-\frac{3}{2}, 1)
\end{array}
\!\!
\right|
X^2
\right], 
\ee 
which has a series expansion analogous to that of $\widetilde{\mathcal{M}}(X)$. 
For a DZ profile truncated at the virial radius, 
$\alpha(X) = \widetilde{\alpha}(X)-X \widetilde{\Sigma}(c)/ \Sigma_{\rm crit}$. 
We give analytic expressions for the lensing potential in Appendix~\ref{appendix:lensing}.

For an axially-symmetric lens, multiple images occur if and only if the central convergence $\kappa_0\equiv\Sigma(0)/\Sigma{\rm crit}>1$ when the surface density does not increases with $X$, while there is only one image when $\kappa_0\leq 1$ \citep[][Section 8]{Schneider1992}. 
If $a \geq 1$, the DZ profile has a singular surface density at the center and there can be multiple images for all masses. 
However if $a < 1$, the surface density is not singular and there can be multiple images only if $\kappa_0> 1$.

\begin{figure*}
	\includegraphics[width=0.33\textwidth,trim={0.cm 0.1cm 0.cm 0.1cm},clip]{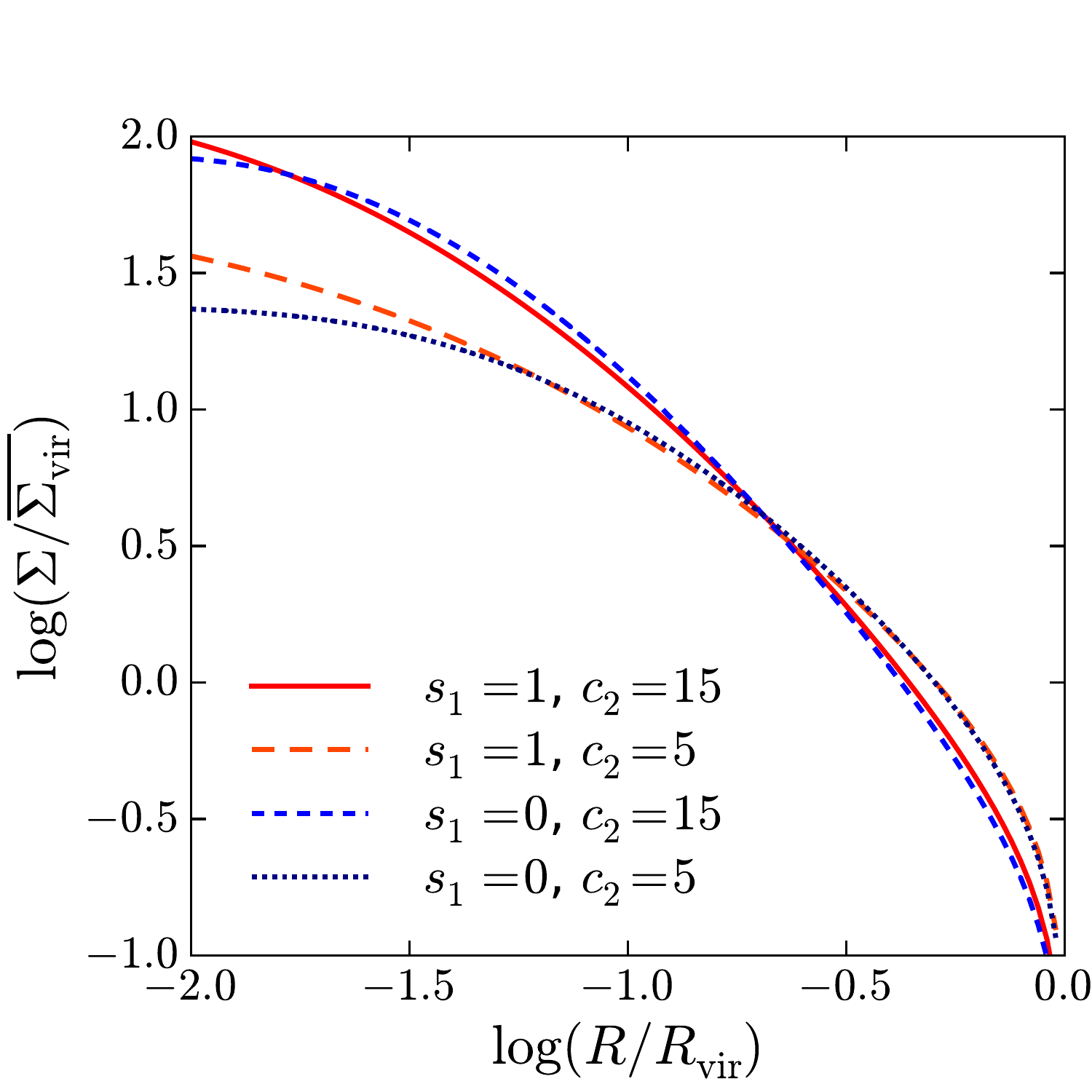}
	\includegraphics[width=0.33\textwidth,trim={0.cm 0.1cm 0.cm 0.1cm},clip]{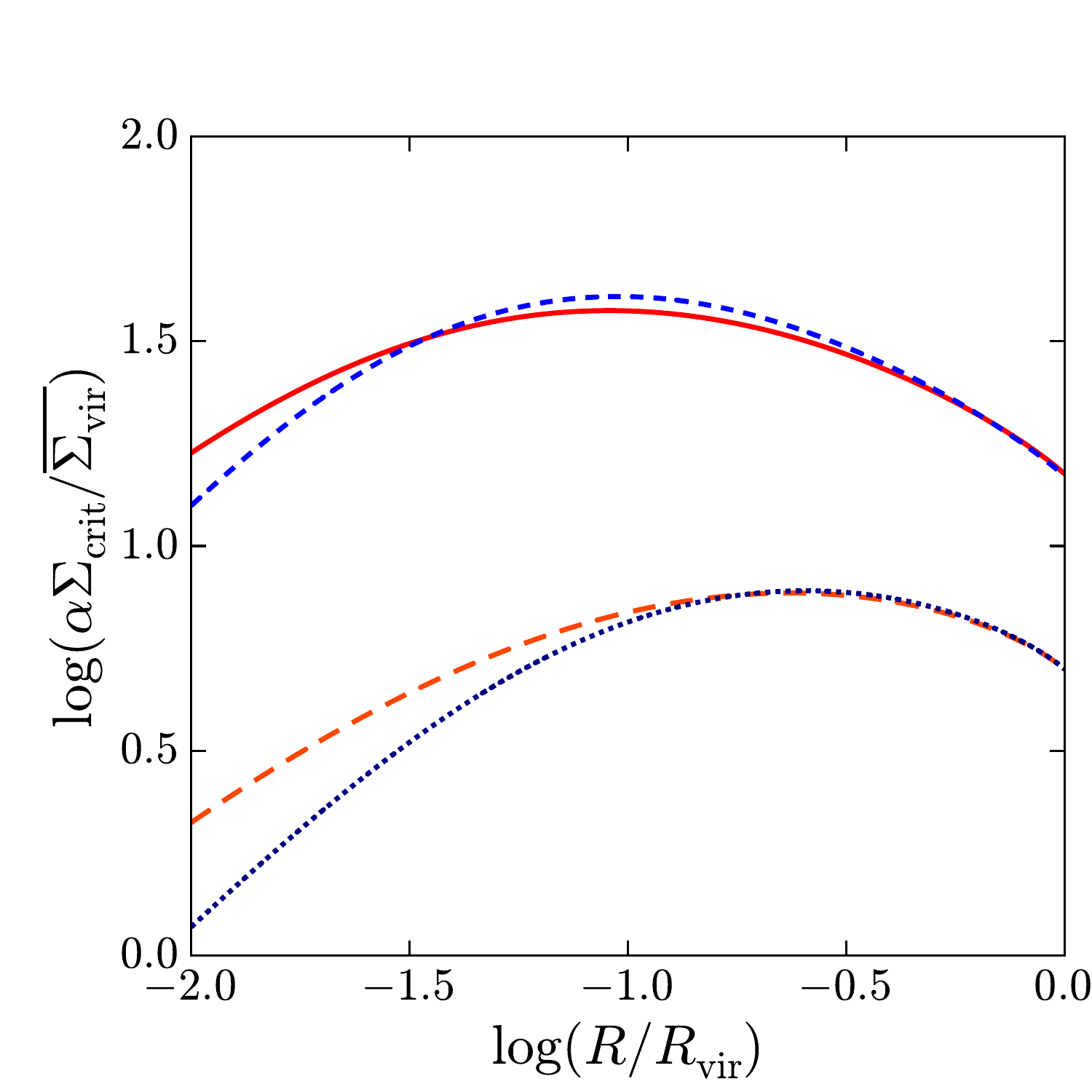}
	\includegraphics[width=0.33\textwidth,trim={0.cm 0.1cm 0.cm 0.1cm},clip]{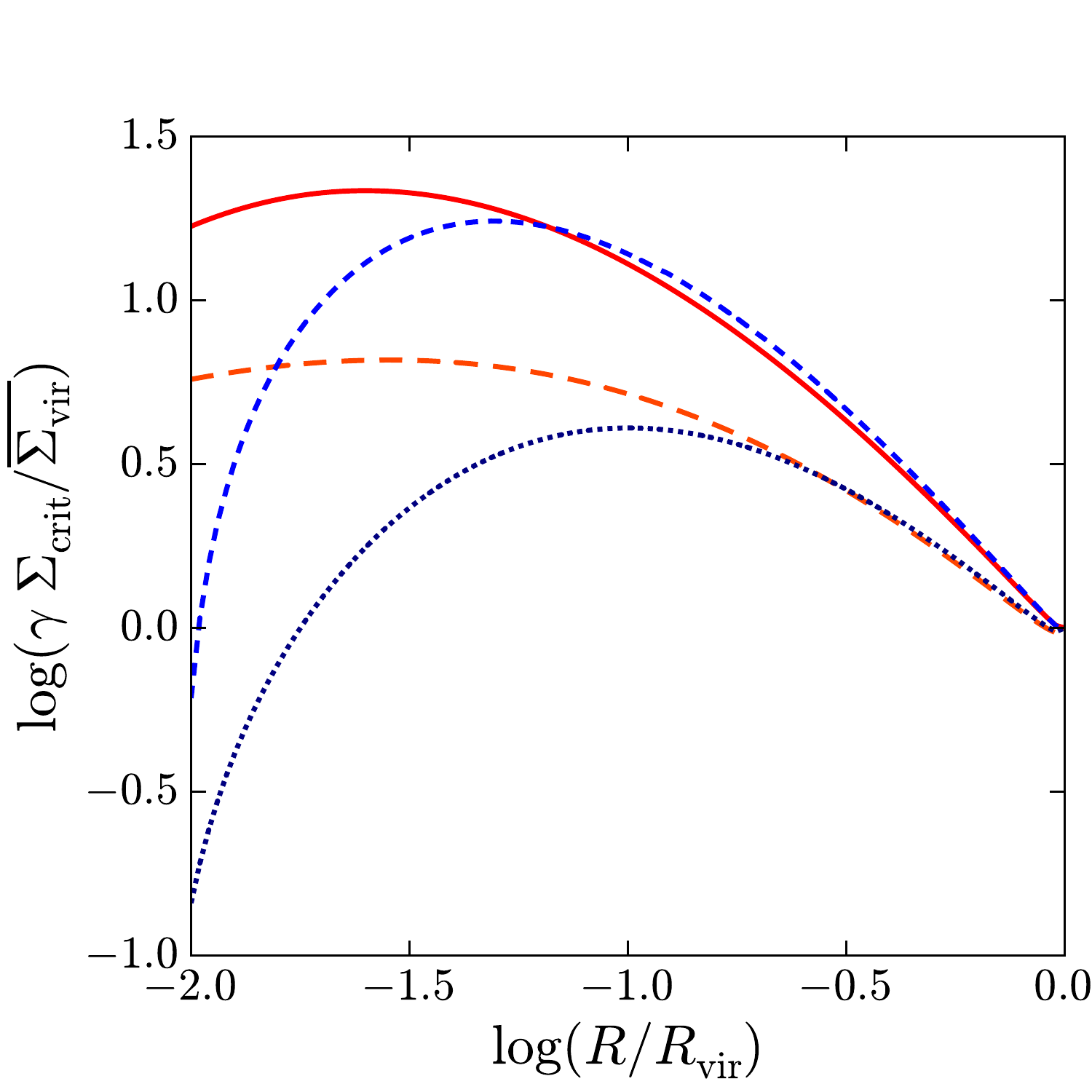}
	\vspace{-0.4cm}
	\caption{Lensing properties of fiducial spherical DZ DM haloes: two-dimensional projected surface density ($\Sigma$), scaled deflection angle ($\alpha$), and lensing shear ($\gamma$) as a function of the projected radius in the lens plane ($R$) for the four DZ haloes truncated at the virial radius with different inner slope ($s_1=0$ or $1$) and concentration ($c_2=5$ or $15$) of Fig.~\ref{fig:examples_dekel}. Eqs.~(\ref{eq:Sigma_H}),  (\ref{eq:lensing_alpha_H}), and (\ref{eq:Sigmab_H}) as well as the series expansions of Appendix~\ref{appendix:lensing} provide analytical expressions for the different profiles. Quantities are normalized by $\overline{\Sigma_{\rm vir}} = \Mvir/\pi \Rvir^2$ and  $\Sigma_{\rm crit}= c^2 D_{\rm S}/4\pi G D_{\rm L} D_{\rm LS}$, with $c$ being here the speed of light and $D_{\rm L}$, $D_{\rm S}$, $D_{\rm LS}$ the distances respectively between the observer and the lens, between the observer and the source, and between the lens and the source. 
	While the deflection angle mainly depends on the concentration $c_2$, the shear towards the center is very sensitive to the inner slope $s_1$. 
	}
	\label{fig:examples_dekel_lensing}
\end{figure*}

\subsubsection{Shear and magnification}

The Jacobian between the unlensed and lensed coordinate sytems depends on the convergence $\kappa$ and on the lensing shear, which for a axially-symmetric lens reads
\be
\label{eq:lensing_shear}
\gamma(X) \equiv \frac{\overline{\Sigma}(X)-\Sigma(X)}{\Sigma_{\rm crit}} 
\ee
with 
\be
\label{eq:lensing_Sigmab}
\overline{\Sigma}(X)=\frac{2}{X^2} \int_0^X x \Sigma(x) dx
\ee
the average surface density within $X$. 
The average surface density for an untruncated DZ profile can be expressed as
\be
\label{eq:Sigmab_H}
\widetilde{\overline{\Sigma}}(X) = \frac{4\sqrt{\pi} \rho_c r_c}{\Gamma(7-2a)} X
H_{3,3}^{2,2} 
\left[
\left.
\!\!\!\!
\begin{array}{c}
	(-6,4), (-\frac{1}{2},1), (0,1)\\
	(-\frac{1}{2},1), (-2a,4), (-\frac{3}{2},1)
\end{array}
\!\!
\right|
X^2
\right]
\ee
in terms of a Fox $H$ function while the average surface density of a DZ profile truncated at the virial radius is $\overline{\Sigma}(X) = \widetilde{\overline{\Sigma}}(X)-\widetilde{\Sigma}(c)$. 
Both have series expansions (Appendix~\ref{appendix:lensing}). 
Eqs.~(\ref{eq:Sigma_H}), (\ref{eq:Sigmab_H}), and the definitions of the convergence $\kappa$ (Eq.~(\ref{eq:lensing_kappa})) and of the shear $\gamma$ (Eq.~(\ref{eq:lensing_shear})) enable to determine the magnification factor
$\mu(X) = [(1-\kappa(X))^2-\gamma^2(X)]^{-1}$
by which the source luminosity is amplified \citep[][Eqs. (5.21) and (5.25)]{Schneider1992}. 
This factor, which is the inverse of the determinant of the Jacobian between the unlensed and lensed coordinate systems, comprises of a term depending on the convergence $\kappa$ that describes the isotropic focussing of the light rays in the lens plane and of a term depending on the shear $\gamma$ that accounts for the anisotropic focusing due to the tangential stretching of the image.

Fig.~\ref{fig:examples_dekel_lensing} displays the radial profiles of some of the lensing properties of the four fiducial DZ haloes of different inner slope and concentration shown in Fig.~\ref{fig:examples_dekel}, assumed to be truncated at the virial radius. We note that the shear $\gamma$ mainly depends on the concentration away from the halo center, with higher concentration leading to more shear, while steeper inner densities induce more shear near the center. 
The quantities expressed in this Section as well as those shown in Fig.~\ref{fig:examples_dekel_lensing} assume spherically-symmetric haloes. Generalizations to elliptical DZ haloes can be obtained by subtituting the projected radius $R$ with an expression depending on the ellipticity of the lens \citep[e.g., ][]{Schneider1992, Golse2002, Meneghetti2003}.

\section{The Dekel-Zhao profile in simulations}
\label{section:results}

\subsection{The NIHAO simulations}

We systematically test the DZ profile on the simulated DM haloes at $z=0$ of the Numerical Investigation of a Hundred Astrophysical Objects project \citep[NIHAO;][]{Wang2015}, which provides a set of about 90 cosmological zoom-in hydrodynamical simulations ran with the improved Smoothed Particle Hydrodynamics (SPH) code \texttt{gasoline2} \citep{Wadsley2017}. 
Each simulation is run at the same resolution with and without baryons, but we focus here on the hydrodynamical simulations including the effects of baryons. 
The simulations assume a flat $\Lambda$CDM cosmology with \cite{Planck2014} parameters, namely $\Omega_m = 0.3175$, $\Omega_r=0.00008$, $\Omega_\Lambda = 1-\Omega_m-\Omega_r = 0.6824$, $\Omega_b = 0.0490$, $H_0 = 67.1~\rm km s^{-1} Mpc^{-1}$, $\sigma_8 = 0.8344$ and $n=0.9624$.

They include a subgrid model describing the turbulent mixing of metals and thermal energy \citep{Wadsley2008}, cooling via hydrogen, helium and other metal lines in a uniform ultraviolet ionizing and heating background \citep{Shen2010} and star formation 
according to the Kennicutt-Schmidt relation when the temperature falls below $15 000~\rm K$ and the density reaches $10.3 ~\rm cm^{-3}$ \citep{Stinson2013}. 
Stars inject energy back to their surrounding intestellar medium (ISM) through ionizing feedback from massive stars \citep{Stinson2013} and supernovae \citep{Stinson2006}. 
During the pre-supernova feedback phase, 13\% of the total stellar luminosity -- which is typically $2 \times 10^{50}~\rm erg$ per $\rm M_\odot$ of the entire stellar population over the 4 Myr preceding the explosion of high-mass stars -- is ejected into the surrounding gas. 
During the supernova feedback phase, stars whose mass is comprised between 8 and 40 $\rm M_{\odot}$ eject 4 Myr after their formation both an energy $E_{\rm SN} = 10^{51}~\rm erg$ and metals into their surrounding ISM according to the blast-wave formalism described in \cite{Stinson2006}. 
Cooling is delayed for 30 Myr inside the blast region to prevent the energy from supernova feedback to be radiated away. 
Without cooling, the added supernova energy heats the surrounding gas, which both prevents star formation and models the high pressure of the blastwave.
AGN feedback is not included.

The NIHAO sample comprises isolated haloes chosen from dissipationless cosmological simulations \citep{Dutton2014} with halo masses between $\rm \log(M_{\vir}/M_\odot)=9.5-12.3$. Their merging histories, concentrations and spin parameters were not taken into account in the selection. 
The virial radius $R_\vir$ is defined as the radius within which the average total density is $\Delta$ times the critical density of the Universe, where $\Delta$ is defined according to \cite{Bryan1998}. The virial mass $M_\vir$ is the total mass enclosed within $R_\vir$. 
The particle masses and force softening lengths are chosen to resolve the DM mass profile below 1\% of the virial radius at all masses in order to resolve the half-light radius of the galaxies. 
Stellar masses, which are calculated within $0.15 R_\vir$, range from $5.10^4$ to $2.10^{11}~\rm M_\odot$, i.e., from dwarfs to Milky Way sized galaxies, with morphologies, colors and sizes that correspond well with observations \citep[e.g.,][]{Wang2015,Stinson2015,Dutton2016a}. 
As shown by \cite{Tollet2016}, \cite{Dutton2016}, \citetalias{Dekel2017}, \citetalias{Freundlich2020}, and \cite{Maccio2020}, NIHAO DM haloes display a variety of inner slopes ranging from steep cusps to flat cores, cores being more prevalent at $z=0$ for stellar masses comprised between $10^7$ and $10^{10}~\rm M_\odot$.

\subsection{Fitting procedure and results}
\label{section:fitting}

\subsubsection{Density profile fits and rotation curves}

We fit the logarithm of the density profile of each simulated halo at $z=0$ according to the DZ parametrisation (Eq.~(\ref{eq:rho32})) through a least-square minimization between $0.01\Rvir$ (the resolution limit) and $\Rvir$. 
Since $\Rvir$ and $\Mvir$ are set, $a$ and $c$ are the only free parameters. We impose the inner logarithmic slope at the resolution limit to be positive, namely $s_1\geq 0$ with $s_1=s(0.01\Rvir)$ expressed in Eq.~(\ref{eq:s1_23}). 
The profile radii $r$ are spaced logarithmically, with $N\sim 100$ radii $r_i$ between $0.01\Rvir$ and $\Rvir$. 
The inner slope $s_1$ and the concentration parameter $c_2$ associated to the fit result can be derived from $a$ and $c$ with Eqs.~(\ref{eq:s1_23}) and (\ref{eq:c2_23}). 
The rms 
\be
\sigma = \sqrt{\frac{1}{N}\sum_{i=1}^N \left(\log {\rho}_i - \log{\rho}_{\rm model}(r_i)\right)^2}
\ee 
of the residuals between the simulated $\log {\rho}$ and the model is used to evaluate the relative goodness of fit in the range $0.01\Rvir-\Rvir$, and we also define $\sigma_{\rm c}$ the rms of the residuals in the central region of the halo between $0.01\Rvir$ and $0.1\Rvir$. 
The residuals themselves can be seen in Appendix~\ref{appendix:figures}.
The absolute value of $\sigma$ (or $\sigma_{\rm c}$) is sensitive to the smoothness of the simulated profile, in particular to the resolution of the simulations, the number of radii used, and the binning procedure for the profile. 
We thus mostly use it to compare the performance of different models in fitting a given target profile. 
We notably note that with profile radii spaced logarithmically, the effective weight assigned to the inner region of the halo is larger than it would have been with linearly-space radii.

\begin{figure*}
	\includegraphics[width=\textwidth]{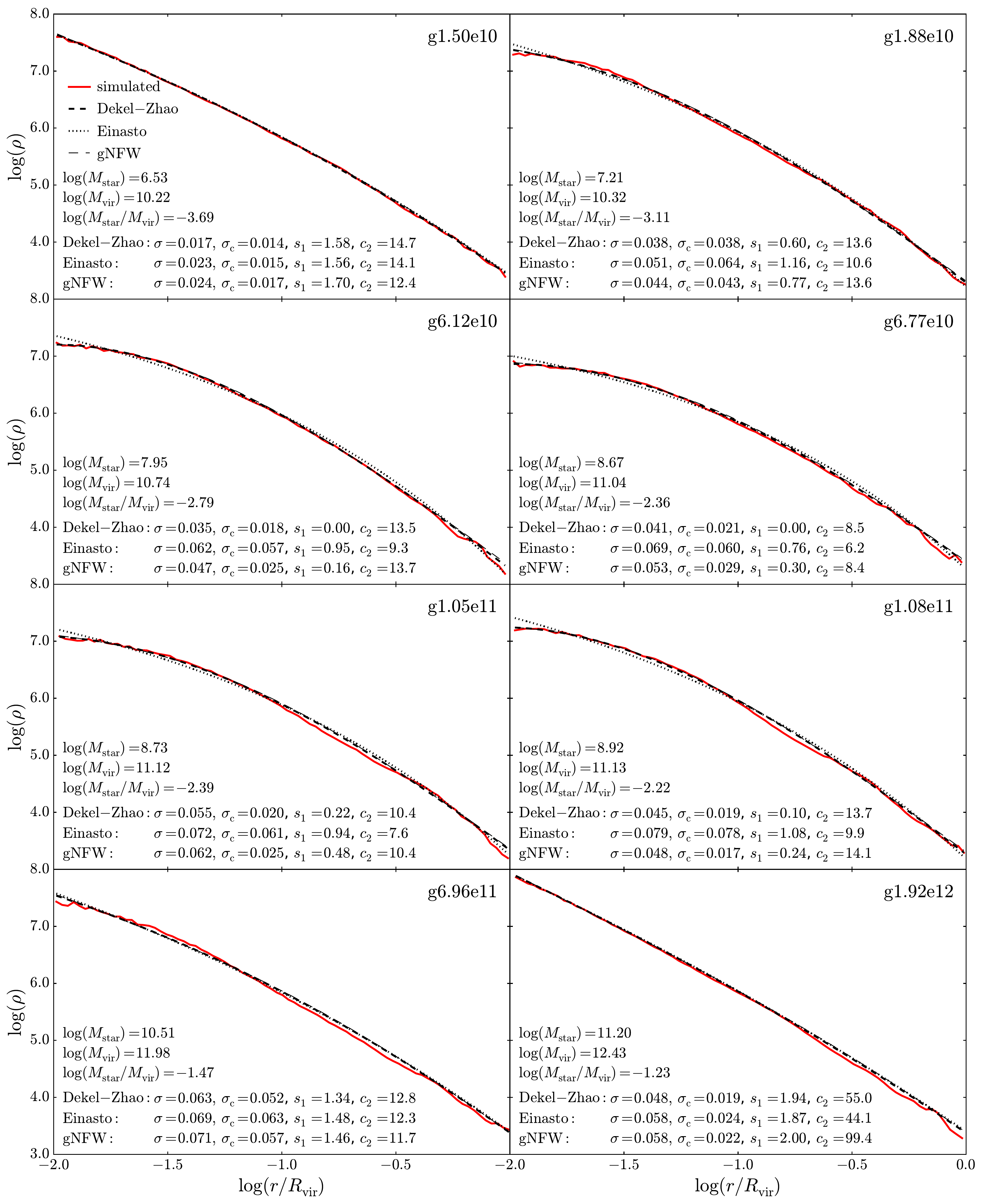}
	\vspace{-0.4cm}
	\caption{Model versus simulated density profiles: the dark matter density profiles at $z=0$ of eight arbitrary NIHAO galaxies with baryons (plain red line) at different masses with their best-fitting DZ, Einasto and gNFW profiles (dashed, dotted, and thin dashed black lines, respectively), for radii covering the range between $0.01 R_{\rm vir}$ and $R_{\rm vir}$. The rms errors $\sigma$ and $\sigma_{\rm c}$ and the best-fit parameters $s_1$ and $c_{2}$ of the different parametrizations are indicated, as well as $M_{\rm star}$, $M_{\rm vir}$, and $M_{\rm star}/M_{\rm vir}$. 
	The gNFW fits follow closely the DZ fits; the Einasto fits do not recover the inner density profiles as well as the others.}
	\label{fig:fits}
\end{figure*}

\begin{figure*}
	\includegraphics[width=\textwidth]{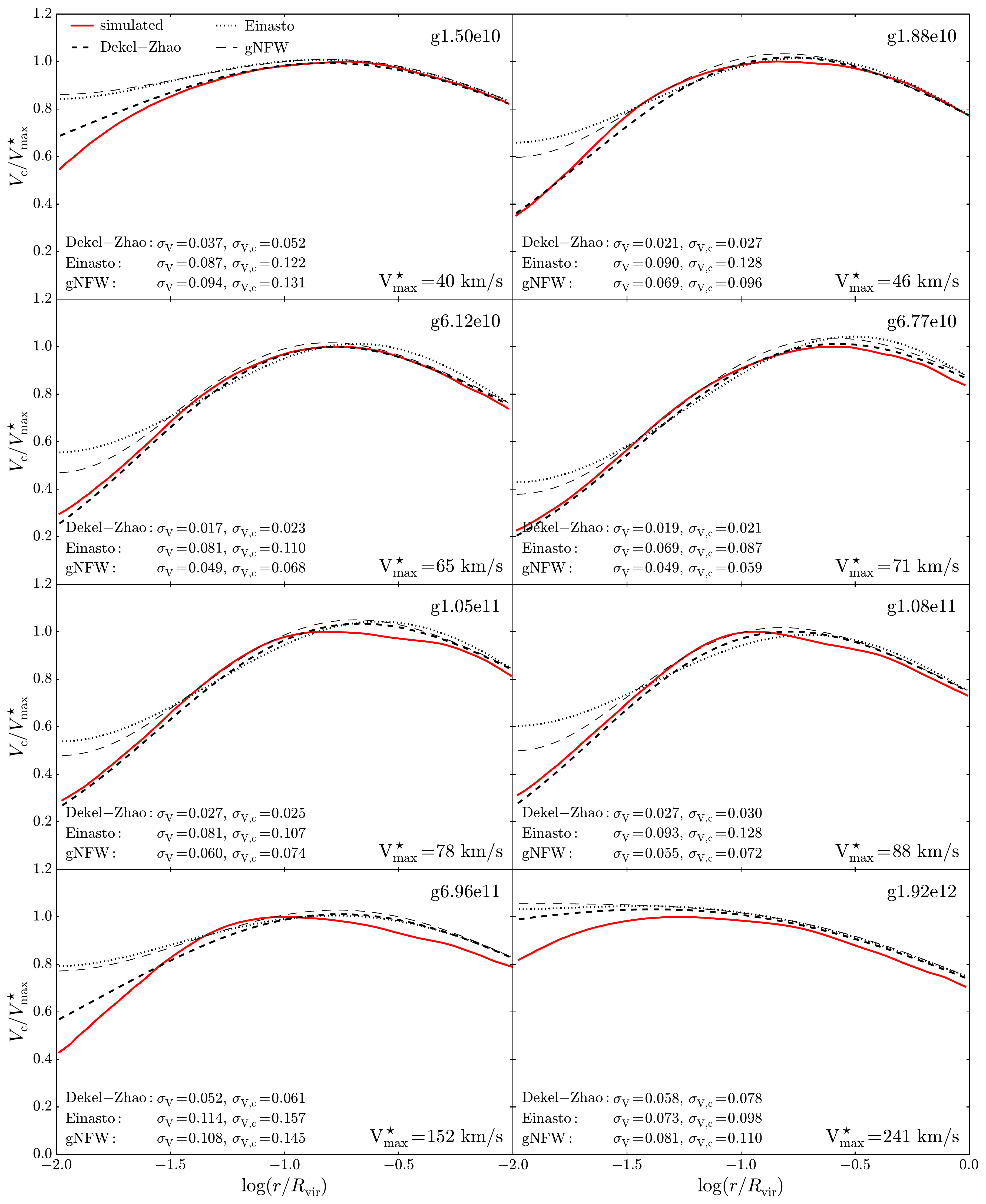}
	\vspace{-0.4cm}
	\caption{Model versus simulated rotation curves: dark matter circular velocity profiles, $V_{\rm c}(r) = \sqrt{GM(r)/r}$, of the eight $z=0$ NIHAO galaxies shown in Fig.~\ref{fig:fits} (plain red line) together with those inferred from the DZ, Einasto and gNFW fits to their density profiles (dashed, dotted, and thin dashed black lines, respectively). The velocity of each galaxy is normalized to its maximum value $V_{\rm max}^\star$, which is an increasing function of mass. The rotation curves inferred from the DZ fits to the density profiles recover better the simulated curves than those inferred from the Einasto and gNFW fits. The residuals of $V_c/V_{\rm max}^\star$ are shown in Appendix~\ref{appendix:figures}.
	}
	\label{fig:fits_Vc}
\end{figure*}

\begin{figure*}
	\includegraphics[height=0.23\textwidth,trim={0.cm 0 1.5cm 1.cm},clip]{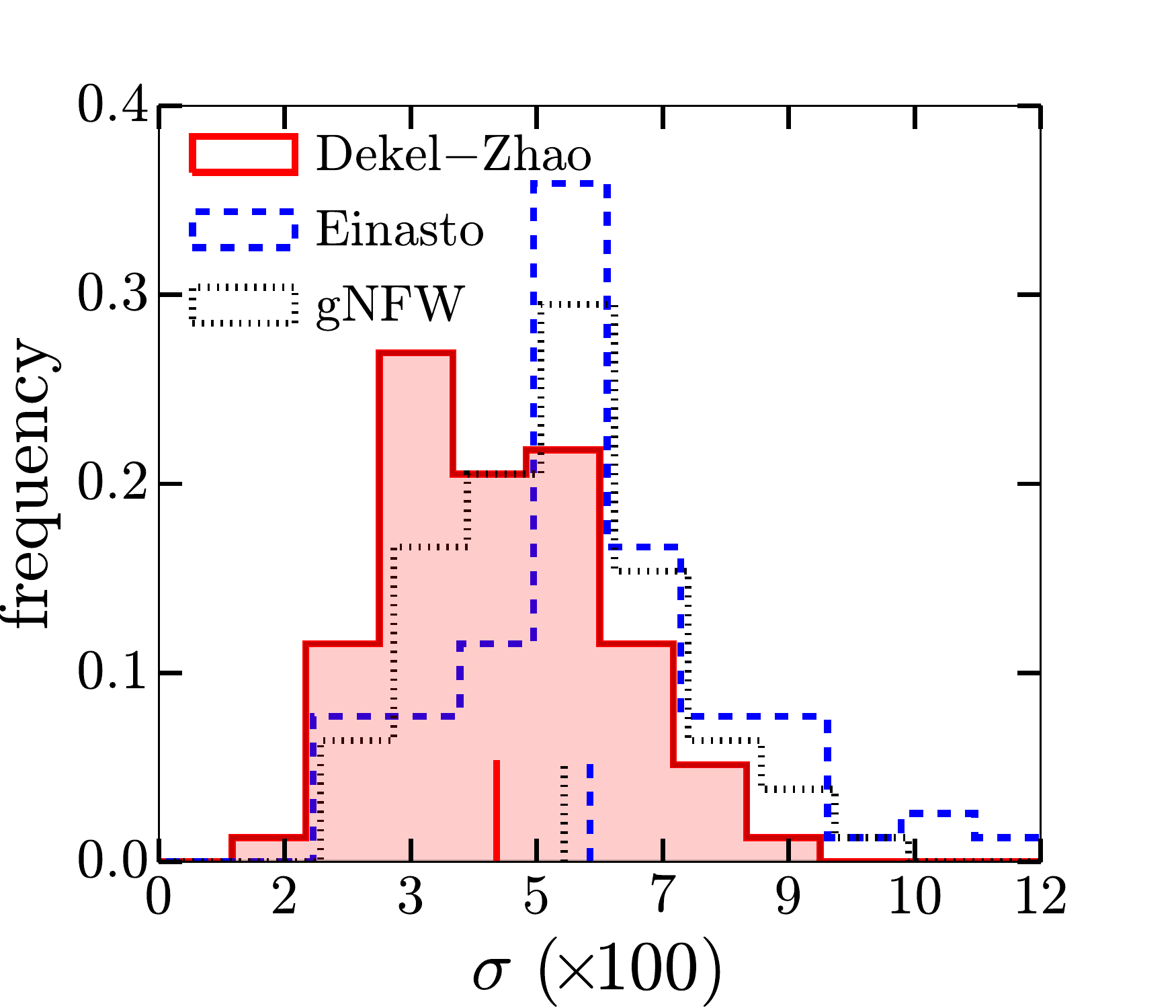}
	\includegraphics[height=0.23\textwidth,trim={1.5cm 0 1.5cm 1.cm},clip]{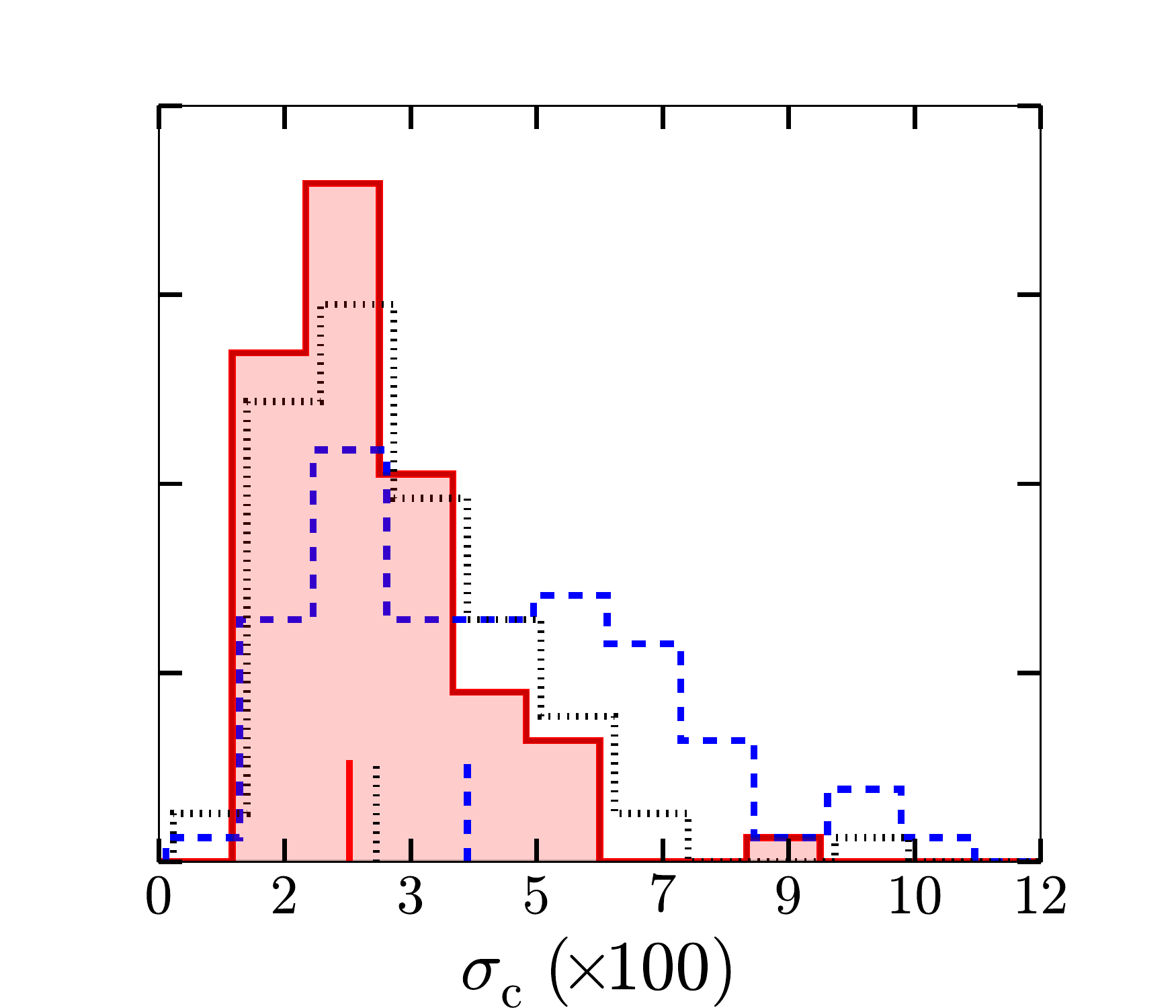}
	\includegraphics[height=0.23\textwidth,trim={1.5cm 0 1.5cm 1.cm},clip]{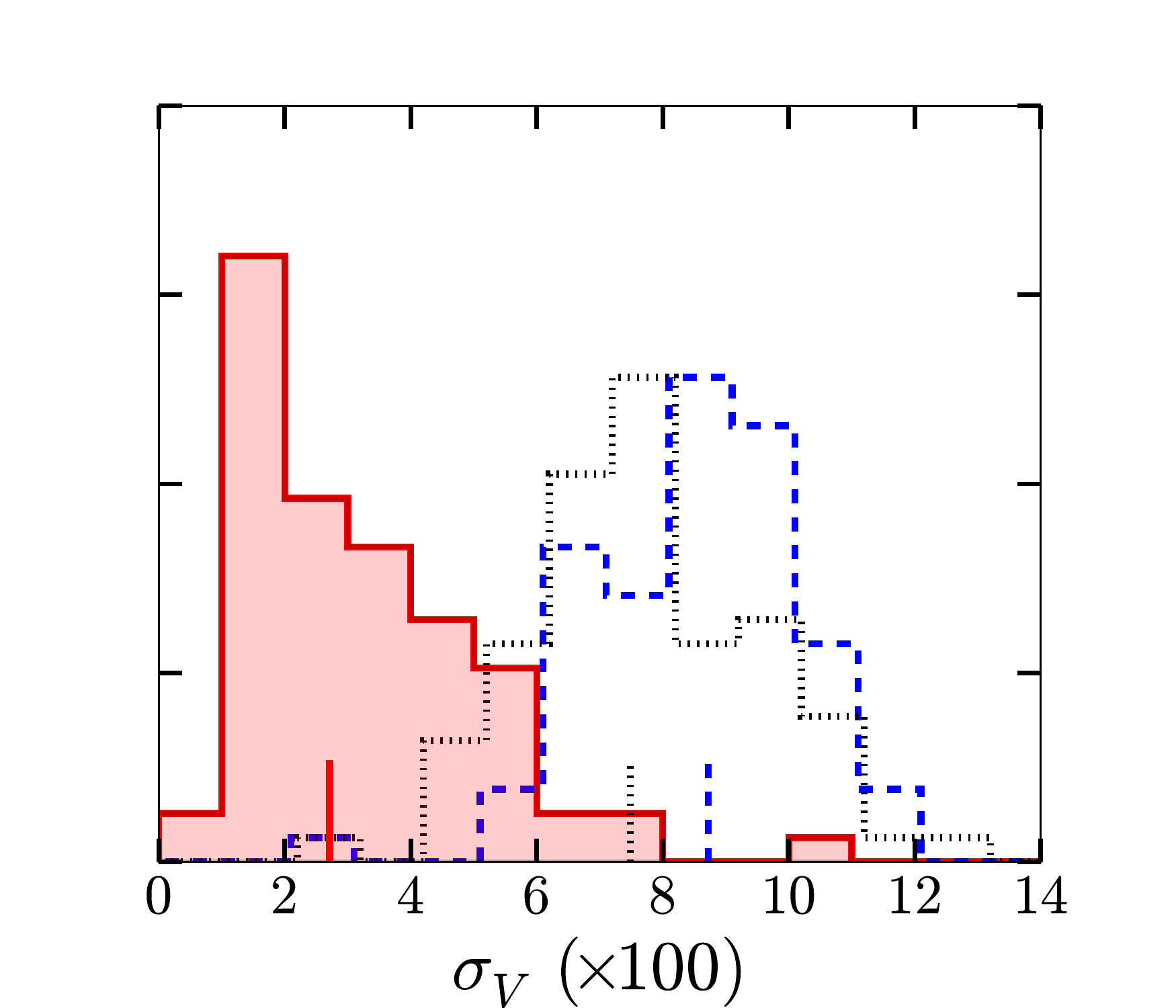}
	\includegraphics[height=0.23\textwidth,trim={1.5cm 0 1.5cm 1.cm},clip]{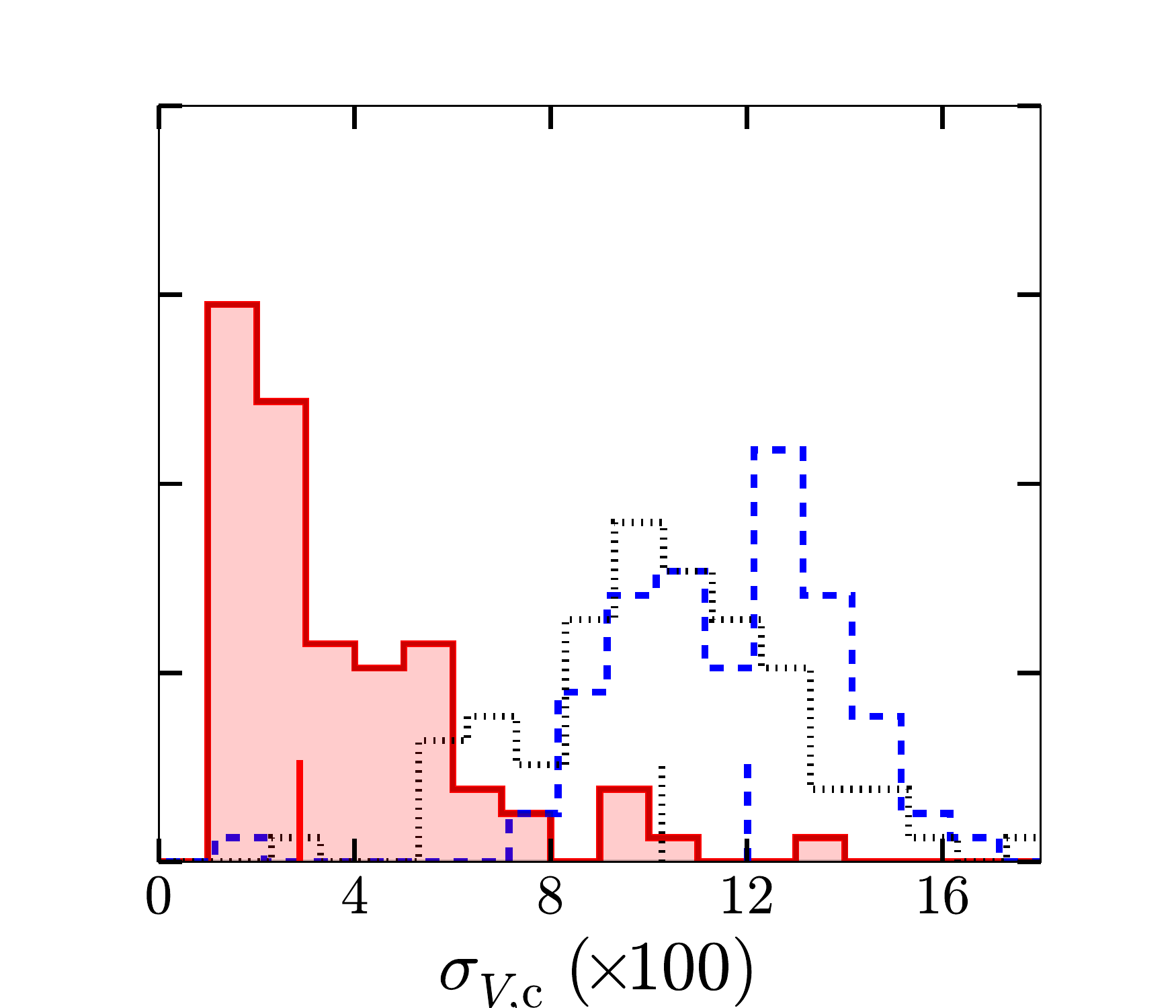}	
	\vspace{-0.1cm}
	\caption{Comparing the model fits in terms of their rms errors: rms errors in $\log\rho$ and $V_c/V_{\rm max}$ of the DZ (plain red line), \protect\citetalias{Einasto1965} (blue dashed line) and gNFW (black dotted line) fits over the ranges $0.01\Rvir-\Rvir$ and $0.01\Rvir-0.1\Rvir$ for all NIHAO galaxies at $z=0$. 
	The median values for the three models, which are highlighted by vertical lines above the x-axis, respectively yield $0.046$, $0.059$, and $0.055$ for $\sigma$, $0.026$, $0.042$, and $0.030$ for $\sigma_{\rm c}$, $0.027$, $0.087$, and $0.075$ for $\sigma_{\rm V}$, $0.029$, $0.120$, and $0.103$ for $\sigma_{\rm V, c}$. 
	The standard deviations respectively yield $0.015$, $0.019$, and $0.015$ for $\sigma$, $0.013$, $0.023$, and $0.015$ for $\sigma_{\rm c}$, $0.018$, $0.016$, and $0.018$ for $\sigma_{\rm V}$, $0.025$, $0.023$, and $0.026$ for $\sigma_{\rm V, c}$. 
	The residuals from which the rms errors are computed are shown in Appendix~\ref{appendix:figures}.
	The DZ profile provides better fits to the DM density profile than the \protect\citetalias{Einasto1965} and gNFW profiles, the difference being particularly striking in the resulting circular velocity profiles through $\sigma_{\rm V}$ and $\sigma_{\rm V, c}$. 
	}
	\label{fig:sigma_hist}
\end{figure*}

Fig.~\ref{fig:fits} displays the DZ fit results to the DM density profile for eight fiducial $z=0$ NIHAO haloes of different masses, simulated with baryons. This selection includes the two haloes studied more specifically in \citetalias{Freundlich2020},  \texttt{g1.08e11} and \texttt{g6.12e10}, but is otherwise arbitrary in each mass range. The best-fit profile parameters $a$ and $c$ as well as the corresponding inner slope $s_1$ and concentration $c_2$ are indicated. The mass-dependence of the DM halo response to baryons described by \cite{DiCintio2014}, \cite{Tollet2016}, and \cite{Dutton2016} is already visible in this figure, with the lowest-mass halo having a relatively steep cusp, haloes with stellar masses between $10^7$ and $10^{10}~\rm M_\odot$ shallower cores, and the two most massive haloes steeper inner slopes.

The figure further compares the fits according to the DZ parametrization with fits according to the \citetalias{Einasto1965} and the generalized NFW with free inner slope (gNFW) parametrizations. 
We recall that the Einasto density profile \citep[][\citetalias{An2013}]{Einasto1965,Navarro2004} can be expressed as 
\be
\label{eq:rho_einasto}
\rho_{\rm Einasto}(r) = \rho_2 \exp \left( -\frac{2}{\nu}\left[ \left( \frac{r}{r_2}\right)^\nu -1\right]\right)
\ee
with $r_2$ the radius where the logarithmic density slope equals $2$, $\rho_2$ the corresponding density and $\nu$ a shape parameter. 
The gNFW profile refers to Eq.~(\ref{eq:rho_abc}) with $b=1$ and $g=3$ (e.g., \citetalias{An2013}), i.e., 
\be
\label{eq:rho_enfw}
\rho_{\rm gNFW}(r) = \frac{\rho_c}{x^a (1+x)^{3-a}}
\ee
with $x=r/r_c$ and $a$ the innermost slope.
These two profiles have two free shape parameters ($c_2=\Rvir/r_2$ and $\nu$ for the Einasto profile, $a$ and $c=\Rvir/r_c$ for the gNFW profile) as is the case for the DZ parametrization. 
As notably indicated by the rms $\sigma$ and $\sigma_{\rm c}$, Einasto fits are significantly worse for shallow inner density slopes than the other two, which seem to follow each other closely. 
This is particularly visible in the inner part of the density profile.

Fig.~\ref{fig:fits_Vc} compares the DM circular velocity profiles of the eight fiducial haloes of Fig.~\ref{fig:fits} with those resulting from the density profile fits. As for the density profile fits, we define $\sigma_{\rm V}$ and $\sigma_{\rm V, c}$ the rms of the residuals between the simulated circular velocity $V_c$ and the model, in the ranges $0.01\Rvir-\Rvir$ and $0.01\Rvir-0.1\Rvir$, respectively. 
Although we note that there may be some $\lesssim 10\%$ offset in the velocity prescription at high masses, the DZ profile fares significantly better than the other two parametrizations in recovering the DM circular velocity profiles, as indicated by the systematically lower values of $\sigma_{\rm V}$ and $\sigma_{\rm V, c}$. The inadequation of the Einasto and gNFW profiles is striking towards the innermost part of the rotation curve. 
Fig.~\ref{fig:sigma_hist} confirms the trends seen in Figs.~\ref{fig:fits} and \ref{fig:fits_Vc} over the whole NIHAO sample at $z=0$ by systematically comparing the rms $\sigma$, $\sigma_{\rm c}$, $\sigma_{\rm V}$, and $\sigma_{\rm V, c}$ distributions of the three two-parameter models. 
We point out that the circular velocities at small radii obtained for the DZ, Einasto, and gNFW profiles are significantly impacted by the behavior of these profiles below the resolution limit of $0.01 R_\vir$.

\subsubsection{Model versus simulated parameters}
\label{subsection:params}

To quantify further the adequation of the different profile parametrizations, we define an inner slope ${s_{1}^\star}$ and a concentration ${c_2^\star}$ directly measured from the simulated density and logarithmic slope profiles. The former is the average slope between $0.01\Rvir$ and $0.02\Rvir$, as notably used by \cite{Tollet2016}; the latter corresponds to the radius where the logarithmic slope equals 2. 
Since the simulated slope profile can be relatively noisy, we smooth it using a Savitsky-Golay filter with maximum window size when measuring ${c_2^\star}$.
Fig.~\ref{fig:definition_s12_c2} in Appendix~\ref{appendix:figures} illustrates how ${s_{1}^\star}$ and ${c_2^\star}$ are obtained from the simulated profiles. 
The definitions of these two quantities each have their own shortcomings, notably as ${s_{1}^\star}$ may in principle be different than the innermost slope at $0.01\Rvir$ and as ${c_2^\star}$ may be affected by the smoothing, but do enable to capture reasonable inner slopes and concentrations.
We use these quantities as references to describe the inner slope and concentration differences between model and simulation, $\Delta s=s_{1, \rm model}-s_1^\star$ and $\Delta c = c_{2,\rm model}-c_2^\star$.

\begin{figure}
	\centering
	\vspace{-0.1cm}
	\includegraphics[height=0.68\linewidth,trim={2.2cm 0cm 4.cm 1.5cm},clip]{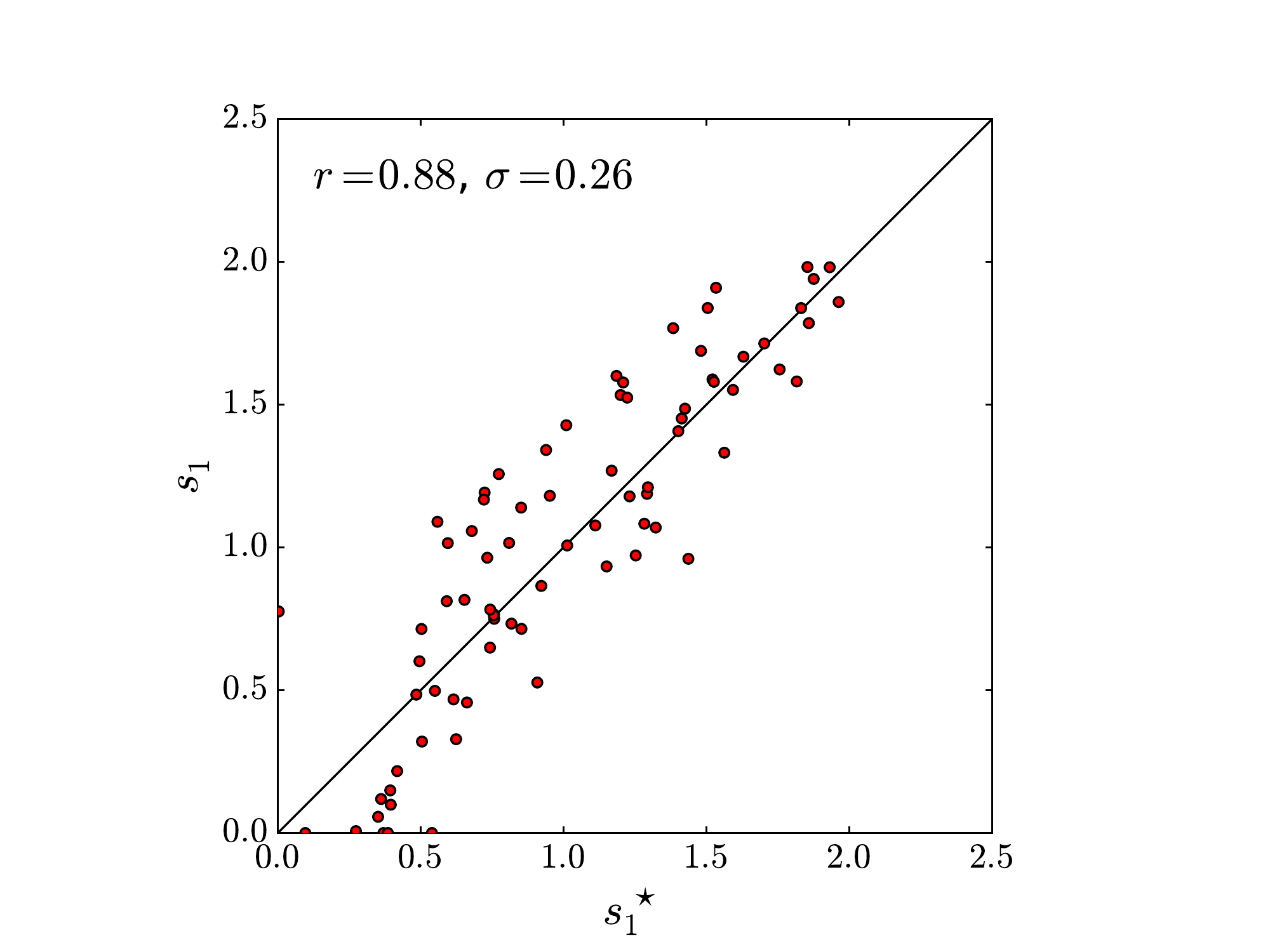}
	\vspace{-0.2cm}
	\\
	\includegraphics[height=0.68\linewidth,trim={2.2cm 0 4.cm 1.5cm},clip]{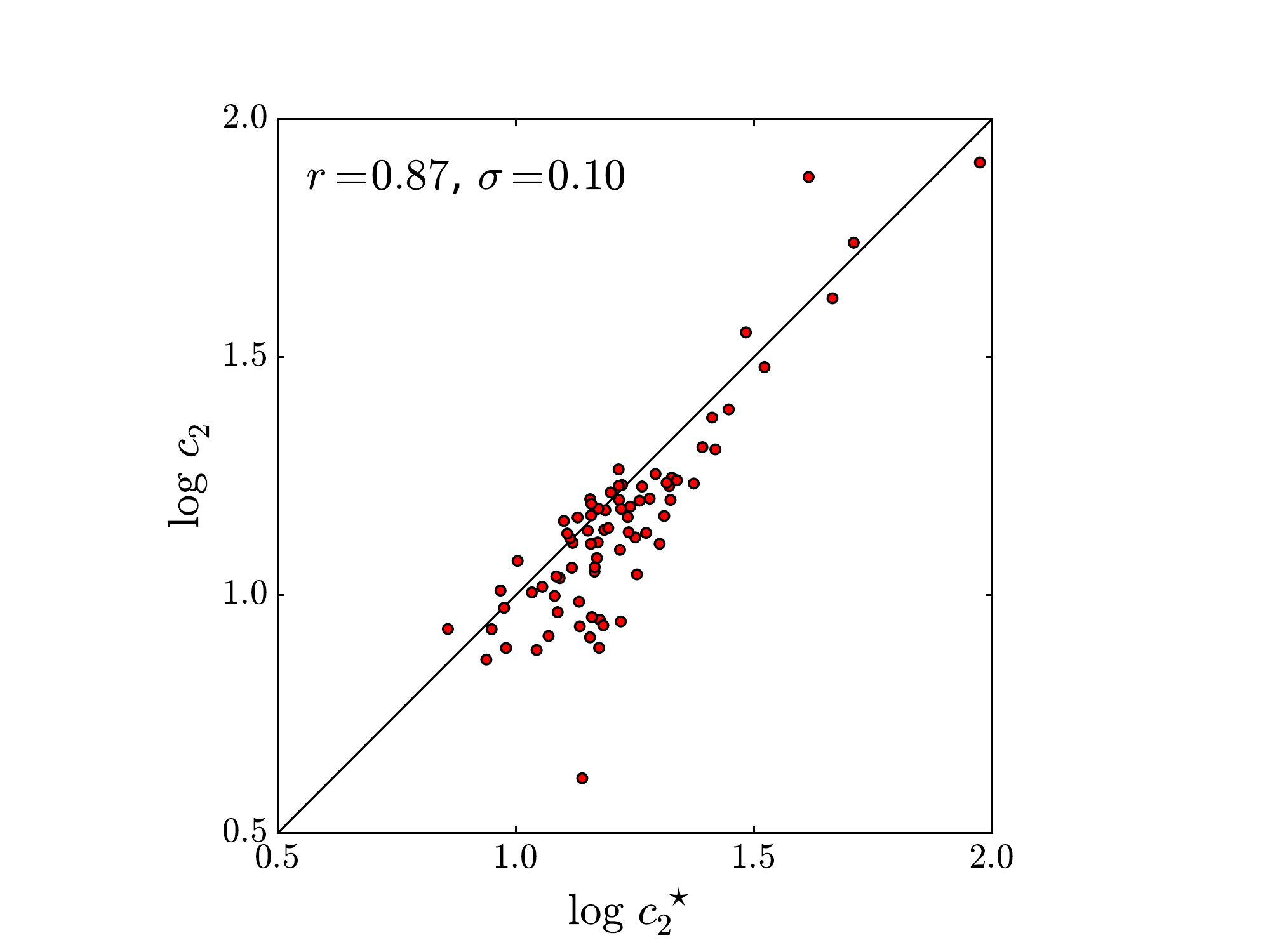}
	\vspace{-0.2cm}
	\caption{Model versus simulated parameters: comparison between the inner slope and concentration stemming from the DZ fits to the density profiles of the $z=0$ NIHAO galaxies with baryons, $s_1$ and $c_2$, and those obtained directly from the simulated profiles, $s_1^\star$ and $c_2^\star$. The plain lines corresponds to a linear least-square fits. The Pearson correlation coefficient ($r$) and the residual scatter ($\sigma$) are indicated. The DZ fits enable to retrieve the inner slope and concentration measured from the simulated profiles.
	}
	\label{fig:comparison_s12_c2}
\end{figure}

\begin{figure*}
	\includegraphics[height=0.23\textwidth,trim={0.cm 0 1.5cm 1.cm},clip]{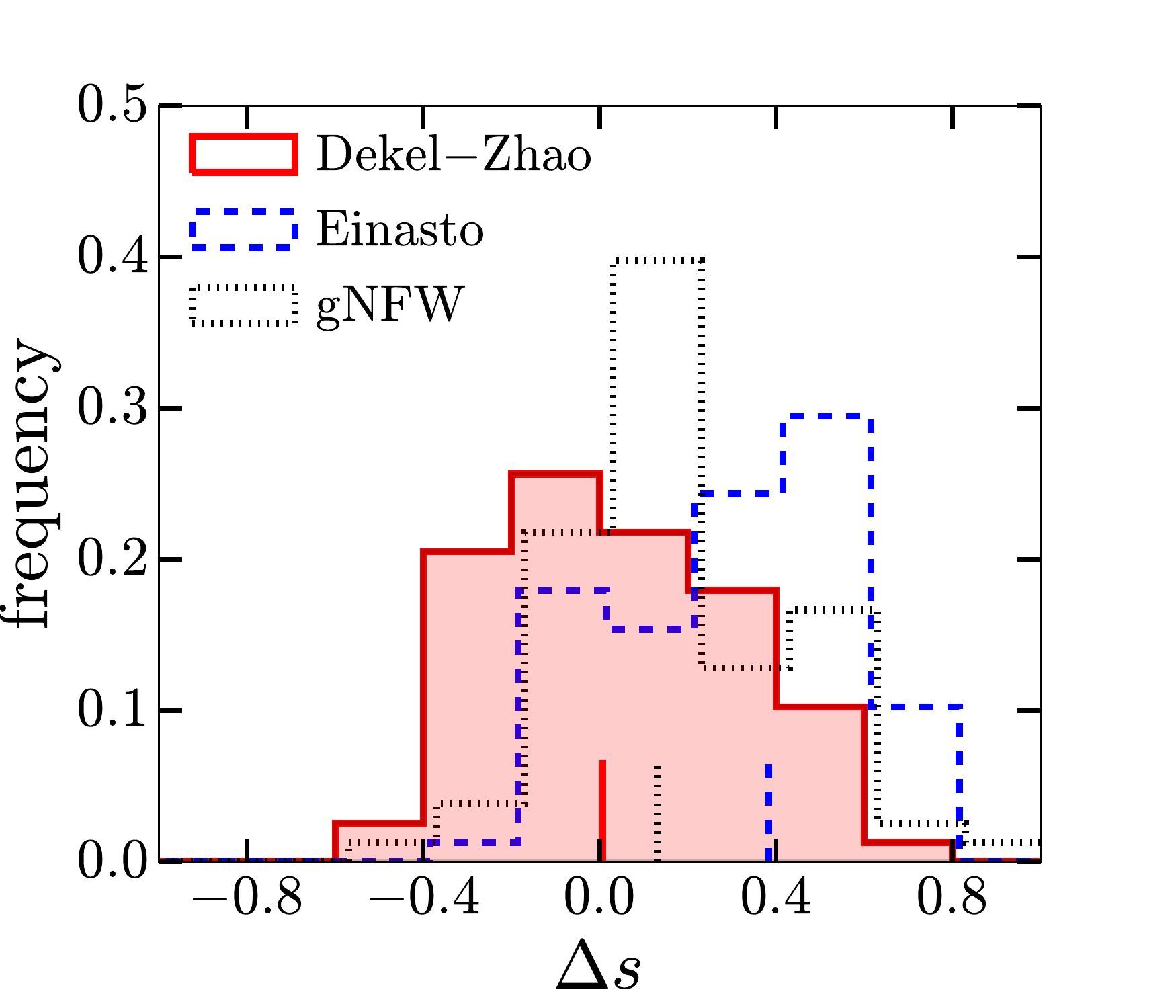}
	\includegraphics[height=0.23\textwidth,trim={1.5cm 0 1.5cm 1.cm},clip]{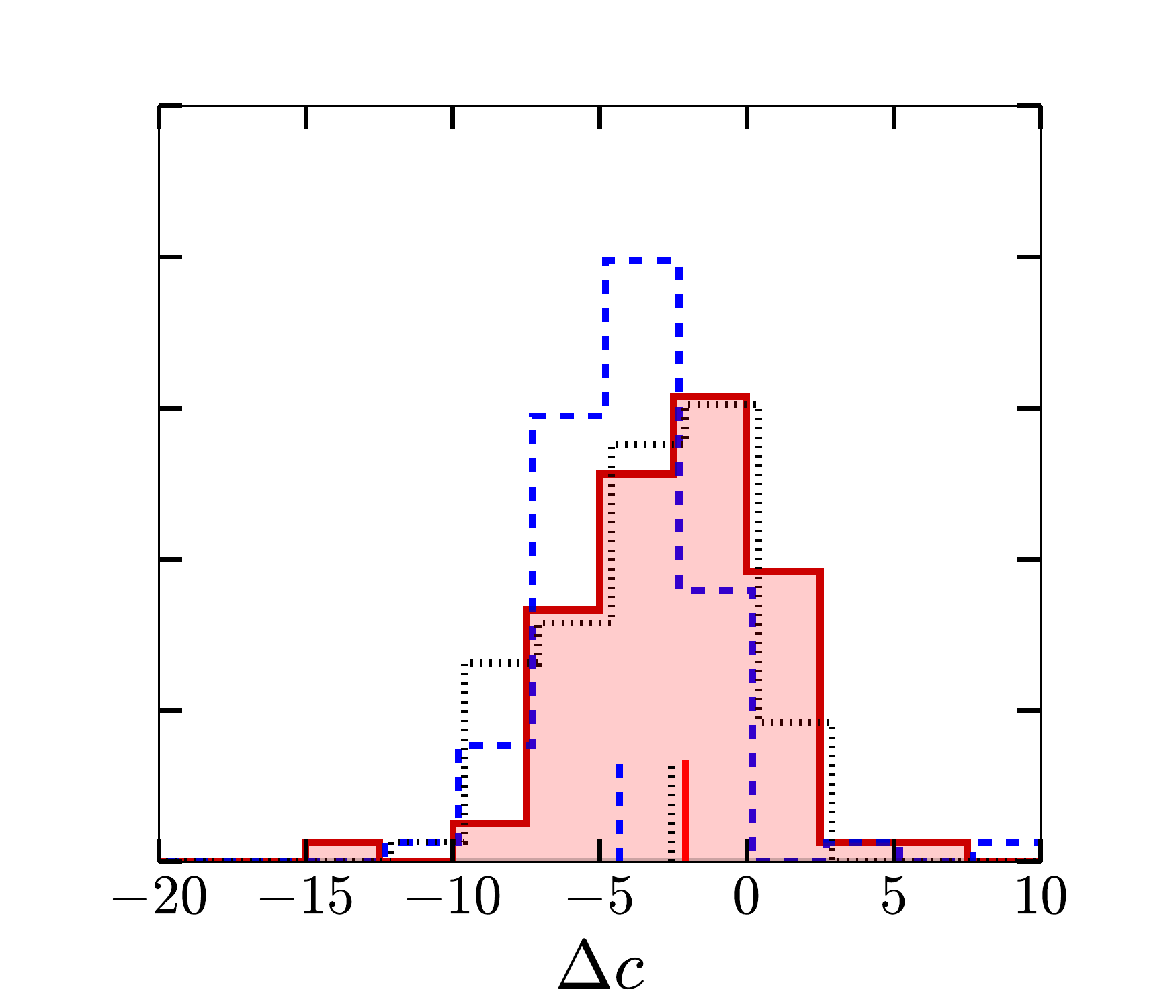}
	\includegraphics[height=0.23\textwidth,trim={1.5cm 0 1.5cm 1.cm},clip]{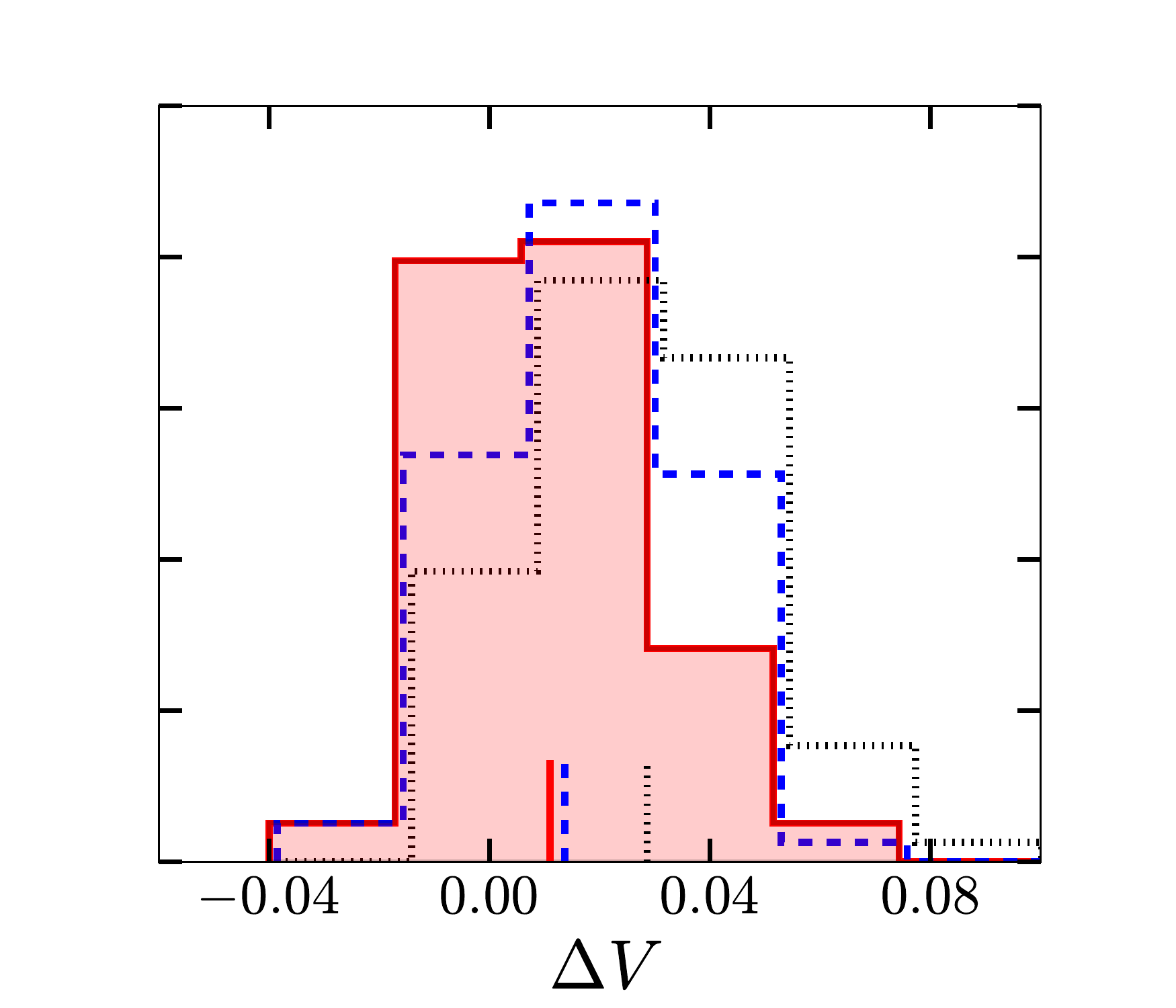}
	\includegraphics[height=0.23\textwidth,trim={1.5cm 0 1.5cm 1.cm},clip]{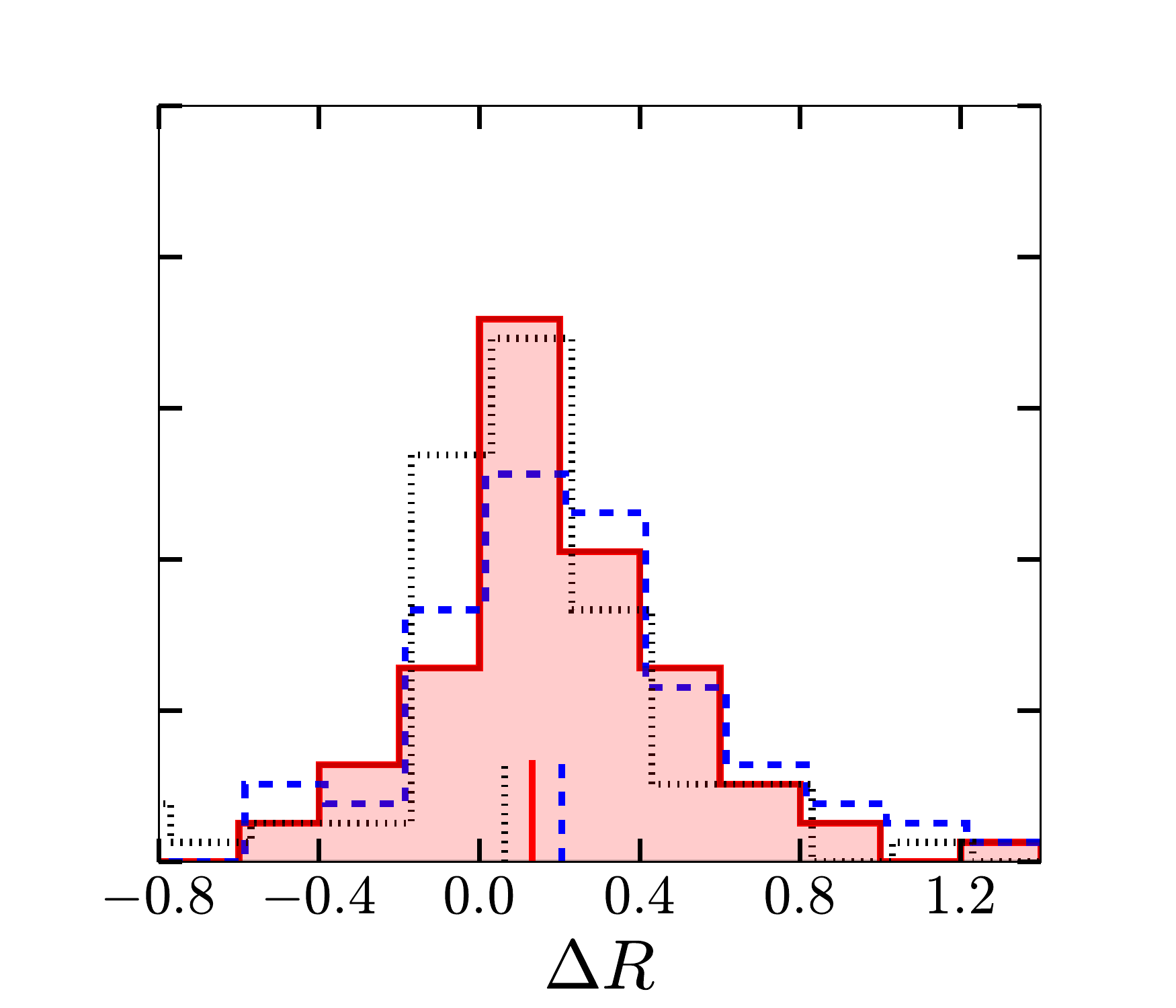}
	\caption{Comparing the model fit parameters: inner slope and concentration differences, $\Delta s = s_{1,\rm model}-s_{1}^\star$ and  $\Delta c = c_{\rm 2, model}-c_{\rm 2}^\star$, as well as the maximum velocity and radius relative differences, $\Delta V= (V_{\rm max, model}-V_{\rm max}^\star)/V_{\rm max}^\star$ and $\Delta R= (R_{\rm max, model}-R_{\rm max}^\star)/R_{\rm max}^\star$, between the DZ (plain red line), \protect\citetalias{Einasto1965} (blue dashed line) and gNFW (black dotted line) fits and the simulated profiles for all $z=0$ NIHAO galaxies simulated with baryons. 
	The median values for the three models, which are highlighted by vertical lines above the x-axis, respectively yield 
	$0.01$, $0.38$, and $0.13$ for $\Delta s$, 
	$-2.1$, $-4.3$, and $-2.6$ for $\Delta c$, 
	$0.011$, $0.014$, and $0.029$ for $\Delta V$, 
	$0.13$, $0.21$, and $0.06$ for $\Delta R$. 
	The standard deviations respectively yield 
	$0.27$, $0.27$, and $0.25$ for $\Delta s$, 
	$5.1$, $4.9$, and $7.3$ for $\Delta c$, 
	$0.018$, $0.018$, and $0.020$ for $\Delta V$, 
	$0.30$, $0.35$, and $0.34$ for $\Delta R$. 
	The DZ profile provides better fits to the DM density profile than the \protect\citetalias{Einasto1965} and gNFW profiles, in particular with $\Delta s$, $\Delta c$, and $\Delta V$ generally closer to zero. We do note however that $R_{\rm max}$ is on average overestimated by $\sim$10\%. 
	}
	\label{fig:comparison_histograms}
\end{figure*}

Fig.~\ref{fig:comparison_s12_c2} compares the inner slopes and concentrations derived from the DZ fits ($s_1$ and $c_2$) with those measured on the simulated profiles ($s_1^\star$ and $c_2^\star$) , highlighting very strong correlations (with Pearson correlation coefficients $r>0.85$) with some scatter ($0.26$ for $s_1$, $0.10$ for $\log c_2$): the DZ fits enable to retrieve the inner slope and concentration measured from the simulated profiles. 
We further define $V_{\rm max}^\star$ and $R_{\rm max}^\star$ the maximum velocity and the corresponding radius on the simulated circular velocity profiles such as those shown in Fig.~\ref{fig:fits_Vc}, as well as $\Delta V = (V_{\rm max, model}-V_{\rm max}^\star)/V_{\rm max}^\star$ and $\Delta R = (R_{\rm max, model}-R_{\rm max}^\star)/R_{\rm max}^\star$ the relative difference between the values derived from the density profile fits and those measured on the simulated profiles. 
Fig.~\ref{fig:comparison_histograms} shows the distributions of $\Delta s$, $\Delta c$, $\Delta V$, and $\Delta R$ for the DZ, Einasto, and gNFW fits for all NIHAO galaxies with baryons at $z=0$. 
The figure shows that the  DZ parametrization provides inner slopes closest to $s_1^\star$ on average while the other two, and in particular the Einasto parametrization, systematically overestimate the inner slope. This can already be seen in Fig.~\ref{fig:fits}, where the Einasto fit is in most cases above the simulated density profile in the innermost part. 
The three parametrizations slightly tend to underestimate the concentration compared to that measured from the slope profile, but we recall that the latter may be affected by the smoothing. In this regard, the Einasto parametrization seems to yield a higher systematic offset than the other two, but the scatters are similar. 
The three parametrizations recover the maximum velocity with a relative error $\lesssim 5\%$, but with a systematic overestimation of $\sim 1\%$ on average for DZ and Einasto, $\sim 3\%$ for gNFW. The maximum radius is less well recovered by all parametrizations, with relative errors spread around a 10\% overestimate in the DZ case (6\% for gNFW, 20\% for Einasto) with a $\sim$30\% standard deviation.

We conclude from this analysis that the DZ parametrization provides significantly better fits  to DM density profiles than Einasto and marginally better than gNFW, and infer better fits than both to the circular velocity profile. It enables to recover the inner density slope $s_1$ with a $\pm 0.27$ scatter but a negligible systematic error, the concentration $c_2$ with a $\pm 5$ scatter and a limited $-2$ systematic offset on average (0.1 dex scatter and -0.05 dex offset in $\log c_2$). It retrieves the maximum velocity $V_{\rm max}$ with a $\pm 2\%$ scatter and a +1\% systematic offset, the corresponding radius with a $\pm 30\%$ scatter and a $+13\%$ offset (0.11 dex scatter and +0.05 dex offset in  $\log R_{\rm max}$).

\subsection{Mass-dependence of the profile parameters}
\label{section:mass_dependence}

\subsubsection{Mass-dependence of $s_1$ and $c_2$}

\begin{figure*}
	\centering
	\includegraphics[width=0.95\textwidth,trim={5.8cm .5cm 2.cm 0.4cm},clip]{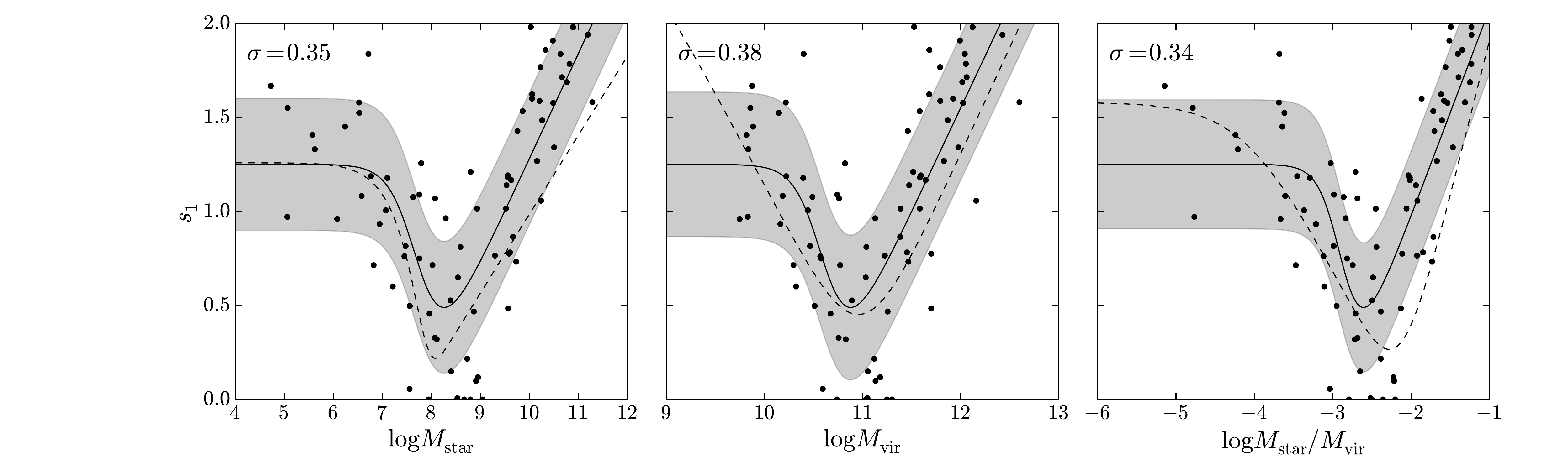}\\
	\includegraphics[width=0.95\textwidth,trim={5.8cm .5cm 2.cm 0.4cm},clip]{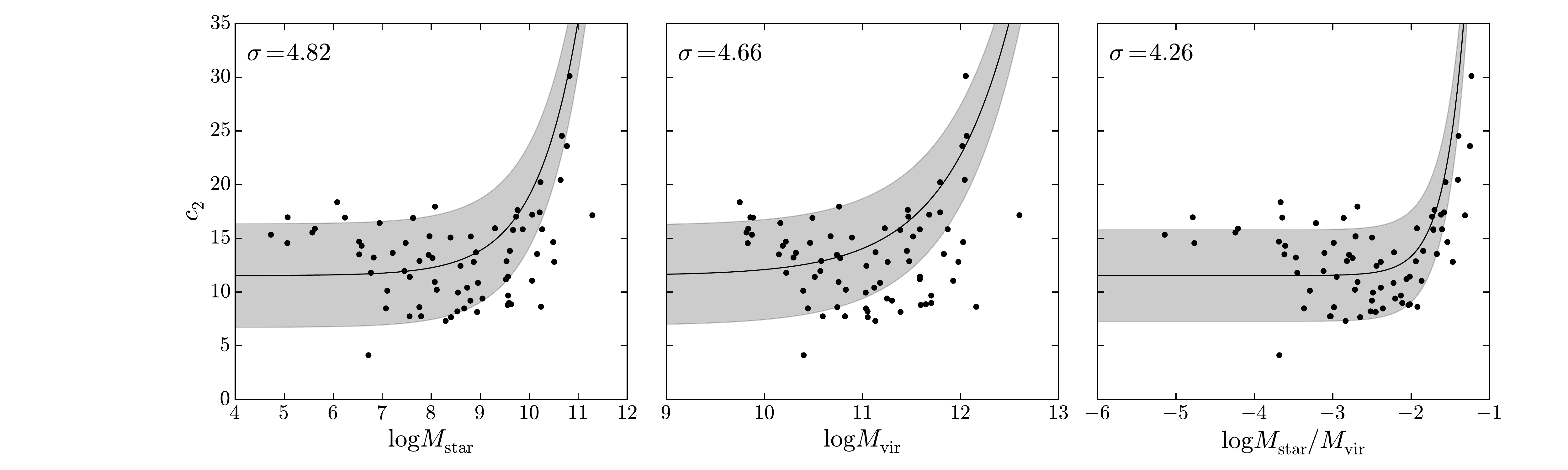}\\
	\vspace{-0.2cm}
	\caption{Mass-dependence of the DZ parameters: the inner slope $s_1$ and the concentration $c_2$ derived from the density profile fits as a function of stellar mass $M_{\rm star}$, halo mass $\Mvir$ and stellar-to-halo mass ratio $M_{\rm star}/\Mvir$.
	The inner slope $s_1$ is fitted using the function proposed by \protect\cite{Tollet2016} specificied in Eq.~(\ref{eq:s1(x)}): the best-fit curve is shown as the plain black line, while the \protect\cite{Tollet2016} fit to $s_1^\star$ is shown as the dashed black line. 
	The concentration $c_2$ is fitted using the function specified in Eq.~(\ref{eq:exp_function}).
	The values of the best-fitting parameters are indicated in Table~\ref{table:fit_rho}. 
	The rms $\sigma$ of the residuals, which is highlighted in gray, is obtained through an iterative process excluding points beyond $3\sigma$: this process does not affect the rms for $s_1$ but does affect that of $c_2$ as it excludes some of the points at high masses. 
	The mass-dependence of $s_1$ is marked by the presence of cores for $M_{\rm star}$ between $10^7$ and $10^{10}~\rm M_\odot$, $M_{\rm vir}$ between $10^{10.5}$ and $10^{11.5}~\rm M_\odot$, and $\log M_{\rm star}/M_{\rm vir}$ between ${-3.5}$ and ${-2}$, adiabatic contraction above. The mass-dependence of $c_2$ also reflects adiabatic contraction at high masses. The tightest relations are those as a function of the stellar-to-halo mass ratio. 
	}
	\vspace{-0.2cm}
	\label{fig:mass-dependence-rho}
\end{figure*}

\begin{figure*}
	\includegraphics[height=0.38\textwidth,trim={0.4cm 0 0.cm 0},clip]{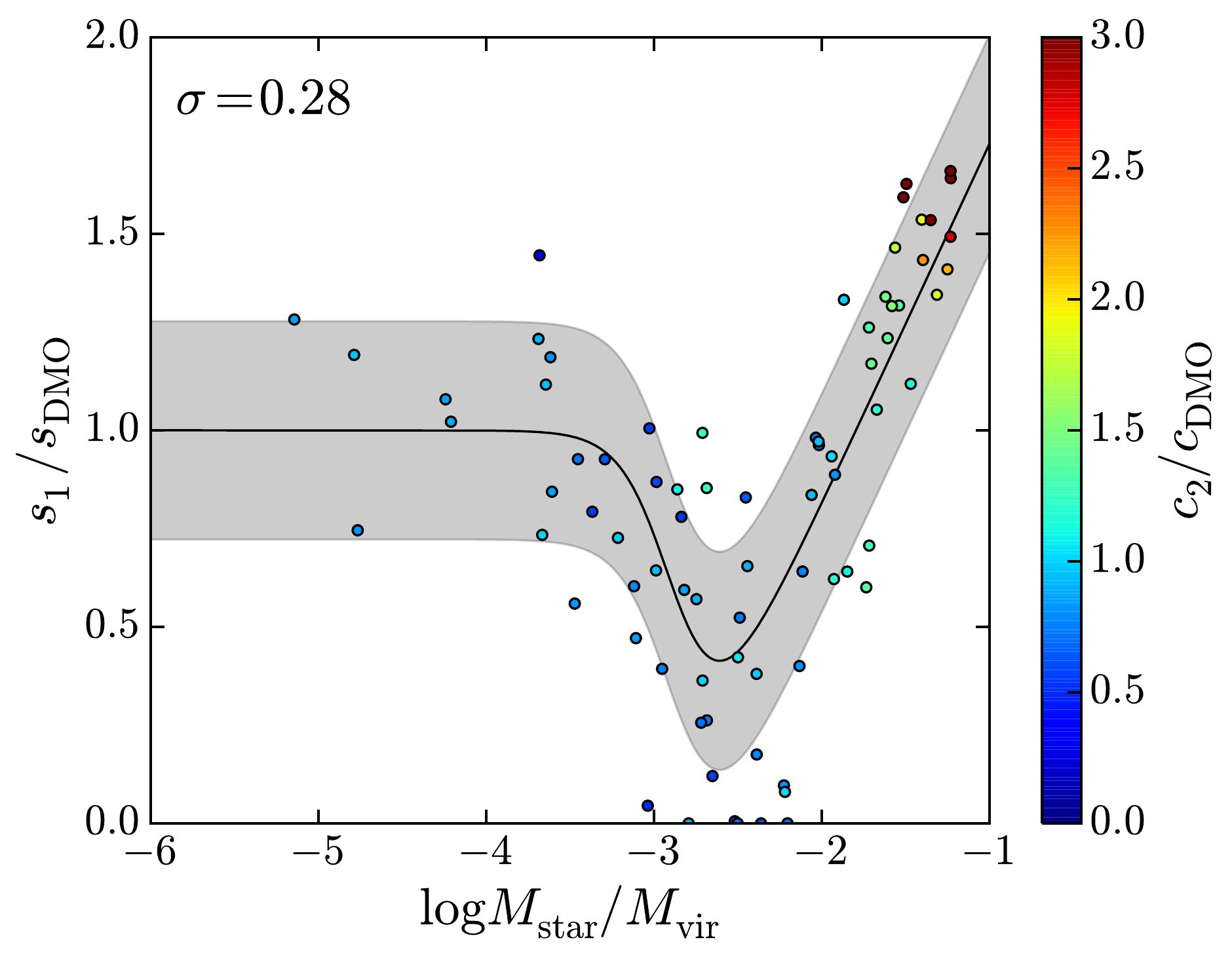}
	\hfill
	\includegraphics[height=0.38\textwidth,trim={0.4cm 0 0.cm 0},clip]{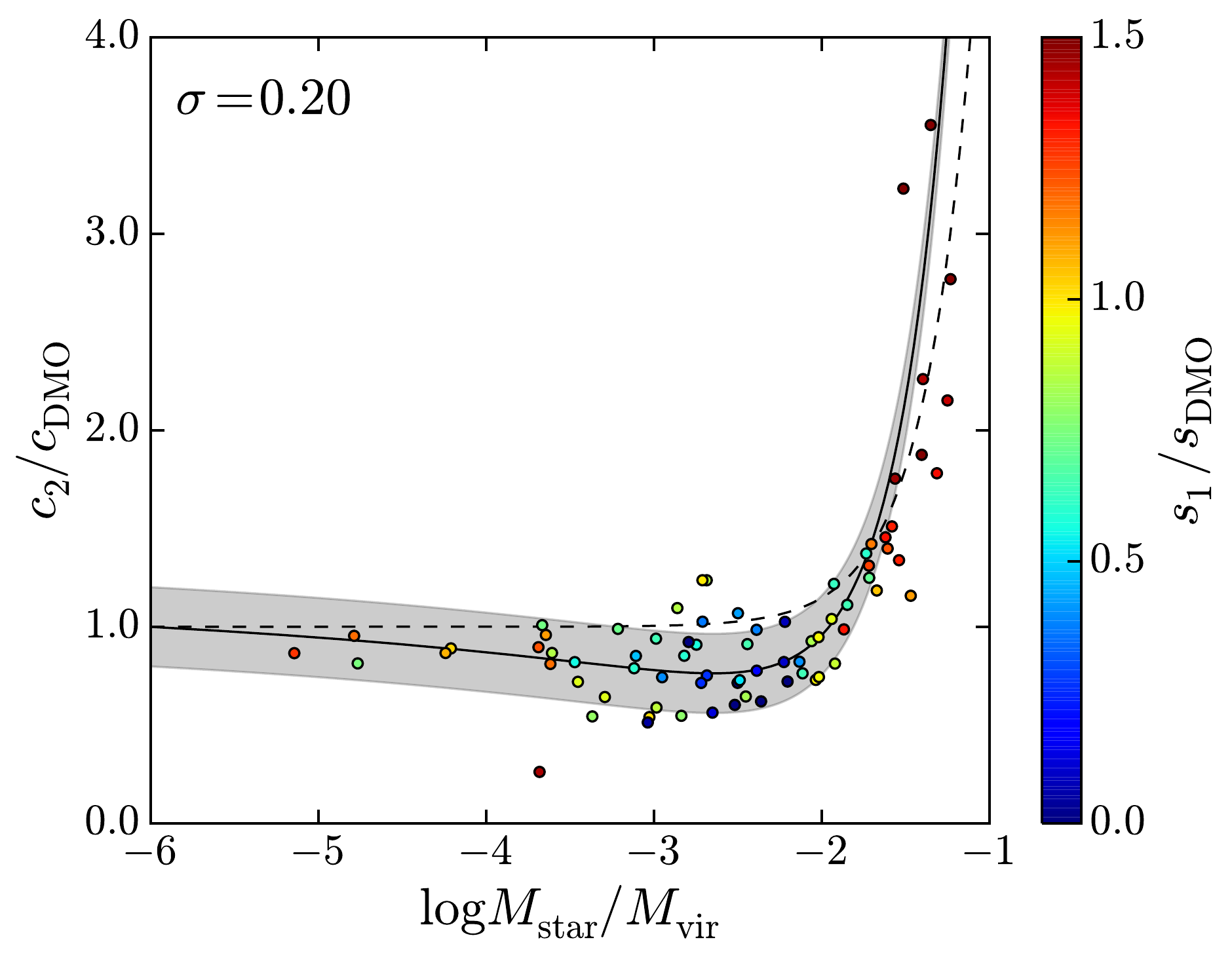}
	\vspace{-0.3cm}
	\caption{Mass-dependence of the DZ parameters at $z=0$ with respect to those of dark-matter-only NFW profiles: the inner slope and concentration $s_1$ and $c_2$ derived from the DZ density profile fits divided by their expected values for NFW haloes of similar halo mass using the \protect\cite{Dutton2014} relation (Eqs.~\ref{eq:c_DMO} and \ref{eq:s1_DMO}), $s_{\rm DMO}$ and $c_{\rm DMO}$, as a function of the stellar-to-halo mass ratio $M_{\rm star}/\Mvir$.
	The best-fit function following Eqs.~(\ref{eq:s1(x)}) and (\ref{eq:exp_function}), whose parameters are indicated in Table~\ref{table:fit_rho}, are shown as plain black lines. 
	The rms $\sigma$ of the residuals, which is highlighted in gray, is obtained through an iterative process excluding points beyond $3\sigma$ as in Fig.~\ref{fig:mass-dependence-rho}.
	The best-fit function for $c_2/c_{\rm DMO}$ obtained by \protect\cite{DiCintio2014b}, whose parameters following Eq.~(\ref{eq:exp_function_c0}) are $c^\prime=1.0$, $c^{\prime\prime}=1.32~10^2$, $\nu=7.83$ (cf. their Eq.~(6)), is indicated as a dashed black line. 
	Colors in each panel correspond to the y-axis of the other panel. 
	Core formation and halo expansion for $\log M_{\rm star}/\Mvir$ between -3.5 and -2 as well as adiabatic contraction are visible both in terms of $s_1/s_{\rm DMO}$ and $c_2/c_{\rm DMO}$.  
	}
	\label{fig:s1_c2_ratios}
\end{figure*}

Fig.~\ref{fig:mass-dependence-rho} shows the dependence of the inner slope $s_1$ and the concentration $c_2$ derived from the DZ density profile fits on the stellar mass $M_{\rm star}$, the halo mass $\Mvir$ and the stellar-to-halo mass ratio $M_{\rm star}/\Mvir$. 
At low stellar mass, halo mass, and stellar-to-halo mass ratio, the halo is dominated by DM and hence follow the NFW slope ($s_1 \approx 1.25$) and concentration ($c_2 \approx 10$); at intermediate mass and stellar-to-halo mass ratio, stellar feedback is strong enough to overcome the gravitational potential and expand the halo; at high mass and stellar-to-halo mass ratio, there is adiabatic contraction of the halo due to the steepening of the gravitational potential. 
Halo expansion occurs for stellar masses between $10^7$ to $10^{10}~\rm M_\odot$, halo masses between $10^{10.5}$ and $10^{11.5} ~\rm M_\odot$, and stellar-to-halo mass ratios between $10^{-3.5}$ and $10^{-2}$. 
As noted by \cite{DiCintio2014}, the range in stellar-to-halo mass ratio where core formation occurs is in agreement with the analytic calculation of \cite{Penarrubia2012} comparing the energy baryons must inject into a DM halo to remove its central cusp and the energy released by Type II supernovae explosions. 
We note that there is a hint of a small drop of the concentration $c_2$ in the range where core formation happens: feedback not only affect the inner part of the DM distribution, but also puffs-up the halo at larger scales. 
We recall that the NIHAO simulations used here do not include AGN feedback. As a consequence, the most massive haloes of the sample are partially overcooled, with $\log(M_{\rm star}/M_{\rm vir})$ close to $-1$. When AGN feedback is included, the stellar mass of the most massive haloes is reduced, their dark matter distribution relaxes, and their inner slope slightly decreases \citep{Blank2019, Maccio2020}.

We try to capture the behaviour of the inner slope $s_1$ as a function of $M_{\rm star}$, $\Mvir$ and $M_{\rm star}/\Mvir$ using the function
\be
\label{eq:s1(x)}
s_1(x) = \frac{s^\prime}{1+\left(\frac{x}{x_0}\right)^\nu} + s^{\prime\prime} \log \left(1+\left(\frac{x}{x_0}\right)^\nu\right)
\ee
where $x_0$, $s^\prime$, $s^{\prime\prime}$, and $\nu$  are ajustable parameters and $\log=\log_{10}$. We impose $s^\prime$ and $s^{\prime\prime}$ to be similar for the three variables $x$, which yields a unique asymptotical value $s^\prime=1.25$ when $x$ goes to zero.
This value corresponds approximately to an NFW cusp in the absence of baryons. Figure~\ref{fig:mass-dependence-rho}  further displays the fitting function obtained by \cite{Tollet2016} for the measured slope $s_1^\star$ between $1\%$ and $2\%$ of the virial radius $\Rvir$ in the same suite of cosmological zoom-in simulations with baryons. 
Motivated by \cite{Dutton2014} and \cite{DiCintio2014b}, we try to capture the behaviour of the concentration $c_2$ as a function of  $M_{\rm star}$, $\Mvir$ and $M_{\rm star}/\Mvir$ using the function 
\be
\label{eq:exp_function_c0}
c_2(x)= c^\prime \left(1+\left(\frac{x}{x_0} \right)^\nu\right)
\ee 
where $x_0$, $c^\prime$, and $\nu$ are adjustable parameters. We impose $c^\prime$ to be similar for the three variables $x$, yielding $c^\prime=11.5$. This asymptotical value when $x$ goes to zero is in accordance with fitting functions for the NFW concentration \citep[e.g., ][]{Dutton2014}. 
The values of the different fitting parameters are indicated in Table~\ref{table:fit_rho}, together with the rms of the residuals ($\sigma$). 
This latter quantity is obtained through an iterative process excluding points beyond $3\sigma$: this process does not affect the rms values of $s_1$, which are equal to the standard deviation of the residuals, but does affect those of $c_2$ as it excludes some of the points at high mass or high mass ratio. The steep exponential rise of $c_2$ indeed leads to artificially high residuals when taking only y-axis errors into account, which is reflected in the standard deviation of the residuals. The value of $\sigma$ obtained by the iterative process and indicated in the figure and the table corresponds to a very good approximation to the standard deviation inferred from the difference between the $16\%$ and $84\%$ quantiles of the residuals (which should be equal to $2\sigma$).

Although the different panels highlight significant scatter, we note that the tightest relations are those as a function of the stellar-to-halo mass ratio: both the inner slope $s_1$ and the concentration $c_2$ react to the presence of baryons. 
The smaller scatter obtained for the stellar-to-halo mass ratio than for the stellar and halo masses is in agreement with the results of \cite{DiCintio2014} and \cite{Tollet2016}, who show that the stellar-to-halo mass ratio (the `integrated star formation efficiency') is the best parameter to capture the effect of baryons on the DM distribution. 
This was also suggested by hydrodynamical simulations of dwarf galaxies \citep[e.g.,][]{Mashchenko2008, Madau2014, Verbeke2015, Read2016}, which showed that core formation occurs above a critical mass depending on the halo mass.

\subsubsection{Comparison with the dark-matter-only parameters}

\begin{table}
	\caption{Best-fitting parameters and relative errors for the slope and concentration relations shown in Figs.~\ref{fig:mass-dependence-rho} and \ref{fig:s1_c2_ratios}. The fitting functions are specified in Eqs.~(\ref{eq:s1(x)}), (\ref{eq:exp_function_c0}), and (\ref{eq:exp_function}). We impose $s^\prime$, $s^{\prime\prime}$, and $c^\prime$ to be the same as a function of the three variables $M_{\rm star}$, $M_{\rm vir}$, and $M_{\rm star}/M_{\rm vir}$  for $s_1$ and $c_2$. The rms $\sigma$ of the residuals within $3\sigma$ is also indicated.}
	\begin{tabular*}{\linewidth}{l@{\extracolsep{\fill}}l@{\extracolsep{\fill}}l@{\extracolsep{\fill}}l@{\extracolsep{\fill}}l@{\extracolsep{\fill}}l}
	\hline
	\hline
	\noalign{\vskip 0.5mm}
	Relation &  $x_0$  & $s^\prime$ & $s^{\prime\prime}$ & $\nu$  & $\sigma$\\
	\hline
	\noalign{\vskip 0.8mm} 
	$s_1(M_{\rm star})$& $5.18~10^7$& $1.25$ & $0.37$& $1.51$& $0.35$ \\
	\noalign{\vskip 0.8mm} 
	$s_1(M_{\rm vir})$& $3.99~10^{10}$& $1.25$& $0.37$& $3.00$& $0.38$\\
	\noalign{\vskip 0.8mm} 
	$s_1\left(\frac{M_{\rm star}}{\Mvir}\right)$& $1.30~10^{-3}$& $1.25$&  $0.37$& $2.98$ & $0.34$\\
	\noalign{\vskip 0.8mm} 
	$\left( \frac{s_1}{s_{\rm DMO}}\right)\left(\frac{M_{\rm star}}{\Mvir}\right)$& $1.30~10^{-3}$& $1$& $0.32$& $2.86$& $0.28$\\
	\noalign{\vskip 0.8mm} 
	\hline
	\hline
	\noalign{\vskip 0.5mm} 
	&  $x_0$ & $c^\prime$ & $\nu$&  $\mu$  & $\sigma$\\
	\hline
	\noalign{\vskip 0.8mm} 
	$c_2(M_{\rm star})$& $2.38~10^{10}$& $11.5$& $0.50$& - &   $4.82$\\
	\noalign{\vskip 0.8mm} 
	$c_2(M_{\rm vir})$ & $1.05~10^{12}$& $11.5$& $0.65$& - &   $4.66$\\
	\noalign{\vskip 0.8mm} 
	$c_2\left(\frac{M_{\rm star}}{\Mvir}\right)$& $3.04~10^{-2}$& $11.5$& $1.67$&  - &   $4.26$\\
	\noalign{\vskip 0.8mm} 
	$\left( \frac{c_2}{c_{\rm DMO}}\right)\left(\frac{M_{\rm star}}{\Mvir}\right)$&  $2.43~10^{-2}$ & $1.14$& $1.37$&  $0.142$ &  $0.20$\\
	\noalign{\vskip 0.8mm} 
	\hline
	\hline
	\vspace{-0.6cm}
	\end{tabular*}
	\label{table:fit_rho}
\end{table}

To isolate the effect of the introduction of baryonic processes on the inner slope $s_1$ and the concentration $c_2$, we normalize these two quantities by their expected NFW values in dark-matter-only simulations, $s_{\rm DMO}$ and $c_{\rm DMO}$. 
Namely, we use the best-fitting relation for the NFW concentration as a function of halo mass \citep[measured using][]{Bryan1998} from \cite{Dutton2014}, 
\be
\label{eq:c_DMO}
\log c_{\rm DMO} = 1.025 - 0.097 \log\left( \frac{h\Mvir}{10^{12}\rm M_\odot} \right)
\ee 
with $h=0.671$ the dimensionless Hubble parameter \citep{Planck2014}, 
the corresponding NFW slope at $0.01\Rvir$ being
\be
\label{eq:s1_DMO}
s_{\rm DMO}= \frac{1+0.03 c_{\rm DMO}}{1+0.01 c_{\rm DMO}}. 
\ee	
Fig.~\ref{fig:s1_c2_ratios} shows the slope and concentration ratios $s_1/s_{\rm DMO}$ and $c_2/c_{\rm DMO}$ as a function of the stellar-to-halo mass ratio $M_{\rm star}/\Mvir$, which is the variable leading to the lowest scatter in Fig.~\ref{fig:mass-dependence-rho}.
Fig.~\ref{fig:s1_c2_ratios} highlights the formation of shallow cores for $\log M_{\rm star}/\Mvir$ between -3.5 and -2 and adiabatic contraction above. Both effects are visible not only in terms of $s_1/s_{\rm DMO}$, but also in terms of $c_2/c_{\rm DMO}$: while the slope ratio decreases from 1 to below 0.5 before increasing above 1.5 as  $M_{\rm star}/\Mvir$ increases, the concentration ratio decreases to $\sim$$0.8$  before sharply rising up to $\sim$$3$. 
This drop in halo concentration for $\log(M_{\rm star}/M_{\rm vir})$ between $-3.5$ and $-2$ had not been seen previously, as highlighted by the dashed line obtained by \cite{DiCintio2014b}, but was also recently reported by \cite{Lazar2020} using the FIRE-2 simulations.

We fit the slope ratio $s_1/s_{\rm DMO}$ as a function of $M_{\rm star}/M_{\rm vir}$ with the function of Eq.~(\ref{eq:s1(x)}) and $s^\prime=1$ to impose an NFW slope when  $M_{\rm star}/\Mvir$ goes to zero. 
The concentration ratio is fitted as a function of $M_{\rm star}/M_{\rm vir}$ with the function of Eq.~(\ref{eq:exp_function})
plus a second power-law term to account for the dip of concentration when $\log M_{\rm star}/\Mvir$ is between -3.5 and -2, namely
\be
\label{eq:exp_function}
\left( \frac{c_2}{c_{\rm DMO}}\right)(x) = c^\prime \left(1+\left(\frac{x}{x_0} \right)^\nu\right) -x^\mu
\ee 
with $x_0$, $c^\prime$ $\nu$, and $\mu$ four adjustable parameters constrained to yield $c_2/c_{\rm DMO}=1$ at $\log M_{\rm star}/\Mvir=-6$. 
Table~\ref{table:fit_rho} lists the best-fit parameters of the functions describing the slope and concentration ratios and the rms of the residuals, which indicates the scatter of the two relations. 
A large part of this scatter has a physical origin related to the individual merger and star formation histories of the simulated galaxies. In particular, we note that the scatter in stellar mass at fixed halo mass is estimated to be between 0.16-0.2 dex at $z=0$ \citep[e.g.,][]{More2009, Reddick2013, Behroozi2013}. The processes responsible for this scatter, such as mergers, star formation, and feedback, are expected to affect DM haloes as well (cf. introduction) and hence the inner slope and the concentration parameter associated to the DZ fits. 
We further note from the colorscale on both panels that the inner slope and concentration ratios $s_1/s_{\rm DMO}$ and $c_2/c_{\rm DMO}$ are correlated.

\section{A mass-dependent profile}
\label{section:mass}

\subsection{Prescriptions}
\label{section:prescriptions}

Section \ref{section:mass_dependence} establishes the DZ profile as a mass-dependent profile, whose shape parameters $s_1$ and $c_2$ (or equivalently, $a$ and $c$) are set by the stellar-to-halo mass ratio $M_{\rm star}/\Mvir$. It further provides fitting functions for the dependences of $s_1$ and $c_2$ on $M_{\rm star}/\Mvir$. As for the \citetalias{DiCintio2014b} profile, it is thus possible to derive the shape of the DM distribution taking into account the effect of baryons for any halo given its stellar or halo mass. While the \citetalias{DiCintio2014b} profile uses four shape parameters including the concentration, the DZ profile describes the DM distribution with only two parameters, with the advantage to have analytic expressions for the gravitational potential and the velocity dispersion
(cf. \citetalias{Zhao1996}, \citetalias{Dekel2017}), the resulting kinetic energy (cf. \citetalias{Freundlich2020}), and lensing properties (cf. Section \ref{section:dekel}). Inspired by the Appendix of \cite{DiCintio2014b}, we provide here prescriptions to derive the DZ DM profile associated to any given halo.

(i) The inputs are the halo mass $\Mvir$ and the stellar mass $M_{\rm star}$. If only one of the two quantities is known, one can use an abundance matching $M_{\rm star}/M_{\rm vir}$ relation to derive the other one \citep[e.g.][]{Moster2013, Behroozi2013, Behroozi2019, Rodriguez-Puebla2017}. 

(ii) Determine the virial radius $\Rvir$ using the overdensity criterion 
\be
\Mvir = \frac{4\pi}{3} \Rvir^3 \Delta \rho_{\rm crit}
\ee 
with $\Delta=18\pi^2 +82 x -39 x^2$ at $z=0$ for $x=\Omega_m - 1$ from \cite{Bryan1998} and $\rho_{\rm crit} = 3H^2/8\pi G$ the critical density of the Universe. With the \cite{Planck2014} parameters, $\Delta = 103.5$ and $\rho_{\rm crit}=124.9~\rm  M_\odot kpc^{-3}$. 

(iii) Compute the inner slope and concentration ratios $s_1/s_{\rm DMO}$ and $c_2/c_{\rm DMO}$ from the stellar-to-halo mass ratio $M_{\rm star}/\Mvir$ using the fitting functions from Eqs.~(\ref{eq:s1(x)}) and (\ref{eq:exp_function}), whose best-fit parameters are indicated in Table~\ref{table:fit_rho}. These functions were obtained in the range $-5\leq \log(M_{\rm star}/\Mvir)\leq -1$ and converge to 1 for smaller values of $\log(M_{\rm star}/\Mvir)$. 

(iv) Obtain the slope $s_1$ and the concentration $c_2$ from the corresponding ratio using the typical concentration $c_{\rm DMO}$ of a DM-only NFW halo from \cite{Dutton2014}, recalled in Eq.~(\ref{eq:c_DMO}), and the corresponding inner slope at $0.01\Rvir$, $s_{\rm DMO}$, expressed in Eq.~(\ref{eq:s1_DMO}). 

(v) Convert $s_1$ and $c_2$ into the DZ parameters $a$ and $c$ using Eqs.~(\ref{eq:a(s1,c2)}) and (\ref{eq:c(s1,c2)}). We recall that these latter parameters are not as physically meaningful as  $s_1$ and $c_2$. 

(vi) Obtain the scale radius $r_c$ and the characteristic density $\rho_c$ entering the expression of the density, $r_c=\Rvir/c$ and $\rho_c=(1-a/3) c^3 \mu \overline{\rho_{\rm vir}}$, with $\mu = c^{a-3} (1+c^{1/2})^{2(3-a)}$ and  $\overline{\rho_{\rm vir}}=3\Mvir/4\pi \Rvir^3 =\Delta \rho_{\rm crit}$. 

(vii) Determine the mass-dependent density profile using Eq.~(\ref{eq:rho32}); the corresponding circular velocity profile using Eqs.~(\ref{eq:rhob}) and (\ref{eq:V2(r)}), with $\rho_c=c^3\mu\overline{\rho_{\rm vir}}$, $b=2$, and $\overline{g}=3$. The gravitational potential profile is obtained from Eq.~(\ref{eq:U32_1}), the velocity dispersion profile from Eq.~(\ref{eq:sigma_dekel}), the projected surface density profile from Eq.~(\ref{eq:Sigma_H}) or its series expansion (Eq.~\ref{eq:appendix_sigma_expansion}), the scaled deflexion angle from Eq.~(\ref{eq:lensing_alpha_H}) or its series expansion (deduced from Eq.~(\ref{eq:appendix_M2D_expansion})), the lensing shear from the average projected surface density of Eq.~(\ref{eq:Sigmab_H}) or its series expansion (Eq.~(\ref{eq:appendix_Sigmab_expansion})). Table~\ref{table:expressions} below summarizes the different analytic expressions available for the DZ profile. 

In the following section, we show that these prescriptions for the DZ profile are in relatively good agreement with simulated density and circular velocity profiles and fare as good as the \citetalias{DiCintio2014b} prescriptions given the stellar and halo masses. 
When fitting rotation curves of galaxies, we however advocate to release the mass-dependent prescription for the concentration $c_2$ and to leave this parameter free \citep[as  advocated for the \protect\citetalias{DiCintio2014b} profile by][]{DiCintio2014b}. This enables to obtain extremely good fits to simulated density and circular velocity profiles (cf. Section~\ref{section:rotation}).

\subsection{Accuracy of the mass-dependent prescriptions}
\label{section:prescription_accuracy}

\begin{figure}
	\centering
	\includegraphics[height=0.68\linewidth,trim={2.2cm 0cm 4.cm 1.5cm},clip]{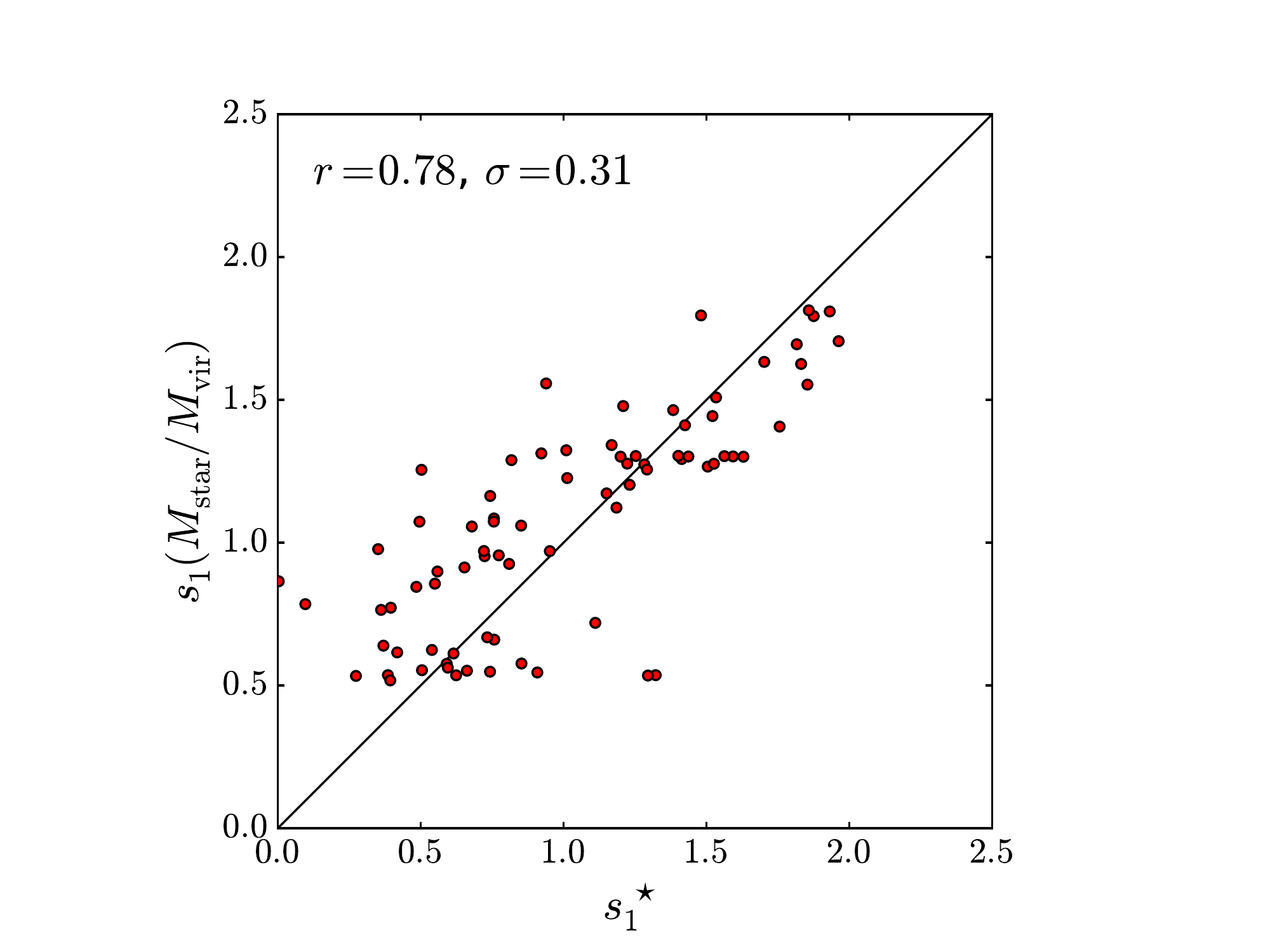}
	\vspace{-0.2cm}
	\hfill
	\includegraphics[height=0.68\linewidth,trim={2.2cm 0 4.cm 1.5cm},clip]{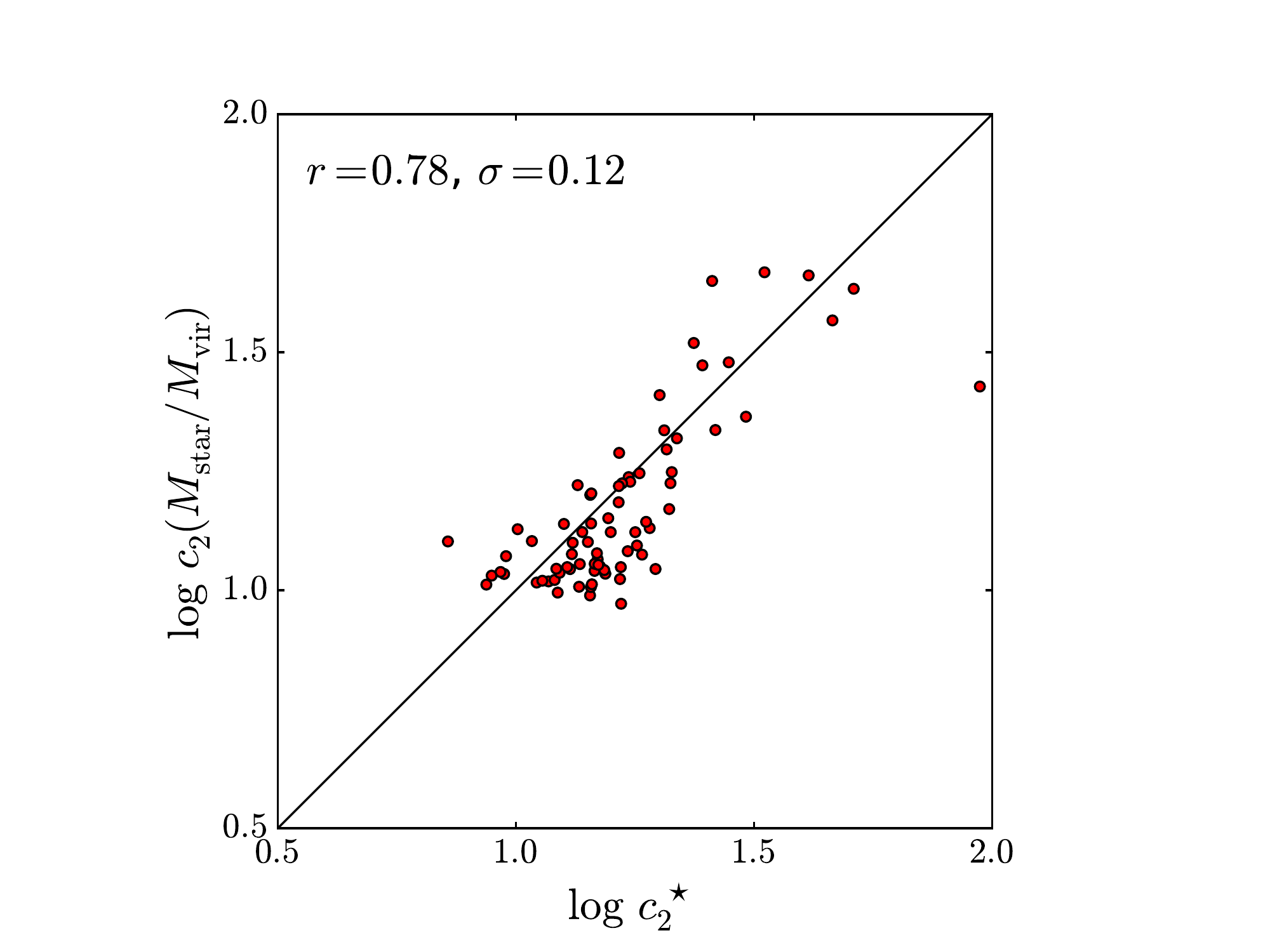}
	\vspace{-0.3cm}
	\caption{Prescripted versus simulated parameters: comparison between the inner slope and concentration stemming from the mass-dependent prescriptions for the $z=0$ NIHAO galaxies with baryons, and those obtained directly from the simulated profiles, $s_1^\star$ and $c_2^\star$. The plain lines corresponds to a linear least-square fits. The Pearson correlation coefficient and the residual scatter are indicated. The prescriptions enable to retrieve the inner slope and concentration. 
	}
	\label{fig:comparison_s12_c2_prescriptions}
\end{figure}

\begin{figure*}
	\includegraphics[width=\textwidth]{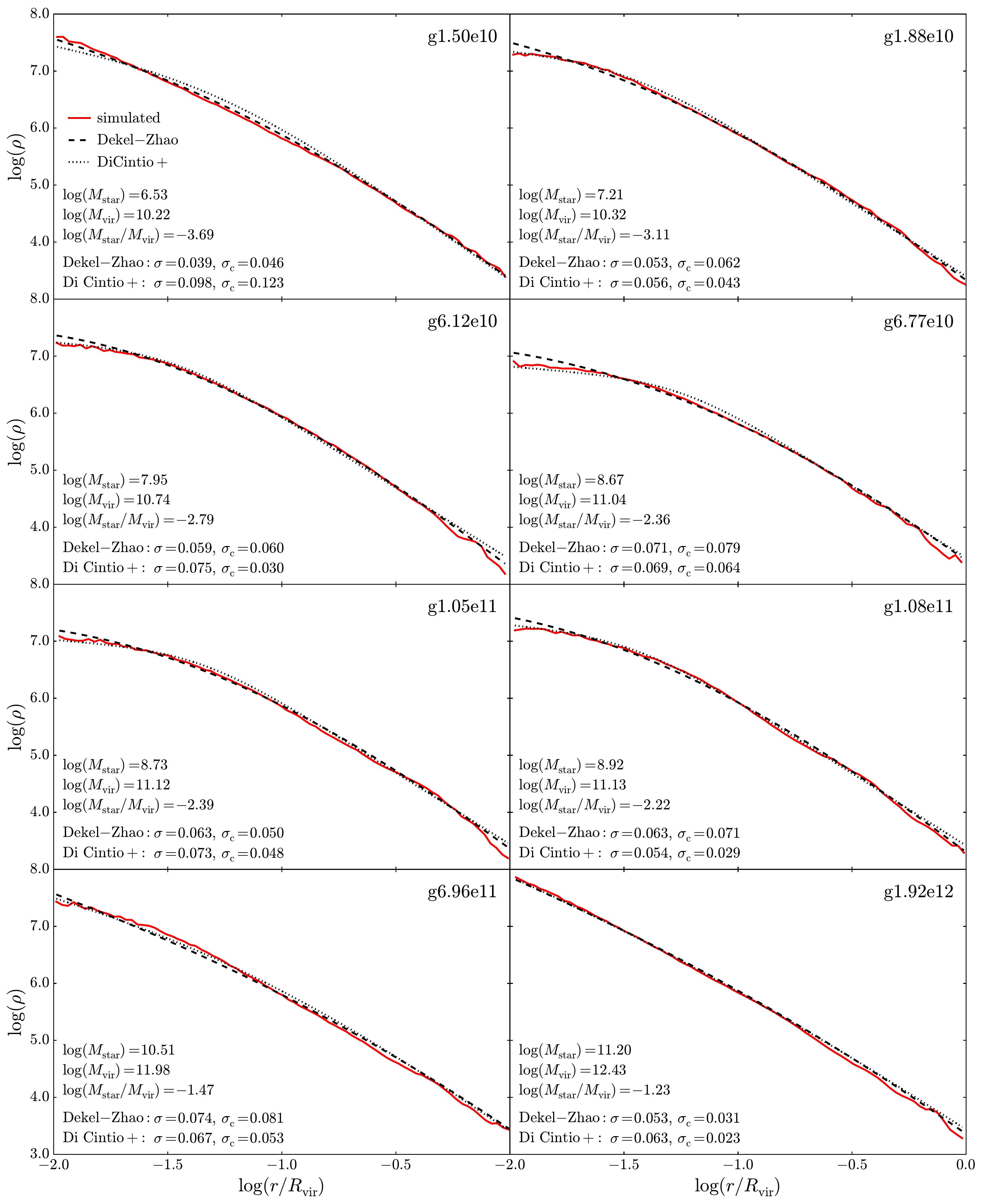}
	\vspace{-0.6cm}
	\caption{Prescripted versus simulated density profiles when fitting rotation curves (leaving the concentration free): the dark matter density profiles at $z=0$ of the 8 arbitrary NIHAO galaxies shown in Fig.~\ref{fig:fits} (plain red line) with their DZ (dashed) and \protect\citetalias{DiCintio2014b} (dotted) one-parameter fit to the rotation curves. 
		For the DZ profile, the inner slope $s_1$ is set by the fitting function of Fig.~\ref{fig:s1_c2_ratios} (cf. Eq.~(\ref{eq:s1(x)}) and Table~\ref{table:fit_rho}) but the concentration $c_2$ is allowed to vary; for the \protect\citetalias{DiCintio2014b} profile, the shape parameters $a$, $b$, $g$ of Eq.~(\ref{eq:rho_abc}) are set by their mass-dependent prescriptions \citep[][Eq.~(3)]{DiCintio2014b}, the scale radius $r_c$ is allowed to vary, and the characteristic density $\rho_c$ is constrained by the halo mass $\Mvir$. 
		The masses $M_{\rm star}$, $M_{\rm vir}$, $M_{\rm star}/M_{\rm vir}$ and the rms errors $\sigma$ and $\sigma_{\rm center}$ are indicated. 
		Both the DZ and the \protect\citetalias{DiCintio2014b} parametrizations provide extremely good fits to the density profiles. 
	}
	\label{fig:Vfit}
\end{figure*}

\begin{figure*}
	\includegraphics[width=\textwidth]{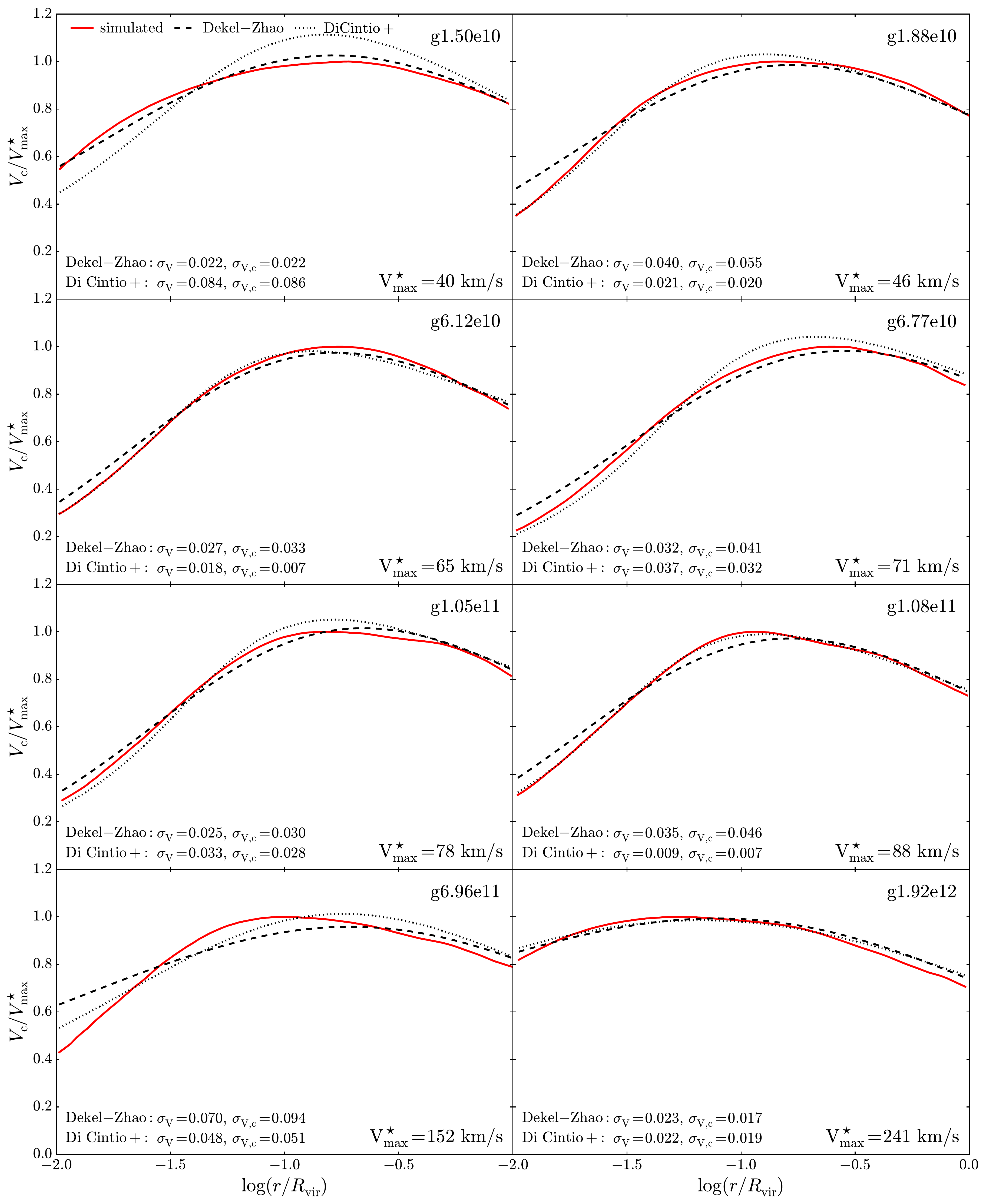}
	\vspace{-0.4cm}
	\caption{Prescripted versus simulated rotation curves when leaving the concentration free: 
		dark matter circular velocity profiles, $V_{\rm c}(r) = \sqrt{GM(r)/r}$, of the eight $z=0$ NIHAO galaxies shown in Fig.~\ref{fig:fits} (plain red line) together with those inferred from the DZ and \protect\citetalias{DiCintio2014b} one-parameter fit to the rotation curves (dashed and dotted lines, respectively). 
		For the DZ profile, the inner slope $s_1$ is set by the fitting function of Fig.~\ref{fig:s1_c2_ratios} (cf. Eq.~(\ref{eq:s1(x)}) and Table~\ref{table:fit_rho}) but the concentration $c_2$ is allowed to vary; for the \protect\citetalias{DiCintio2014b} profile, the shape parameters $a$, $b$, $g$ of Eq.~(\ref{eq:rho_abc}) are set by their mass-dependent prescriptions \citep[][Eq.~(3)]{DiCintio2014b}, the scale radius $r_c$ is allowed to vary, and the characteristic density $\rho_c$ is constrained by the halo mass $\Mvir$. 
		The velocity of each galaxy is normalized to its maximum value $V_{\rm max}^\star$, which is an increasing function of mass. 
		Both the DZ and the \protect\citetalias{DiCintio2014b} parametrizations provide extremely good fits to the rotation curves, with differences below $10\%$ that are well within observational errors. 
	}
	\label{fig:Vfit_Vc}
\end{figure*}

\begin{figure*}
	\includegraphics[height=0.23\textwidth,trim={0.cm 0 1.5cm 1.cm},clip]{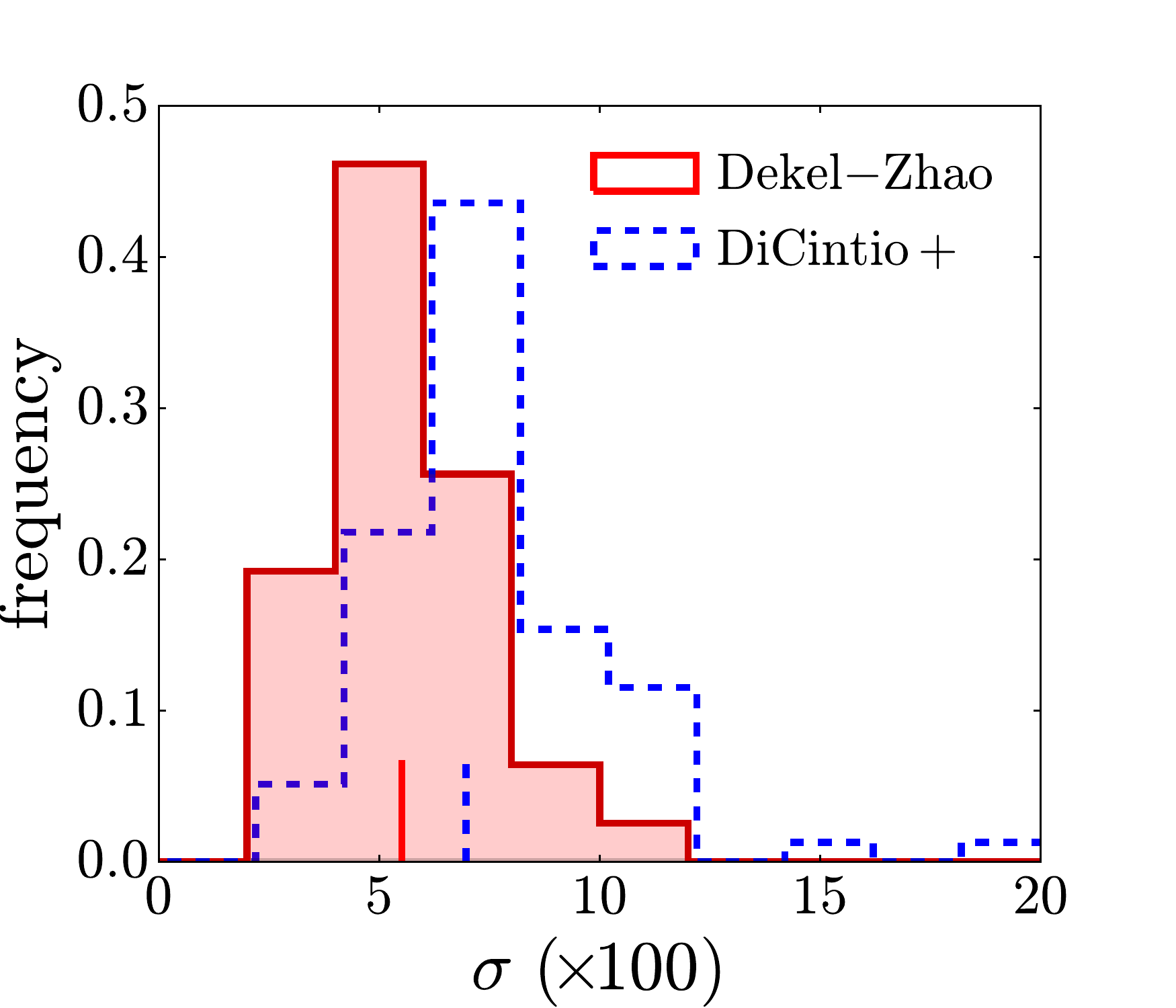}
	\includegraphics[height=0.23\textwidth,trim={1.5cm 0 1.5cm 1.cm},clip]{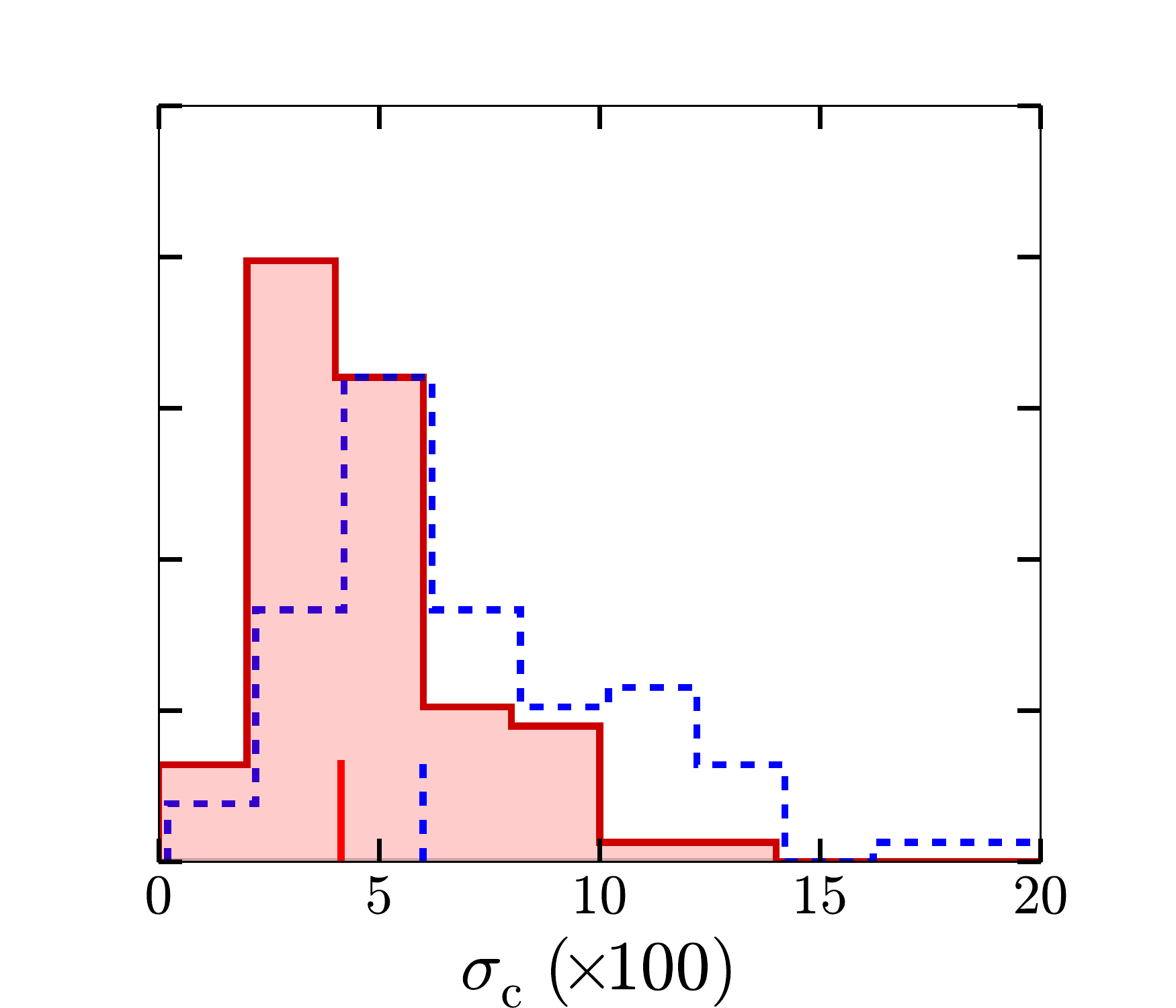}
	\includegraphics[height=0.23\textwidth,trim={1.5cm 0 1.5cm 1.cm},clip]{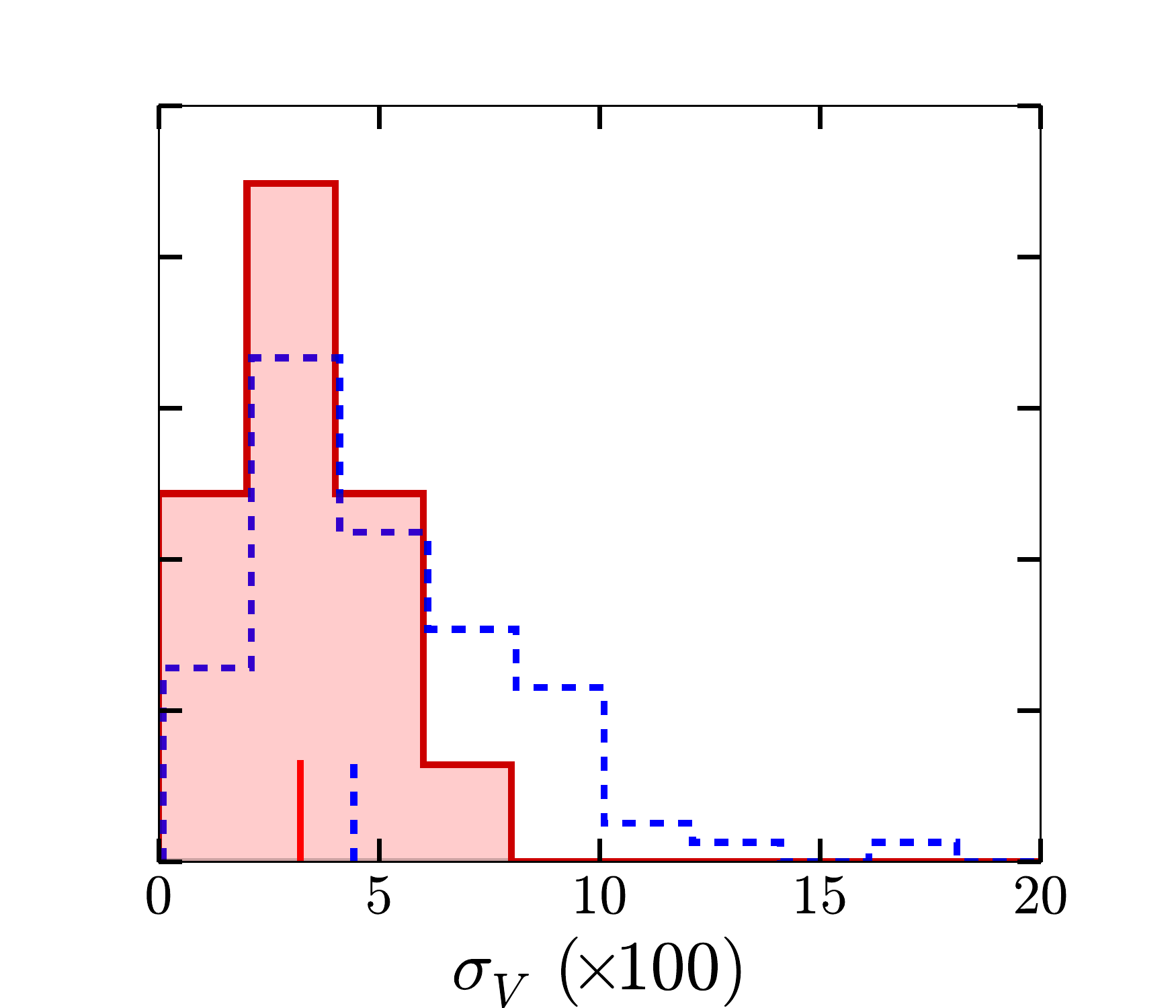}
	\includegraphics[height=0.23\textwidth,trim={1.5cm 0 1.5cm 1.cm},clip]{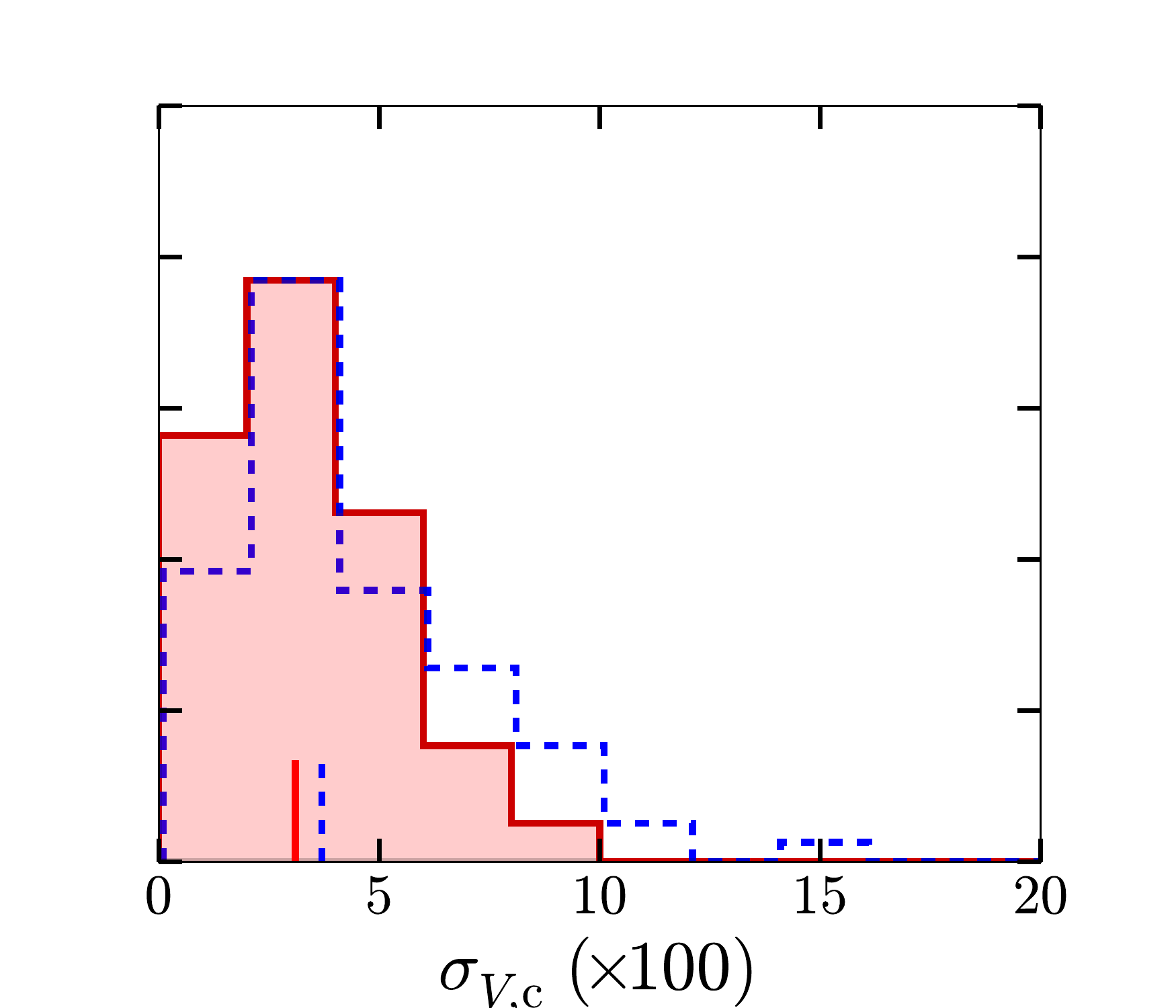}	
	\caption{Comparing the current DZ and \protect\citetalias{DiCintio2014b} rotation curve fits in terms of their rms errors: rms errors in $\log\rho$ and $V_c/V_{\rm max}$ of the current (plain red line) and \protect\citetalias{DiCintio2014b} (blue dashed line) one-parameter fit to the rotation curves over the ranges $0.01\Rvir-\Rvir$ and $0.01\Rvir-0.1\Rvir$ for all NIHAO galaxies at $z=0$. 
	The median values for the two prescriptions, which are highlighted by vertical lines above the x-axis, respectively yield 
		$0.055$ \& $0.070$ for $\sigma$, 
		$0.041$ \& $0.060$  for $\sigma_{\rm center}$, 
		$0.032$ \& $0.044$  for $\sigma_{\rm V}$, 
		$0.031$ \& $0.037$ for $\sigma_{\rm V, center}$. 
	The standard deviations respectively yield 
		$0.017$ \& $0.026$ for $\sigma$, 
		$0.022$ \& $0.035$  for $\sigma_{\rm center}$, 
		$0.015$ \& $0.030$  for $\sigma_{\rm V}$, 
		$0.019$ \& $0.027$ for $\sigma_{\rm V, center}$. 
	The DZ profile provides marginally better fits than the \protect\citetalias{DiCintio2014b} profile. 
	}
	\label{fig:Vfit_sigma_hist}
\end{figure*}

\begin{figure*}
	\vspace{-0.4cm}
	\includegraphics[height=0.23\textwidth,trim={0.cm 0 1.5cm 1.cm},clip]{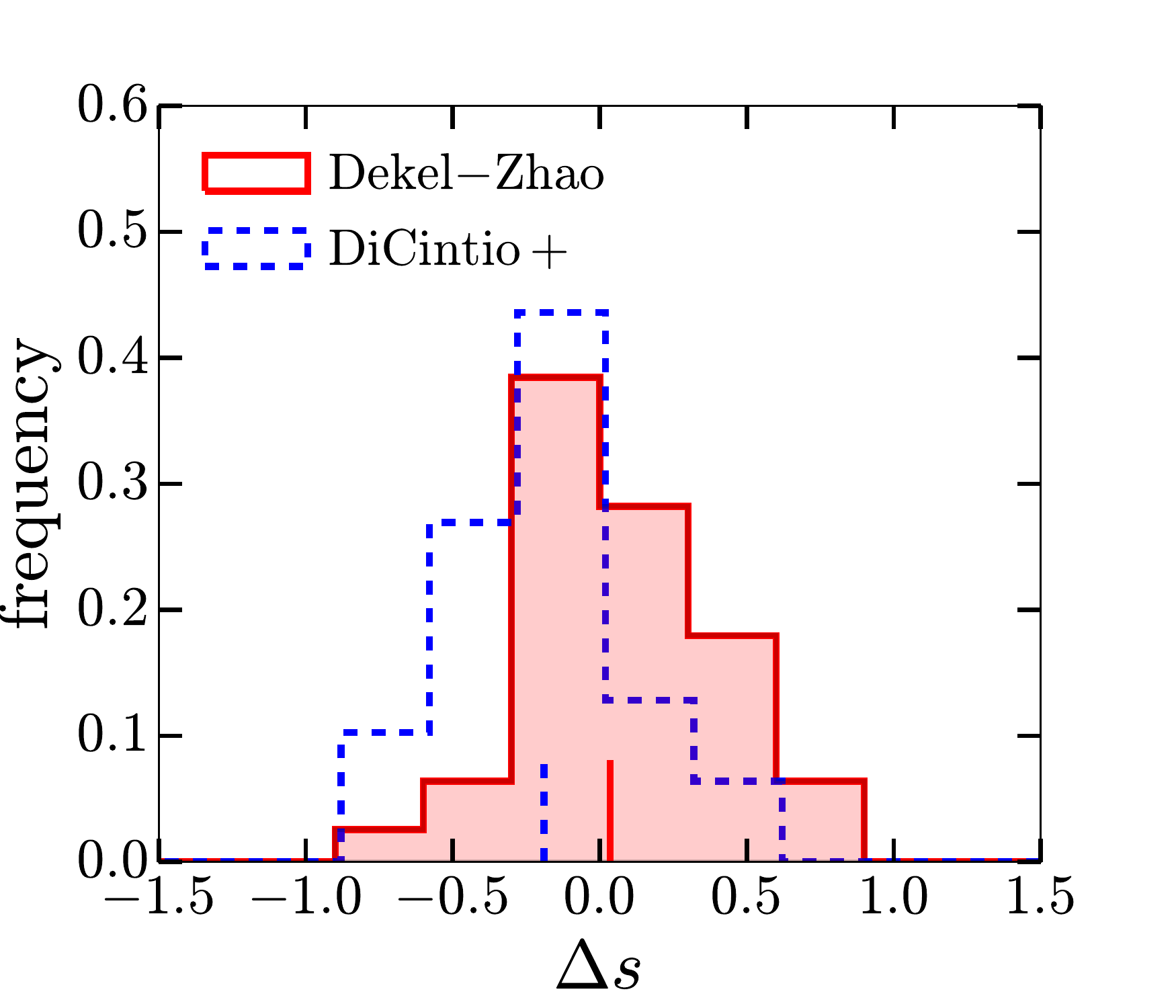}
	\includegraphics[height=0.23\textwidth,trim={1.5cm 0 1.5cm 1.cm},clip]{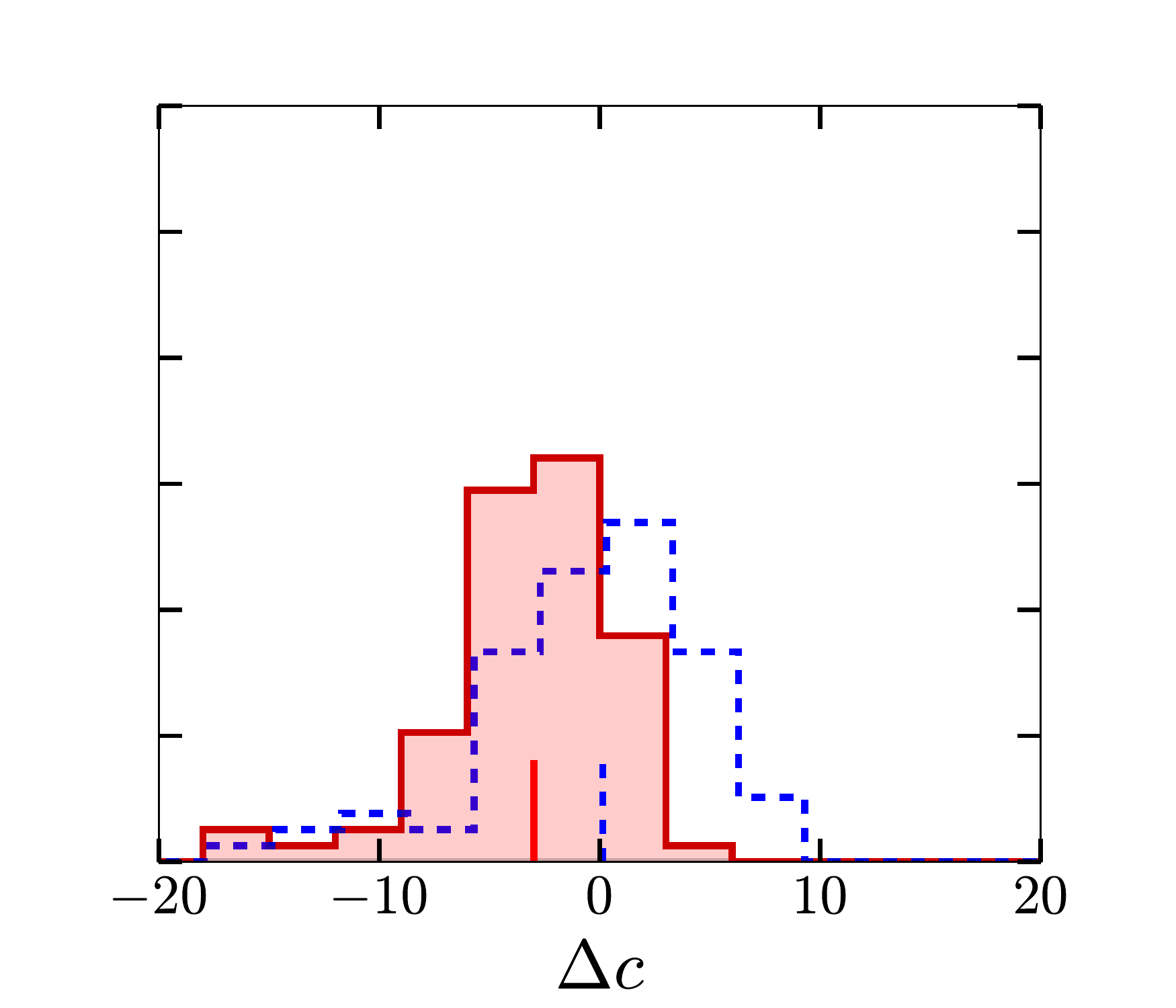}
	\includegraphics[height=0.23\textwidth,trim={1.5cm 0 1.5cm 1.cm},clip]{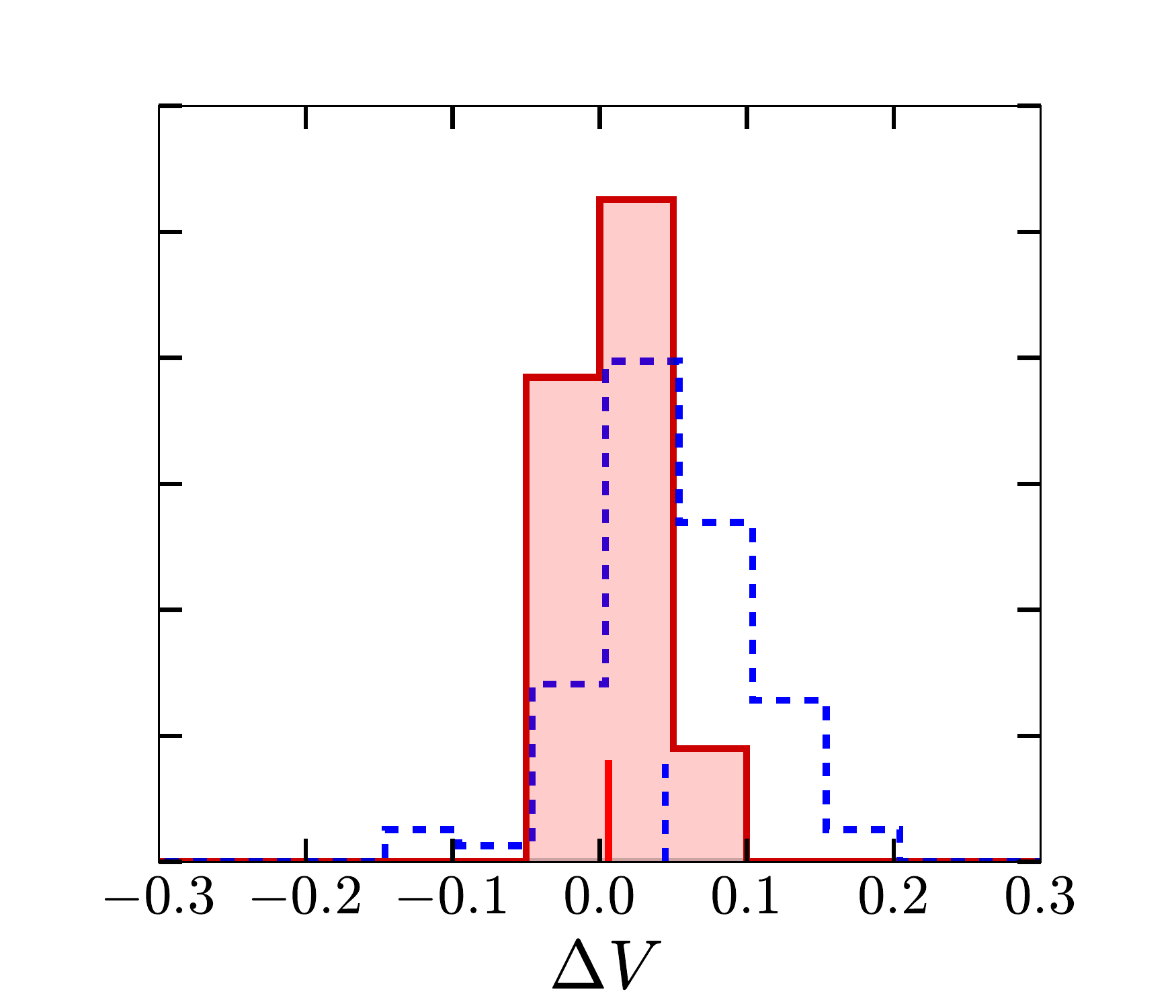}
	\includegraphics[height=0.23\textwidth,trim={1.5cm 0 1.5cm 1.cm},clip]{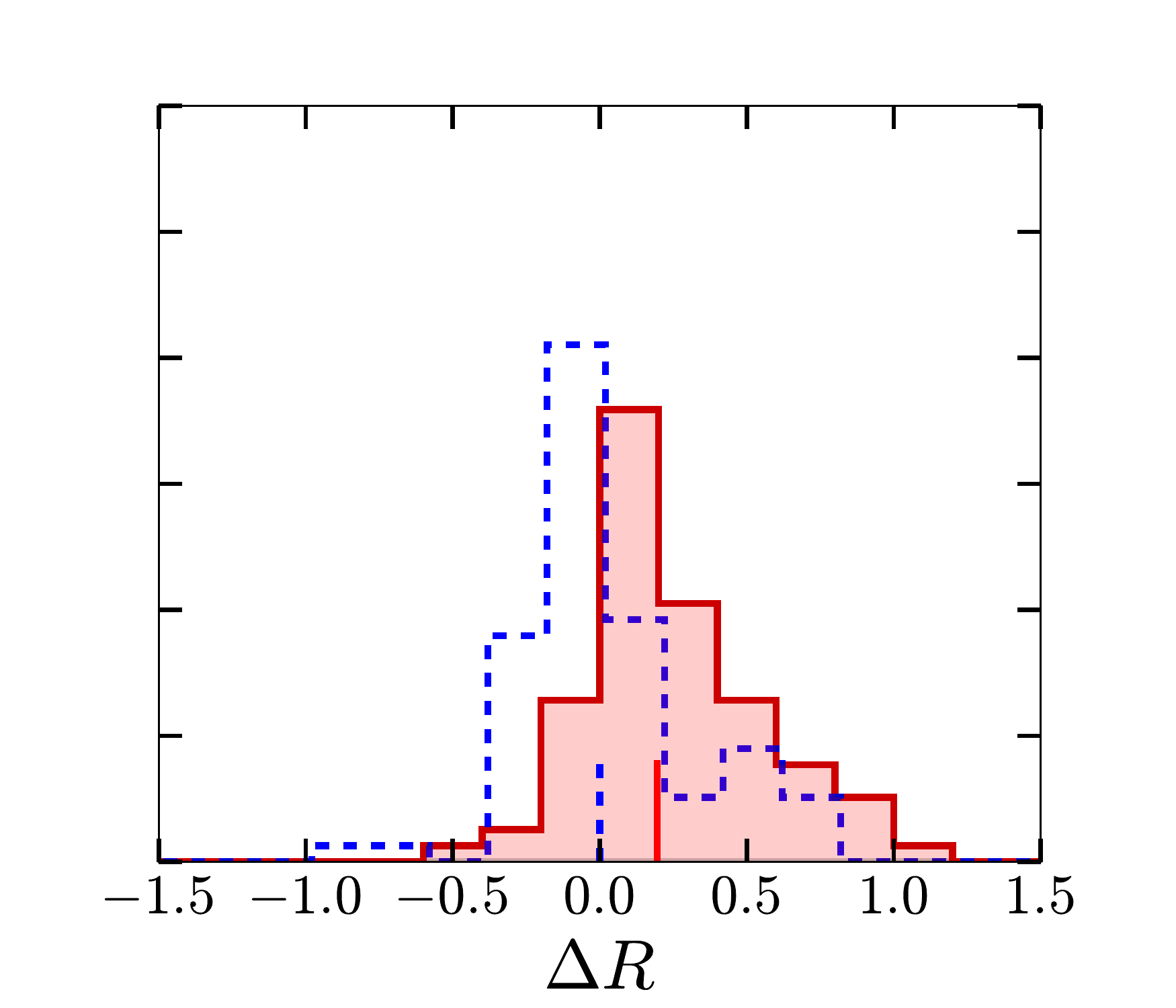}
	\caption{Comparing the current DZ and \protect\citetalias{DiCintio2014b} rotation curve fit parameters: 
		inner slope and concentration differences, $\Delta s = s_{1,\rm model}-s_{1}^\star$ and $\Delta c = c_{\rm 2, model}-c_{\rm 2}^\star$, as well as the maximum velocity and radius relative differences, $\Delta V= (V_{\rm max, model}-V_{\rm max}^\star)/V_{\rm max}^\star$ and $\Delta R= (R_{\rm max, model}-R_{\rm max}^\star)/R_{\rm max}^\star$, between  the 
		DZ (plain red line) and the \protect\citetalias{DiCintio2014b} (blue dashed line) one-parameter fit to the rotation curves and the simulated profiles for all $z=0$ NIHAO galaxies simulated with baryons. 
		The median values for the two prescriptions, which are highlighted by vertical lines above the x-axis, respectively yield 
		$0.04$ \& $-0.19$ for $\Delta s$, 
		$-3.0$ \& $0.1$ for $\Delta c$, 
		$0.01$ \& $0.05$ for $\Delta V$,
		$0.20$ \& $0.00$ for $\Delta R$. 
		The standard deviations respectively yield 
		$0.31$ \& $0.29$ for $\Delta s$,
		$8.6$ \& $9.0$ for $\Delta c$, 
		$0.03$ \& $0.06$  for $\Delta V$, 
		$0.30$ \& $0.30$  for $\Delta R$. 
		The mass-dependent prescription for the inner logarithmic slope $s_1$ enables extremely good fits to rotation curves of simulated galaxies when the concentration $c_2$ is allowed to vary. 
	}
	\label{fig:Vfit_comparison_histograms}
\end{figure*}

Fig.~\ref{fig:comparison_s12_c2_prescriptions} compares the inner logarithmic slope and the concentration stemming from the mass-dependent prescriptions of Section \ref{section:prescriptions}  with $s_1^\star$ and $c_2^\star$ determined directly from the simulated profiles (cf. Section~\ref{section:fitting}). Although the Pearson correlation coefficients are slighly lower than those of Fig.~\ref{fig:comparison_s12_c2}, the inner slope and concentration are well recovered. 
Overall, these mass-dependent prescriptions enable to retrieve the inner slope $s_1$ with a $\pm 0.31$ scatter and a negligible systematic error and the concentration $c_2$ with a $\pm 9$ scatter and a small $-1.5$ systematic offset (0.12 dex scatter and $-0.05$ dex offset in $\log c_2$). 
As further shown in Figs.~\ref{fig:prescriptions_sigma_hist} and \ref{fig:prescriptions_comparison_histograms}, these prescriptions retrieve the maximum velocity $V_{\rm max}$ with a $\pm 9\%$ scatter and a $+3\%$ offset, and the corresponding radius $R_{\rm max}$ with a $\pm 31\%$ scatter and a $+12\%$ offset ($0.14$ dex scatter and $+0.05$ dex offset in $\log R_{\rm max}$). 
The scatters and offsets in $\Delta s$, $\Delta c$, $\Delta V$, and $\Delta R$ are comparable to those described for the fits in Section~\ref{section:fitting} but the rms errors and the discrepancies between prescripted and simulated profiles are significantly higher ($\sigma$ progressing on average from $0.046$ to $0.080$, $\sigma_V$ from $0.027$ to $0.072$, and similar trends for $\sigma_{\rm center}$ and $\sigma_{\rm V, center}$), especially at high stellar-to-halo mass ratio: while the overall scatters and offsets are preserved,  discrepancies arise on a case by case basis. The difference between the parametrized and the simulated rotation curves can be as high as $20\%$. As discussed in the following section, releasing the constraint on the concentration when fitting rotation curves enables to significantly improve the fits.

In Appendix~\ref{appendix:prescriptions}, we further compare the current mass-dependent prescriptions with those of  \cite{DiCintio2014b}. For this other mass-dependent profile, the four shape parameters entering Eq.~(\ref{eq:rho_abc}) -- namely  $a$, $b$, $g$, and the concentration parameter associated to the scale radius $r_c$ -- are expressed as a function of the stellar-to-halo mass ratio while the scale density $\rho_c$  is deduced from the halo mass, since the enclosed mass associated to the profile must verify
$M(\Rvir)=\Mvir$. 
Both for the current and \citetalias{DiCintio2014b}  prescriptions, all the parameters describing the profiles are set given the stellar and halo masses. 
Appendix~\ref{appendix:prescriptions} shows that the prescriptions of Section~\ref{section:prescriptions} provide equally good (or even marginally better) fits to the simulated density and velocity profiles than the \citetalias{DiCintio2014b} prescriptions. 
We caution however that while the current prescriptions stem from the NIHAO sample itself, the \citetalias{DiCintio2014b} prescriptions were obtained from a smaller sample of 10 simulated galaxies \citep[the \mbox{MaGICC} sample;][]{Brook2012, Stinson2013}, such that the slightly better accuracy of the current prescriptions is most likely due to the different nature and size of the simulations used. 
We thus prefer to conclude that the accuracy of the two prescriptions are comparable. 
An update of the  \citetalias{DiCintio2014b} prescriptions with the NIHAO simulations  would indeed slightly increase their accuracy within the current sample, but is left for future work -- especially as it would only lead to small differences and as the \citetalias{DiCintio2014b} prescriptions are widely used as they are.

\subsection{Modelling rotation curves}
\label{section:rotation}

To fit circular velocity profiles, \cite{DiCintio2014b} use their prescriptions for the three shape parameters $a$, $b$, $g$ describing the density profile (cf. Eq.~(\ref{eq:rho_abc})) but leave the scale radius $r_c$ and the scale density $\rho_c$ as free parameters.
The right panel of Fig.~\ref{fig:s1_c2_ratios} showing the concentration ratio $c_2/c_{\rm DMO}$ as a function of the stellar-to-halo mass ratio $M_{\rm star}/\Mvir$ (as well as Figs. \ref{fig:prescriptions} and \ref{fig:prescriptions_Vc}) highlights the difficulty to account for the scatter in concentration at high stellar-to-halo mass ratio, which leads to significant discrepancies in the density and velocity profiles derived from the current mass-dependent prescriptions in the domain where $c_2/c_{\rm DMO}$ increases exponentially. 
This motivates to release the mass constraint on the concentration $c_2$ when modelling rotation curves of galaxies with the DZ profile, thus treating it as a two-parameter profile (the two parameters being $\Mvir$ and $c_2$ given the stellar mass $M_{\rm star}$). This is similar to what is advocated for the \citetalias{DiCintio2014b} profile. 
When applied to simulated haloes whose mass $\Mvir$ is known, enforcing $M(\Rvir)=\Mvir$ effectively leaves one free parameter ($r_c$ or its associated concentration).

Figs.~\ref{fig:Vfit} and \ref{fig:Vfit_Vc} show the density and circular velocity profiles resulting from one-parameter fits to the rotation curves using the DZ profile and its current mass-dependent prescription for the inner logarithmic slope $s_1$ for the eight fiducial NIHAO haloes shown in Figs.~\ref{fig:fits}, together with the corresponding \citetalias{DiCintio2014b} one-parameter fits and the simulated profiles. 
The inner slope $s_1$ of the DZ profile is set by the fitting function of Fig.~\ref{fig:s1_c2_ratios} (cf. Eq.~(\ref{eq:s1(x)}) and Table~\ref{table:fit_rho}) given the stellar and halo masses while its concentration $c_2$ is allowed to vary. 
The shape parameters $a$, $b$, $g$ of the \citetalias{DiCintio2014b} profile (Eq.~(\ref{eq:rho_abc})) are set by their mass-dependent prescriptions \citep[][Eq.~(3)]{DiCintio2014b}, while the scale radius $r_c$ is allowed to vary. The \citetalias{DiCintio2014b} characteristic density $\rho_c$ is constrained by the halo mass $\Mvir$. 
Fig.~\ref{fig:Vfit_sigma_hist} further shows the distributions of the rms of the residuals in density and velocity between model and simulation within the whole NIHAO sample, while Fig.~\ref{fig:Vfit_comparison_histograms} shows the  corresponding distributions of $\Delta s$, $\Delta c$, $\Delta V$, and $\Delta R$.
Both the DZ and the \citetalias{DiCintio2014b} profiles provide extremely good fits to the density and circular velocity profiles, with rms values comparable to those obtained from the two-parameter fits of Section~\ref{section:fitting} and much smaller than those obtained in Section~\ref{section:prescription_accuracy}.
The one-parameter DZ fits to the rotation curves enable to retrieve the inner slope $s_1$ with a $\pm 0.3$ scatter and a negligible systematic error (as in the previous Section \ref{section:prescription_accuracy}, since $s_1$ is set by its mass-dependent prescription), the concentration $c_2$ with a $\pm 8.6$ scatter and a small $-3.0$ systematic offset (0.1 dex scatter and $-0.08$ dex offset in $\log c_2$), the maximum velocity $V_{\rm max}$ with a $3\%$ scatter and a $+0.6\%$ offset, and the corresponding radius $R_{\rm max}$ with a $\pm 30\%$ scatter and a $+20\%$ offset ($0.11$ dex scatter and $+0.08$ dex offset in $\log R_{\rm max}$). These scatters and offsets are comparable to those obtained previously, except for $\Delta V$ where they are significantly smaller. 
The differences between the parametrized and the simulated rotation curves are below $10\%$ at any radius and for any galaxy, i.e., well within observational errors.
In contrast, the NFW profile used for DM haloes is in contrast unable to describe such rotation curves in the presence of baryons, with differences as high as $50\%$ in the intermediate mass range where core formation occurs \citep{DiCintio2014b}.

\section{Conclusion}
\label{section:conclusion}

Baryonic processes affect the dark matter haloes in which galaxies are embedded, their inner density profiles ranging from steep NFW-like cusps as in DM-only simulations \citep{NFW1996,NFW1997} at low stellar masses, flat cores in the stellar mass range between $10^7$ and $10^9~\rm M_\odot$, and cusps steeper than NFW at higher stellar masses \citep[e.g., ][]{DiCintio2014b, Tollet2016, Dutton2016}. 
In the present article, we study a parametrisation of DM haloes that enables to describe this variety of halo responses to baryonic processes with a variable inner logarithmic slope $s_1$ and a variable concentration parameter $c_2$. This parametrization, which we refer  to here as the Dekel-Zhao (DZ) profile, is a specific case of the Zhao family of double power-law models (Eq.~(\ref{eq:rho_abc}), \citetalias{Zhao1996}) in which the outer logarithmic slope is set to $g=3.5$ and the exponent describing the transition between the inner and outer regions to $b=2$. 
As shown by \citetalias{Zhao1996} and \citetalias{An2013}, it allows analytic expressions for the gravitational potentiel and the velocity dispersion, which we recall in Section~\ref{section:potential_velocity} (Eqs.~(\ref{eq:U32_1}) and (\ref{eq:sigma_dekel})). 
Using three pairs of haloes at different masses with and without baryons at $z=0$, taken from the NIHAO suite of hydrodynamical cosmological zoom-in simulations \citep{Wang2015}, \citetalias{Dekel2017} show that this parametrization yields excellent fits to the density and circular velocity profiles of DM haloes ranging from steep cusps to flat cores, notably capturing cores better than the NFW and \cite{Einasto1965} profiles. In \citetalias{Freundlich2020}, we further derive the kinetic energy associated to this DZ profile and show that it fits well with the simulated quantity.

In the present article, we extend the work done by \citetalias{Zhao1996}, \citetalias{An2013}, \citetalias{Dekel2017} and \citetalias{Freundlich2020} by gathering most analytic expressions obtained for the DZ profile (Sections~\ref{section:dekel_general} and \ref{section:dekel+}), by deriving  additional analytic expressions for its lensing properties in terms of Fox $H$ functions (Section~\ref{section:lensing}) and by testing this profile over the whole NIHAO suite of simulations at $z=0$ (Section~\ref{section:results}). 
We also provide analytic expressions in terms of the maximum circular velocity and radius $V_{\rm max}$ and $r_{\rm max}$ (Appendix~\ref{appendix:cmax}), a second-order Taylor expansion of the distribution function (Appendix~\ref{appendix:DF}), expressions for the velocity dispersion and the kinetic energy in the presence of an additional baryonic component (Appendix~\ref{appendix:additional}), and series expansions of the lensing properties (Appendix~\ref{appendix:lensing}). 
Table~\ref{table:expressions} summarizes the analytic expressions available for the DZ profile. 
The systematic test on the NIHAO simulations enables us to quantitatively show that the DZ profile provides better fits to the density and circular velocity profiles of DM haloes than the other two-parameter Einasto and generalized NFW with variable inner slope profiles, in particular in the innermost regions (Section~\ref{section:fitting}).

\begin{table}
	\caption{Analytic expressions for the DZ profile, which depends on two shape parameters -- $a$ and $c$, or equivalently, the inner slope $s_1$ and the concentration $c_2$.}
	\begin{tabular*}{\linewidth}{l@{\extracolsep{\fill}}l}
		\hline
		\hline
		\noalign{\vskip 0.5mm} 
		Quantity & Equation\\
		\hline
		\noalign{\vskip 1.1mm} 
		Density & $\displaystyle \rho(r) = \frac{\rho_c}{x^{a}\left(1+x^{1/2}\right)^{2(3.5-a)}}$ with $\displaystyle x=\frac{r}{r_c}$\\
		\noalign{\vskip 0.6mm} 
		\hline
		\noalign{\vskip 0.5mm} 
		Characteristic radius & $r_c=R_{\rm vir}/c$\\
		Characteristic density & $\rho_c=(1-a/3)\overline{\rho_c}$\\
		Characteristic av. density & $\overline{\rho_c}= c^3\mu \overline{\rho_{\rm vir}}$\\
		Average virial density & $\overline{\rho_{\rm vir}}=3M_{\rm vir}/4\pi R_{\rm vir}^3 = \Delta \rho_{\rm crit}$\\ 
		\noalign{\vskip 0.5mm} 
		Mass factor & $\mu = c^{a-3}(1+c^{1/2})^{2(3-a)}$\\
		Inner slope $s_1$ from $a$, $c$& Eq.~(\ref{eq:s1_23})\\
		Concentration $c_2$ from $a$, $c$& Eq.~(\ref{eq:c2_23})\\
		Parameter $a$ from $s_1$, $c_2$ & Eq.~(\ref{eq:a(s1,c2)})\\
		Parameter $c$ from $s_1$, $c_2$ & Eq.~(\ref{eq:c(s1,c2)})\\
		Core radius $r_{\rm core}$ & Eq.~(\ref{eq:rcore})\\
		Half-mass radius and $r_f$& Eq.~(\ref{eq:rf})\\
		\hline
		\noalign{\vskip 0.5mm} 
		Maximum velocity radius $r_{\rm max}$& Eq.~(\ref{eq:rc_rmax}) with $b=2$ and $\overline{g}=3$\\
		Maximum velocity $V_{\rm max}$ & Eq.~(\ref{eq:vmax}) with $b=2$ and $\overline{g}=3$\\
		Concentration $c_{\rm max}$ from $a$, $c$ & Eq.~(\ref{eq:cmax_23})\\
		Parameter $a$ from $s_1$, $c_{\rm max}$ & Eq.~(\ref{eq:a(s1,cmax)})\\
		Parameter $c$ from $s_1$, $c_{\rm max}$ & Eq.~(\ref{eq:c(s1,cmax)})\\
		\hline
		\noalign{\vskip 0.5mm} 
		Average density & Eq.~(\ref{eq:rhob}) with $b=2$ and $\overline{g}=3$\\
		Enclosed mass & Eq.~(\ref{eq:M(r)})\\
		Circular velocity & Eq.~(\ref{eq:V2(r)})\\
		Gravitational force & Eq.~(\ref{eq:F(r)})\\
		Logarithmic slope & Eq.~(\ref{eq:s}) with $b=2$ and $\overline{g}=3$\\
		Gravitational potential & Eq.~(\ref{eq:U32_1})\\
		Velocity dispersion & Eqs.~(\ref{eq:sigma_dekel}), (\ref{eq:appendix_sigma_dekel2}) and (\ref{eq:appendix_sigma_dekel3})\\
		Surface density & Eqs.~(\ref{eq:Sigma_H}) and (\ref{eq:appendix_sigma_expansion})\\
		Average surface density & Eqs.~(\ref{eq:Sigmab_H}) and (\ref{eq:appendix_Sigmab_expansion})\\
		Projected mass & Eqs.~(\ref{eq:M2D_H}) and (\ref{eq:appendix_M2D_expansion})\\
		Deflection angle &  Eq.~(\ref{eq:lensing_alpha_H}) and from Eq.~(\ref{eq:appendix_M2D_expansion})\\
		Lensing shear & Eq.~(\ref{eq:lensing_shear})\\
		Lensing potential &Eqs.~(\ref{eq:lensing_potential}) and (\ref{eq:appendix_psi_expansion})\\
		Distribution function & Eqs.~(\ref{eq:eddington2}) and (\ref{eq:fE_int2}) (integral forms)\\
		\noalign{\vskip 0.5mm} 
		\hline
		\noalign{\vskip 0.5mm} 
		$s_1(M_{\rm star}/M_{\rm vir})$ & Eqs.~(\ref{eq:s1(x)}) and (\ref{eq:s1_DMO}), Table~\ref{table:fit_rho}\\
		$c_2(M_{\rm star}/M_{\rm vir})$ & Eqs.~(\ref{eq:exp_function}) and (\ref{eq:c_DMO}), Table~\ref{table:fit_rho}\\
		\noalign{\vskip 0.5mm} 
		\hline
		\hline
		\vspace{-0.8cm}
	\end{tabular*}
	\label{table:expressions}
\end{table}

\begin{figure*}
	\includegraphics[width=0.9\textwidth,trim={.cm 2.cm 0.cm 2.3cm},clip]{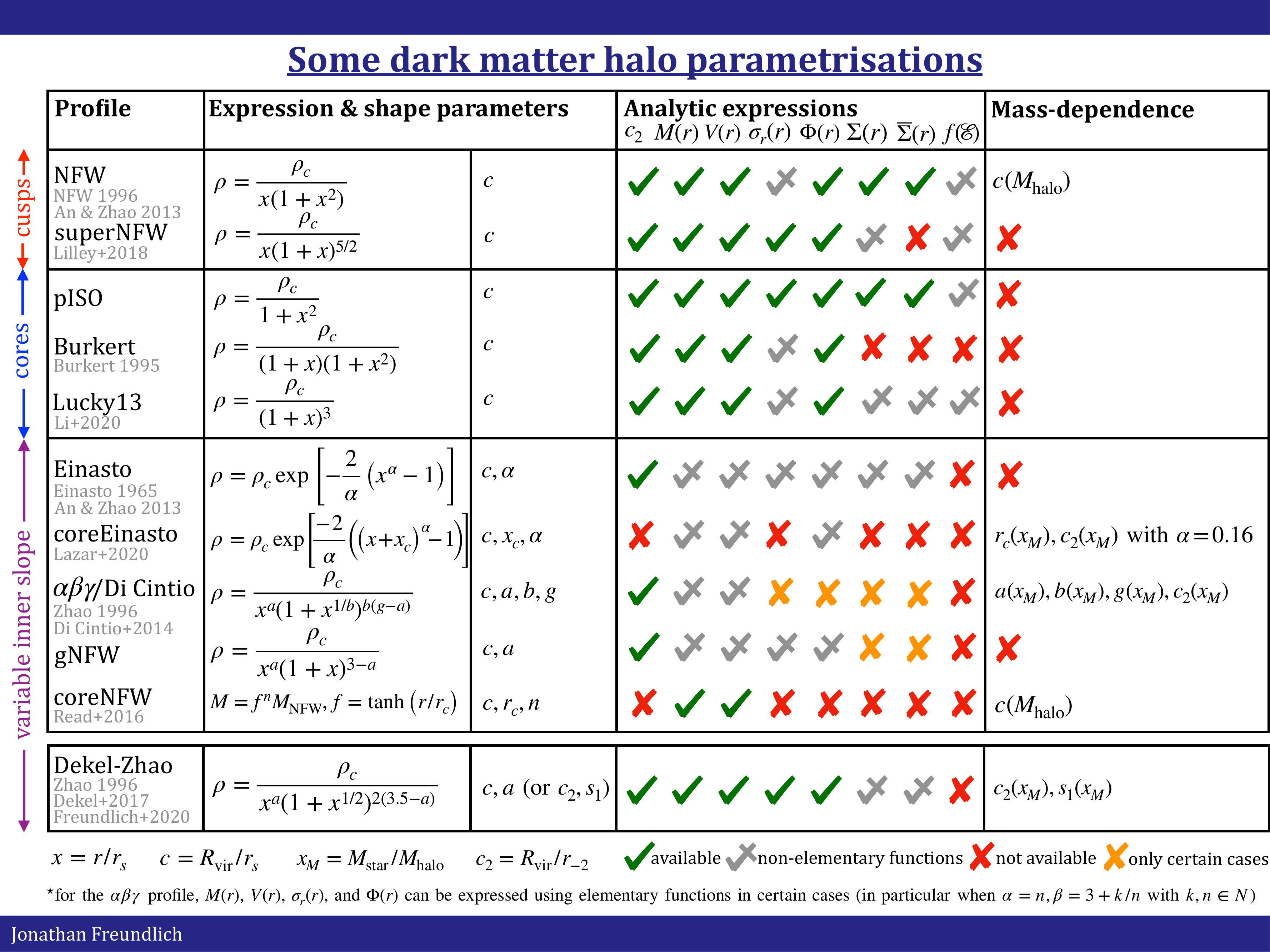}
	\vspace{-0.2cm}
	\caption{The Dekel-Zhao (DZ) profile against other existing parametrisations of DM halo density profiles. 
			For each parametrisation, we indicate the analytic expression of the density (or mass), its shape parameters, whether analytic expressions for the concentration ($c_2$) where the logarithmic density slope equals $2$ in absolute value, enclosed mass ($M$), circular velocity ($V$), radial velocity dispersion ($\sigma_r$), gravitational potential ($\Phi$), projected surface density ($\Sigma$), average surface density ($\overline{\Sigma}$), and distribution function ($f$) are available to the best of our knowledge, and whether the shape parameters have been expressed as functions of the stellar and halo masses ($M_{\rm star}$ and $M_{\rm halo}$) using numerical simulations. 
			The projected surface densities $\Sigma$ and $\overline{\Sigma}$ enable to define lensing properties such as the convergence, the shear and the magnification. 
			The parametrisations listed alongside the DZ profile include 
			the NFW \protect\citep[e.g.][\citetalias{An2013}]{NFW1996,NFW1997, Lokas2001, Evans2006, Eliasdottir2007} and 
			``superNFW'' \protect\citep{Lilley2018} cuspy profiles, 
			the pseudo-isothermal (pISO),
			\protect\cite{Burkert1995}, and 
			``Lucky13'' \protect\citep{Li2020} cored profiles, and 
			the Einasto \protect\citep[e.g., ][\citetalias{An2013}]{Einasto1965, Retana-Montenegro2012, Dutton2014}, 
			``core-Einasto'' \protect\citep{Lazar2020}, 
			double power-law $\alpha\beta\gamma$ \protect\citep[e.g., \citetalias{Zhao1996}, \citetalias{An2013},][]{DiCintio2014b}, 
			generalized NFW \protect\citep[gNFW, e.g., ][]{Umetsu2011, Mamon2019}, and 
			``core-NFW'' \protect\citep{Read2016} profiles with flexible inner slope, more suited to describe the diversity of DM halo shapes in the presence of baryons. 
			The ``core-Einasto'' profile can become a two-parameter profile by fixing its parameter $\alpha$ \citep{Lazar2020}, but limited to fitting cored profiles in a certain mass range. 
			For the double power-law $\alpha\beta\gamma$ profile (Eq.~(\ref{eq:rho_abc})), $M(r)$, $V(r)$, $\sigma_r(r)$, and $\Phi(r)$ can be expressed using elementary functions in certain cases, in particular within the family of profiles with $b=n$ and $g=3+k/n$ where $k$, $n$ are natural integers (\citetalias{Zhao1996}, \citetalias{An2013}). 
			The \protect\citetalias{DiCintio2014b} profile corresponds to a double power-law profile whose shape parameters are set by the stellar-to-halo mass ratio \protect\citep{DiCintio2014b}. In this case, only $c_2$, $M(r)$, and $V(r)$ have analytic expressions (using non-elementary functions for the latter two). 
			The mass-dependent prescriptions as a function of halo mass for the NFW and Einasto profiles stem from dark matter only simulations \citep[e.g.,][]{Dutton2014}.
			The DZ profile is a double-law profile with $b=2$ and $g=3.5$.
			We show in Section~\ref{section:fitting} that it provides better fits to simulated density profiles and rotation curves than the Einasto and gNFW profiles, with the same number of free parameters, and in Section~\ref{section:rotation} that its mass-dependent prescriptions are as accurate as the  \protect\cite{DiCintio2014b} prescriptions. 
			The DZ profile stands out amongst the parametrisations with variable inner slope for its available analytic expressions and its mass-dependent prescriptions as a function of the stellar-to-halo mass ratio, taking into account the effect of baryons. 
	}
	\label{fig:profiles_compare}
\end{figure*}

But most importantly, this test enables us to describe the mass dependence of the inner slope $s_1$ and concentration parameters $c_2$ associated with the DZ profile (Section~\ref{section:mass_dependence}) and to establish it as a mass-dependent profile (Section~\ref{section:mass}) on par with the double power-law \citetalias{DiCintio2014b} profile proposed by \cite{DiCintio2014b} -- with the advantage to have analytic expressions for many of its properties and only two shape parameters instead of four.  
We show that both $s_1$ and $c_2$ correlate with stellar and halo mass, especially with the stellar-to-halo mass ratio $M_{\rm star}/\Mvir$, and we provide fitting functions for the corresponding relations. 
The inner logarithmic slope $s_1$ corresponds to the NFW slope for $\log (M_{\rm star}/\Mvir)\leq -4$, to flatter inner density profiles for $\log (M_{\rm star}/\Mvir)$ between $-3.5$ and $-2$, and to steeper-than-NFW inner density profiles  for  $\log (M_{\rm star}/\Mvir)>-2$ (Fig.~\ref{fig:s1_c2_ratios}, Eq.~(\ref{eq:s1(x)}), and Table~\ref{table:fit_rho}). 
The concentration $c_2$ similarly corresponds to the NFW concentration at low $M_{\rm star}/\Mvir$, becomes slightly ($\sim$20$\%$) smaller than the NFW concentration for $\log (M_{\rm star}/\Mvir)$ between $-3.5$ atnd $-2$, and increases exponentially compared to NFW for  $\log (M_{\rm star}/\Mvir)>-2$ (Fig.~\ref{fig:s1_c2_ratios}, Eq.~(\ref{eq:exp_function}), and Table~\ref{table:fit_rho}). 
In terms of stellar mass, the range for core formation and halo expansion corresponds to $10^7$ to $10^{10}~\rm M_\odot$.

The DZ profile thus enables to follow the expansion of the halo due to baryons in the mass range with $\log (M_{\rm star}/\Mvir)$ between $-3.5$ and $-2$ not only in terms of inner logarithmic slope as for the \citetalias{DiCintio2014b} profile but also in terms of concentration -- i.e., at larger radii than those concerned by the inner slope. 
With the fitting functions of $s_1$ and $c_2$ as functions of $M_{\rm star}/\Mvir$, the DM distribution in haloes ranging from dwarfs to Milky-Way-like in stellar mass is set by the stellar and halo masses.  
We show that the mass-dependent DZ profile thus established is as accurate as the multi-parameter \citetalias{DiCintio2014b} profile to describe density and circular velocity profiles of DM haloes (Section~\ref{section:prescription_accuracy}), in particular when the concentration parameter is left free  (Section~\ref{section:rotation}). 
In Fig.~\ref{fig:profiles_compare}, we compare the DZ profile with existing parametrisations of DM halo density profiles, emphasizing on the number of parameters, the availability of analytic expressions, and the availability of mass-dependent prescriptions derived from simulations. Amongst the parametrisations with variable inner slop, the DZ profile stands out for its available analytic expressions and its mass-dependent prescriptions as a function of the stellar-to-halo mass ratio, taking into account the effect of baryons.

We caution that this study relies on a specific suite of hydrodynamical cosmological simulations \citep[NIHAO; ][]{Wang2015}, which is notably characterised by a  strong stellar feedback implementation with a blast-wave formalism and  delayed cooling and no AGN feedback. We note that the \citetalias{DiCintio2014b} profile was proposed using a previous suite of simulations \citep[MaGICC; ][]{Brook2012, Stinson2013} with a similar implementation. 
Other simulation suites with different feedback schemes \citep[e.g., ][]{Mashchenko2008, Teyssier2013, Madau2014, Verbeke2015, Read2016} suggest a similar behaviour of the inner density profile of DM haloes as a function of the stellar-to-halo mass ratio. 
This behaviour can be understood in theoretical terms as a competition between outflows induced by feedback and the confinement imposed by the halo gravity \citep[e.g., ][\citetalias{Freundlich2020}]{Dekel1986, Read2005, Penarrubia2012, Pontzen2012, Dutton2016, El-Zant2016}. As such, the halo response to baryonic processes may not necessarily depend on the details of the feedback implementation as long as outflows are well-reproduced in the simulations. 
These outflows are expected to affect the stellar and gaseous components of galaxies, such that the good agreement of NIHAO galaxies with observations in terms of morphologies, color, sizes and rotation curves \citep{Wang2015, Stinson2015, Dutton2016a, Dutton2017, Obreja2019, Santos-Santos2019} may reflect outflows comparable to those of actual galaxies \citep{Tollet2019} and of other simulation suites reproducing the aforementioned observables. 
We leave detailed tests of the DZ profile in other simulation suites with different feedback implementations for future work.

The accuracy of the DZ profile to describe the DM distributions of simulated haloes makes it a useful tool to study the evolution of DM density profiles, to model rotation curves of galaxies, to parametrize gravitational lenses, and to implement in semi-analytical models of galaxy formation and evolution. 
The analytic expressions for the gravitational potential, the velocity dispersion and the lensing properties can notably be used to model core formation in DM haloes from outflow episodes resulting from feedback, as in \citetalias{Freundlich2020}, to model gravitational lenses, to generate halo potentials or initial conditions for simulations, to compare different DM distributions in semi-analytical models \citep{Jiang2020}, and to quantify simulated and observed rotation curves of galaxies without numerical integrations. 
%


\section*{Acknowledgements}


We thank the referee, HongSheng Zhao, for a detailed and constructive report.
We acknowledge A. Wasserman and N. Bouch\'e for providing observational incentives for this work. We thank G. Mamon, A. Burkert, F. Combes, K. Kaur, K. Sarkar, A. Zitrin, F. Lelli, B. Famaey, and K. Malhan for stimulating discussions; D. Maoz for his support.
This work has received funding from the European Research Council (ERC) under the European Union's Horizon 2020 research and innovation programme PE9 ERC-2018-ADG. 
This work was partly supported by the grants France-Israel PICS,
I-CORE Program of the
PBC/ISF 1829/12, BSF 2014-273, NSF AST-1405962,
GIF I-1341-303.7/2016, and DIP STE1869/2-1 GE625/17-1.
NIHAO simulations were carried out at the Gauss Centre for Super-computing e.V. (\url{www.gauss-centre.eu}) at the GCS Supercomputer SuperMUCat Leibniz Supercomputing Centre (\url{www.lrz.de}) and on the High Performance Computing resources at New York University Abu Dhabi. 
We used the software \texttt{pynbody} \citep{Pontzen2013} for our analyses.



\section*{Data availability}

We provide codes to implement the DZ profile at \url{https://github.com/JonathanFreundlich/Dekel_profile}. The simulation data underlying this article will be shared on reasonable request to the corresponding author. 



\bibliographystyle{mnras}
\bibliography{freundlich_cuspcore} 




\appendix

\clearpage

\section{Profile parameters in terms of $r_{\rm max}$ and $V_{\rm max}$}
\label{appendix:cmax}

In certain situations, it may be useful to express the density profile of Eq.~(\ref{eq:rho}) -- or the average density profile of Eq.~(\ref{eq:rhob}) --in terms of the radius $r_{\rm max}$ at which the circular velocity peaks and $V_{\rm max}=V(r_{\rm max})$ the maximum velocity instead of $R_{\rm vir}$ and $M_{\rm vir}$. 
This can be achieved by writing $r_c$ and $\overline{\rho_c}$ in terms of these two parameters. 
The circular velocity (Eq.~(\ref{eq:V2(r)})) peaks at $r_{\rm max}$ such that 
\be
\label{eq:rc_rmax}
r_c=\left( \frac{\overline{g}-2}{2-a} \right)^b r_{\rm max}
\ee
while the expression of the enclosed mass (Eq.~(\ref{eq:M(r)})) yields
\be
\label{eq:vmax}
V_{\rm max}^2 = \frac{GM(r_{\rm max})}{r_{\rm max}}= \frac{G \mu M_\vir}{r_{\rm max}} \left( \frac{(2-a)^{3-a}}{(\overline{g}-2)^{3-\overline{g}} (\overline{g}-a)^{\overline{g}-a}}\right)^b 
\ee
and hence
\be
\label{eq:rhoc_rVmax}
\overline{\rho_c} = \frac{3 \mu M_\vir}{4\pi r_c^3} = \frac{3 V_{\rm max}^2}{4\pi G r_{\rm max}^2} \left( \frac{(2-a)^a (\overline{g}-a)^{\overline{g}-a}}{(\overline{g}-2)^{\overline{g}} } \right)^b. 
\ee
In particular, the virial radius of subhaloes is difficult to define in practice so semi-analytical models of satellite evolution may prefer to express subhalo properties in terms of $r_{\rm max}$ and $V_{\rm max}$ \citep[e.g.,][]{Jiang2020}.

\begin{figure}
	\centering
	\includegraphics[width=1\linewidth,trim={0.3cm 0.cm 1cm .5cm},clip]{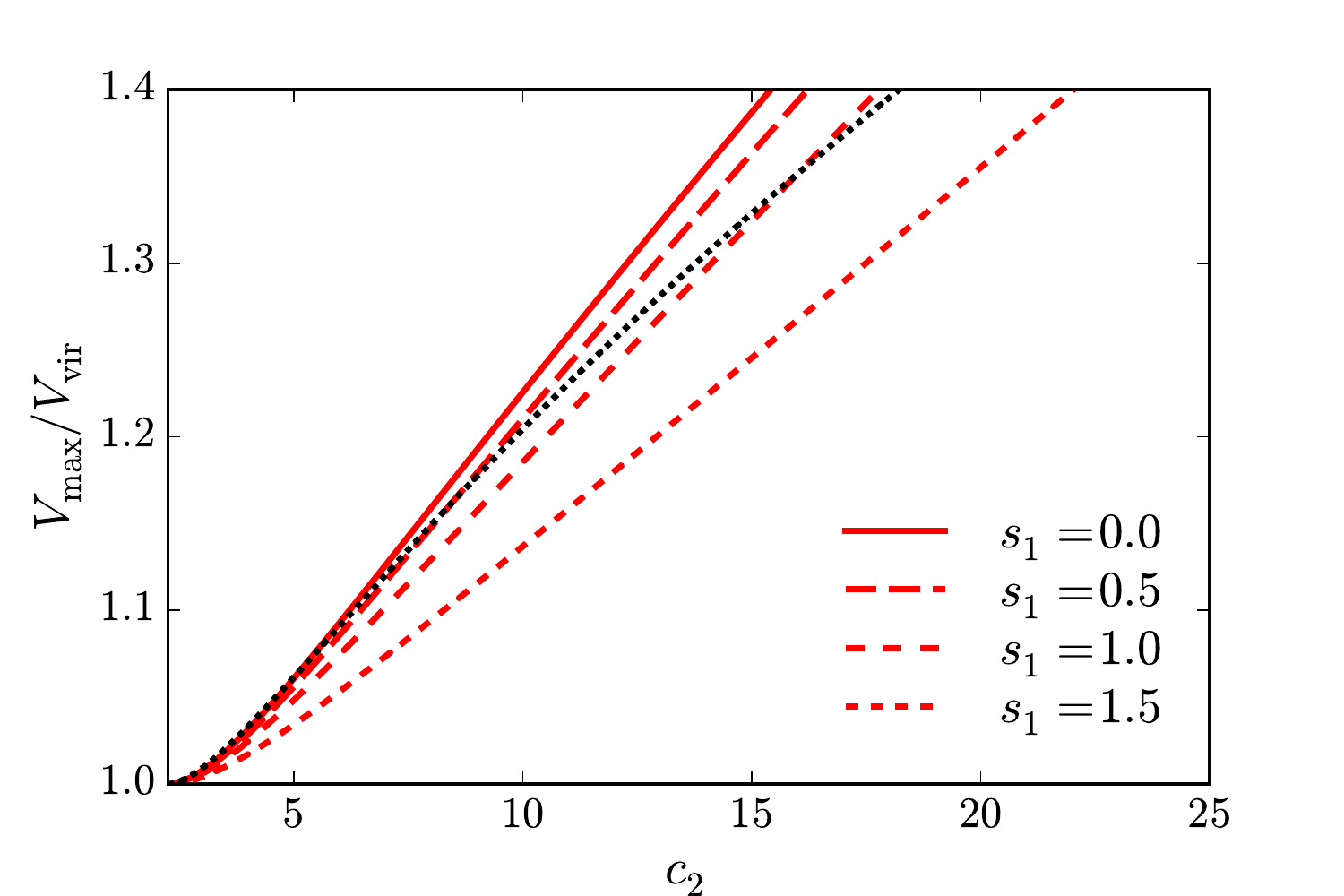}
	\vspace{-0.4cm}
	\caption{Relation between the ratio of the maximum circular velocity $V_{\rm max}$ to the circular velocity at the virial radius $V_{\rm vir}$ and the concentration $c_2$ for DZ profiles with different values of the inner slope $s_1$. The black dotted line corresponds to the NFW relation, where the velocity ratio only depends on the concentration. The lower limit of the x-axis corresponds to $c_2=2.25$, for which $r_{\rm max}=\Rvir$ and $V_{\rm max} = V_\vir$.}
	\label{fig:vmax_c2}
\end{figure}

Given $\Rvir$ and $\Mvir$, the values of $r_{\rm max}$ and $V_{\rm max}$ further enable to retrieve the shape parameters of a DZ halo. 
For a DZ profile with $\overline{g}=3$ and $b=2$, we indeed have
\be
\label{eq:rmax/rvir}
\frac{r_{\rm max}}{\Rvir} = \frac{(2-a)^2}{c} = \frac{2.25}{c_2}
\ee
and
\be
\label{eq:Vmax/Vvir}
\left(\frac{V_{\rm max}}{V_{\rm vir}}\right)^2 = c^{a-2} (1+c^{1/2})^{6-2a}(2-a)^{4-2a}(3-a)^{2a-6}
\ee
with $V_{\rm vir}=GM_{\rm vir}/R_{\rm vir}$, which can also be expressed as a function of $s_1$ and $c_2$ using Eqs.~(\ref{eq:a(s1,c2)}) and (\ref{eq:c(s1,c2)}). While the $V_{\rm max}/V_{\rm vir}$ velocity ratio only depends on the concentration for an NFW halo \citep[e.g., ][]{Prada2012}, it depends on the two shape parameters for a DZ profile. 
Fig.~\ref{fig:vmax_c2} shows how this ratio varies with the concentration $c_2$ for different values of $s_1$ (red lines) and compares it with the NFW relation (black dotted line). 
Having the values of $r_{\rm max}/R_{\rm vir}$ and $V_{\rm max}/V_{\rm vir}$ enables to retrieve the shape parameters by numerically solving Eqs.~(\ref{eq:rmax/rvir}) and (\ref{eq:Vmax/Vvir}), as illustrated in the figure.

Radius $r_{\rm max}$ also defines a concentration $c_{\rm max}\equiv R_{\rm vir}/r_{\rm max}$, which is 
\be
\label{eq:cmax}
c_{\rm max} =c \left(\frac{\overline{g}-2}{2-a}\right)^b
\ee
for the general double power-law profile of Eq.~(\ref{eq:rhob}) and 
\be
\label{eq:cmax_23}
c_{\rm max}
=\frac{c}{\left(2-a\right)^2} 
\ee
for a DZ profile with $\overline{g}=3$ and $b=2$. In this  case, there are bijections between the couples ($a$,$c$), ($s_1$, $c_2$), and ($s_1$, $c_{\rm max}$), with 
\be
\label{eq:a(s1,cmax)}
a = \frac{s_1 -2 \left( 3.5-s_1\right)\left(r_1/R_\vir\right)^{1/2}c_{\rm max}^{1/2} }{1 - \left(3.5-s_1\right)\left(r_1/R_\vir\right)^{1/2}c_{\rm max}^{1/2} }
\ee
and
\be
\label{eq:c(s1,cmax)}
c= \left( \frac{s_1-2}{\left(3.5-s_1\right)\left(r_1/R_\vir \right)^{1/2} -c_{\rm max}^{-1/2}} \right)^2.
\ee
The couple ($s_1$, $c_{\rm max}$) is thus equivalent to the other two couples to describe the DZ profile (cf. Section~\ref{section:shape}).

\section{The velocity dispersion as a sum of elementary functions}
\label{appendix:sigmar_sum}

Eq.~(\ref{eq:sigma_dekel}) expressing the velocity dispersion in a DZ halo involves incomplete beta functions, i.e.,  non-elementary functions. 
Following Eqs. (19) and (A.9)$-$(A.11) of \citetalias{Zhao1996}, this equation can also be expressed (see also \citetalias{Dekel2017}, Eq. (A.10)) as the following sum: 
\be \label{eq:appendix_sigma_dekel2} 
\sigma_r^2(r)
= 2 c\mu \frac{GM_{\rm vir}}{R_{\rm vir}}  x^a (1+x^{1/2})^{2(3.5-a)}
\sum_{i=0}^{8} \frac{(-1)^i 8!}{i!(8-i)!}  \frac{1-\chi^{4(1-a)+i}}{4(1-a)+i},
\ee
which only involves elementary functions, since $\chi=x^{1/2}/(1+x^{1/2})$. 
Alternatively, noticing that $\mathcal{B}(a,b,x) = a^{-1} x^a (1-x)^{b-1}+a^{-1}(b-1) \times
\mathcal{B}(a+1,b-1,x)$,
we can deduce that 
\be
\mathcal{B}(a,9,x) = \sum_{i=0}^8 \frac{8!}{i!} \frac{\Gamma(a)}{\Gamma(a+9-i)} x^{a+8-i}(1-x)^i,
\ee
where $\Gamma$ denotes the usual gamma function. 
This enables to write the local kinetic energy also as
\begin{multline}
\label{eq:appendix_sigma_dekel3}
\sigma_r^2(r)
= 2 c\mu \frac{GM_{vir}}{R_{vir}}  x^a (1+x^{1/2})^{2(3.5-a)}
\Bigg[  
\frac{8! \Gamma(4(1-a))}{\Gamma(4(1-a)+9)}
\\
-\sum_{i=0}^{8} \frac{8!}{i!} \frac{\Gamma(4(1-a))}{\Gamma(4(1-a)+9-i)} \chi^{4(1-a)+8-i}(1-\chi)^i
\Bigg].
\end{multline}
The different expressions for $\sigma_r^2$ are formally equivalent, but our numerical implementation using \texttt{scipy} (cf. \url{https://github.com/JonathanFreundlich/Dekel_profile}) results in significantly larger numerical errors with Eq.~(\ref{eq:appendix_sigma_dekel2}) than with Eqs.~(\ref{eq:sigma_dekel}) and (\ref{eq:appendix_sigma_dekel3}).

\section{Potential and velocity dispersion with an additional mass}
\label{appendix:additional}

In the presence of baryons or more generally when the DZ profile only describes part of the mass distribution, the enclosed mass entering the integral  defining the gravitational potential, 
\begin{equation}
\label{eq:Ur}
\displaystyle U_{\rm tot}(r)  
= -\int_r^{\Rvir} \frac{G M(y)}{y^2} dy - \frac{G\Mvir}{\Rvir}
\end{equation}
for a halo truncated at the virial radius $\Rvir$,
and that defining the radial velocity dispersion (Eq.~(\ref{eq:sigmar2-0})) should include all components 
We indicate here analytical expressions for the potential and the radial velocity dispersion of a DZ DM halo in the presence of an additional component, focussing on the following cases: (1)  a radial power-law between the total mass and the dark matter mass, as assumed in \citetalias{Freundlich2020}; (2) an additional point mass at the center of the halo; (3) an additional sphere of constant density; and (4) an additional isothermal sphere. 
In this effect, we consider $U_m$ and $\sigma_m$ the contributions of the additional mass to the potential and radial velocity dispersion, such that 
\be
U_{\rm tot}=U+U_m
\ee
and 
\be
\sigma_{r, \rm tot}^2=\sigma_{r}^2+\sigma_m^2
\ee
where $U$ and $\sigma_r$ are those of the DZ profile (Eqs.~(\ref{eq:U32_1}) and (\ref{eq:sigma_dekel})).

\subsection{Power law multi-component halo}
\label{appendix:power-law}

To account for the difference between the total enclosed mass $M_{\rm tot}$ and the enclosed dark matter mass $M$, we can model their ratio as a power law 
\be
\label{eq:Mratio}
\frac{M_{\rm tot}}{M} = X_M \left(\frac{r}{R_\vir} \right)^{-n}, 
\ee 
where $X_M$ and $n$ are ajustable parameters as in \citetalias{Freundlich2020}. 
With this parametrization, the gravitational potential from Eq.~(\ref{eq:Ur}) becomes
\begin{align}
U_{\rm tot}(r) &= -X_M V_\vir^2 \left( 1+\mu c^{n+1} \int_x^c \frac{1}{z^{a+n-1} (1+z^{1/2})^{2(3-a)}} dz  \right) \\
&= -X_M V_\vir^2 \left( 1+2\mu c^{n+1} \int_\chi^{\chi_c} \zeta^{3-2n-2a}(1-\zeta)^{1+2n}d\zeta \right)\\
U_{\rm tot}(r) &= -X_M V_\vir^2 \left( 1+2\mu c^{n+1} \left[ \mathcal{B}(4-2n-2a,2+2n,\zeta) \right]_\chi^{\chi_c}\right), 
\end{align}
which reverts to Eq.~(\ref{eq:U32_1}) when $n=0$ and $X_M=1$. These calculations use the variable change $\zeta = z^{1/2}/(1+z^{1/2})$, highlighted for Eq.~(\ref{eq:U32}), and the incomplete beta function as in Eq.~(\ref{eq:sigma_dekel}). 
The radial velocity dispersion per unit mass from Eq.~(\ref{eq:sigmar2-0}) similarly becomes 
\begin{align} 
\sigma_{r,\rm tot}^2 (r) 
&= \mu X_M c^{n+1}  V_\vir^2  \frac{{\rho_c}}{\rho(r)}
\int_x^\infty \frac{y^{1-2a-n}}{(1+y^{1/2})^{13-4a}} \rmd y
\\
& = 2\mu X_M c^{n+1}  V_\vir^2 \frac{{\rho_c}}{\rho(r)}
\int_\chi^1 \zeta^{3-4a-2n} (1-\zeta)^{8+2n} \rmd \zeta
\\
\sigma_{r, \rm tot}^2 (r) 
& 
\label{eq:appendix_Kmulti}
= 2 \mu X_M c^{n+1} 
V_\vir^2\frac{{\rho_c}}{\rho(r)}
\Big[ \mathcal{B}(4-4a-2n,9+2n,\zeta)\Big]_\chi^1, 
\end{align}
which reverts to Eq.~(\ref{eq:sigma_dekel}) when $n=0$ and $X_M=1$.

\subsection{Additional point mass}

The contribution of an additional mass $m$ at the center of the halo to the gravitational potential is
\be
U_{m} = - \frac{Gm}{r}. 
\ee
The corresponding contribution to the kinetic energy assuming isotropy can be derived from Eq.~(\ref{eq:sigmar2-0}), with $\rho(r^\prime)$ the DZ density expressed in Eq.~(\ref{eq:rho32}) and $M(r^\prime)=m$: 
\begin{align}
\sigma_{m}^2 (r) 
& \displaystyle =  \frac{G m}{\rho(r)}\int_{r}^{\infty}\rho(\rp) r^{-2}\rmd\rp\\
& \displaystyle = \frac{Gmc}{R_{\rm vir}} \frac{{\rho_c}}{{\rho(r)}} 
\int_{x}^{\infty} \frac{1}{y^{2+a}(1+y^{1/2})^{2(3.5-a)}} dy\\
& \displaystyle = \frac{2Gmc}{R_{\rm vir}} \frac{{\rho_c}}{{\rho(r)}} 
\int_{\chi}^{1} \zeta^{-3-2a} (1-\zeta)^8 d\zeta\\
\label{eq:appendix_Km}
\displaystyle \sigma_{m}^2 (r) 
& \displaystyle =  \frac{2Gmc}{R_{\rm vir}} \frac{{\rho_c}}{{\rho(r)}} 
\Big[ \mathcal{B}(-2-2a,9,\zeta)\Big]_\chi^1.
\end{align}

\subsection{Additional uniform sphere}

The mass distribution of a uniform sphere of total mass $m$ and radius $r_m$ can be written as
\be
m(r) = \left\{ 
\begin{array}{ll}
	\displaystyle m &{\rm if}~r\geq r_m\\
	\displaystyle m\left(\frac{r}{r_m}\right)^3 &{\rm if}~r\leq r_m. 
\end{array}
\right.
\ee
Its contribution to the gravitational potential according to Eq.~(\ref{eq:Ur}) is 
\be
U_m(r) = \left\{ 
\begin{array}{ll}
	\displaystyle-\frac{Gm}{r} 								&{\rm if}~r\geq r_m\\
	\displaystyle-\frac{Gm}{2r_m}\left[3-\left(\frac{r}{r_m}\right)^2  \right] 		&{\rm if}~r\leq r_m. 
\end{array}
\right.
\ee
If $r\geq r_m$, its contribution to the radial velocity dispersion corresponds to that of a point mass $m$ as in Eq.~(\ref{eq:appendix_Km}), i.e., 
\be
\label{eq:Km_uniform_geq}
\sigma_{m}^2 (r)  = \frac{2Gmc}{R_{\rm vir}} \frac{{\rho_c}}{{\rho(r)}} 
\Big[ \mathcal{B}(-2-2a,9,\zeta)\Big]_\chi^1. 
\ee
However, if $r\leq r_m$, Eq.~(\ref{eq:sigmar2-0}) yields
\be
\label{eq:Km_uniform_leq}
\sigma_{m}^2 (r) = \frac{G}{\rho(r)} \left[ \int_r^{r_m} \rho(\rp) m\left( \frac{\rp}{r_m}\right)^3 {\rp}^{-2}d\rp +\int_{r_m}^{+\infty} \rho(\rp) m {\rp}^{-2} d\rp\right], 
\ee
where the second term corresponds to the contribution of a point mass $m$ evaluated at $r_m$. Hence if $r\leq r_m$, 
\begingroup\makeatletter\def\f@size{8.7}\check@mathfonts
\def\maketag@@@#1{\hbox{\m@th\normalsize\normalfont#1}}%
\begin{align}
\sigma_{m}^2 (r) & = \!\frac{2Gmc}{R_{\rm vir}} \frac{{\rho_c}}{{\rho(r)}} 
\Bigg(
\frac{1}{2 x_m^3}\!\! \int_x^{x_m} \!\!\!\!\!\frac{dz}{z^{a-1}(1+z^{1/2})^{7-2a}}
+\Big[ \mathcal{B}(-2\!-\!2a,9,\zeta)\Big]_{\chi_m}^1
\Bigg)\\
& = \!\frac{2Gmc}{R_{\rm vir}} \frac{{\rho_c}}{{\rho(r)}} 
\Bigg(
\frac{1}{x_m^3} \int_\chi^{\chi_m}\!\! \zeta^{3-2a}(1-\zeta)^2 d\zeta 
+\Big[ \mathcal{B}(-2-2a,9,\zeta)\Big]_{\chi_m}^1
\Bigg)\\
\sigma_{m}^2 (r) &= \!\frac{2Gmc}{R_{\rm vir}} \frac{{\rho_c}}{{\rho(r)}} 
\Bigg(
\frac{1}{x_m^3} \Big[ \mathcal{B}(4-2a,3,\zeta)\Big]_{\chi}^{\chi_m}
+\Big[ \mathcal{B}(-2-2a,9,\zeta)\Big]_{\chi_m}^1
\Bigg)
\end{align}
\endgroup
where $x_m=r_m/r_c$ and $\chi_m=\chi(x_m)$.

\subsection{Additional singular isothermal sphere}

The mass distribution of a singular isothermal sphere of total mass $m$ and radius $r_m$ can be written as
\be
m(r) = \left\{ 
\begin{array}{ll}
	\displaystyle m &{\rm if}~r\geq r_m\\
	\displaystyle m\left(\frac{r}{r_m}\right) &{\rm if}~r\leq r_m. 
\end{array}
\right.
\ee
Its contribution to the gravitational potential according to Eq.~(\ref{eq:Ur}) is 
\be
U_m(r) = \left\{ 
\begin{array}{ll}
	\displaystyle-\frac{Gm}{r} 								&{\rm if}~r\geq r_m\\
	\displaystyle-\frac{Gm}{r_m}\Bigg[1+\ln\left(\frac{r_m}{r}\right) \Bigg] 		&{\rm if}~r\leq r_m. 
\end{array}
\right.
\ee
If $r\geq r_m$, its contribution to the kinetic energy corresponds to that of a point mass $m$ as in Eqs.~(\ref{eq:appendix_Km}) and (\ref{eq:Km_uniform_geq}). 
If $r\leq r_m$, Eq.~(\ref{eq:Km_uniform_leq}) yields 
\begingroup\makeatletter\def\f@size{8.7}\check@mathfonts
\def\maketag@@@#1{\hbox{\m@th\normalsize\normalfont#1}}%
\begin{align}
\sigma_{m}^2 (r) & = \!\frac{2Gmc}{R_{\rm vir}} \frac{{\rho_c}}{{\rho(r)}} 
\Bigg(
\frac{1}{2 x_m}\!\! \int_x^{x_m} \!\!\!\!\!\frac{dz}{z^{a+1}(1+z^{1/2})^{7-2a}}
+\Big[ \mathcal{B}(-2\!-\!2a,9,\zeta)\Big]_{\chi_m}^1
\Bigg)\\
& = \!\frac{2Gmc}{R_{\rm vir}} \frac{{\rho_c}}{{\rho(r)}} 
\Bigg(
\frac{1}{x_m}\! \int_\chi^{\chi_m}\!\! \zeta^{-1-2a}(1-\zeta)^6 d\zeta 
+\Big[ \mathcal{B}(-2-2a,9,\zeta)\Big]_{\chi_m}^1
\Bigg)\\
\sigma_{m}^2 (r) &= \!\frac{2Gmc}{R_{\rm vir}} \frac{{\rho_c}}{{\rho(r)}} 
\Bigg(
\frac{1}{x_m} \Big[ \mathcal{B}(-2a,7,\zeta)\Big]_{\chi}^{\chi_m}
+\Big[ \mathcal{B}(-2-2a,9,\zeta)\Big]_{\chi_m}^1
\Bigg). 
\end{align}
\endgroup

\section{Notes on the distribution function}
\label{appendix:DF}

The ergodic distribution function for a spherical density distribution can be recovered by Eddington's formula \citep[][Eq. 4.46]{Eddington1916, BinneyTremaine2008}, 
\be
\label{eq:eddington}
f(\mathcal{E}) = \frac{1}{\sqrt{8}\pi} 
\left[ \int_0^{\mathcal{E}} \frac{d\Psi}{\sqrt{\mathcal{E} - \Psi}} \frac{d^2\nu}{d\Psi^2} + \frac{1}{\sqrt{\mathcal{E}}} \left( \frac{d\nu}{d\Psi}\right)_{\Psi=0} \right], 
\ee
where the relative potential  $\Psi=-U$  is defined with respect to the potential at infinity and $\nu(r)$ is the probability density. 
For a DZ profile truncated at $\Rvir$, the probability density corresponding to the mass density is
\be
\nu(r)= \frac{(3-a)\mu c^3}{4\pi \Rvir^3} \frac{1}{x^a(1+x^{1/2})^{2(3.5-a)}}
\ee 
within $\Rvir$ and zero outside, with 
\be
\frac{d\nu}{d\Psi} = \frac{d\nu}{dx} \frac{dx}{d\Psi} = \frac{(3-a)c^2}{4\pi G \Rvir^2 \Mvir} \frac{a+3.5x^{1/2}}{x^2(1+x^{1/2})^2}
\ee
within $\Rvir$ and zero at $\Psi=0$ when $x\rightarrow +\infty$ so the second term of Eq.~(\ref{eq:eddington}) vanishes, 
\be
\label{eq:dxdnudpsi}
\frac{d}{dx}\left( \frac{d\nu}{d\Psi} \right) 
=
- \frac{(3-a)c^2}{4\pi G \Rvir^2 \Mvir} 
\frac{2a + (5.25 + 3a)x^{1/2} + 8.75x}{x^3(1+x^{1/2})^3}
\ee
within $\Rvir$, and Eddington's formula can be rewritten 
\be
\label{eq:eddington2}
f(\mathcal{E}) = \frac{1}{\sqrt{8}\pi} 
\int_{\Psi^{-1}(\mathcal{E})}^{c} \frac{1}{\sqrt{\mathcal{E} - \Psi(x)}} \left|\frac{d}{dx}\left( \frac{d\nu}{d\Psi} \right)\right| dx, 
\ee
introducing $\Psi^{-1}$ the inverse function of $\Psi$. The distribution function $f(\mathcal{E})$ can be evaluated numerically from this equation,
using the analytic expression of $\Psi(r)=-U(r)$ and Eq.~(\ref{eq:dxdnudpsi}) while evaluating $\Psi^{-1}(\mathcal{E})$ numerically.
\citetalias{Zhao1996} gives an explicit analytic expression of $f(\mathcal{E})$ when $a=1.5$ in their Eq.~(27), but there is sadly no simple analytic expressions for $\Psi^{-1}(\mathcal{E})$ and $f(\mathcal{E})$ in the general case. 
It can however be shown (cf. below) and seen in the bottom right panel of Fig.~\ref{fig:examples_dekel} that the slope of $f(\mathcal{E})$ primarily depends on $c_2$ and only weakly on $s_1$ for $\mathcal{E}_0/V_{\rm vir}^2\lesssim 5$ while the trend reverses when $\mathcal{E}$ approaches its maximum value $\Psi(0)$. 
The differences between the distributions functions shown in  Fig.~\ref{fig:examples_dekel} mainly reflect the different gravitational potentials: stars in the deeper potential of the more concentrated ($c_2=15$) haloes are distributed through larger volumes of phase space. 
We note that the DZ distribution function is always divergent as $\mathcal{E}$ tends to the central $\Psi(0)$.

The integral of Eq.~(\ref{eq:eddington2}) expressing the distribution function $f(\mathcal{E})$ from Eddington's formula can be integrated by parts twice to obtain 
\be
\label{eq:fE_int2}
f({\mathcal{E}})=A\sqrt{{\mathcal{E}}-\Psi(c)}-B{\left({\mathcal{E}}-\Psi(c)\right)^{3/2}}-F[{\mathcal{E}},{\mathcal{E}}]
\ee
with 
\be
A=\frac{-1}{\sqrt{2}\pi}\frac{g(c)}{\Psi'(c)}, 
\ee
\be
B=\frac{\sqrt{2}}{3\pi}\left(\frac{g'(c)\Psi'(c)-g(c)\Psi''(c)}{(\Psi'(c))^3}\right), 
\ee
and
\be
F[u,v]=\frac{\sqrt{2}}{3\pi}\int_{\Psi(c)}^u{\left(v-y\right)^{3/2}}h''(y)\;\mathrm{d}y
\ee 
where 
\be
g(x) \equiv \left| \frac{d}{dx}\left( \frac{d\nu}{d\Psi} \right)  \right| 
\ee
is expressed in Eq.~(\ref{eq:dxdnudpsi}), 
and
\be
h(y)=\frac{g\left(\Psi^{-1}(y)\right)}{\Psi'\left(\Psi^{-1}(y)\right)}.
\ee
Derivatives are indicated by primes. 
Eq.~(\ref{eq:fE_int2}) not only enables to minimize numerical errors when estimating $f(\mathcal{E})$, since the double integration by parts removes the $\sqrt{\mathcal{E}-\Psi(x)}$ denominator of the integrand of Eq.~(\ref{eq:eddington2}), but also to expand $f(\mathcal{E})$ as a series around any energy $\mathcal{E}_0$. 
In particular, if we define $\delta=\mathcal{E}-\mathcal{E}_0$, we can expand $F$ for $u=v={\mathcal{E}_0} +\delta$ around $\delta=0$. One gets at the first order 
\begin{align}
\label{eq:F[E0,E0]}
 F[{\mathcal{E}_0} + \delta,{\mathcal{E}_0} +\delta]
\simeq
F[{\mathcal{E}_0} ,{\mathcal{E}_0} ]+
\delta \left(\frac{\partial  F}{\partial u}+\frac{\partial F}{\partial v}\right)_{{\mathcal{E}_0} ,{\mathcal{E}_0}} 
\!\!\!\!
\simeq F[{\mathcal{E}_0} ,{\mathcal{E}_0} ]+C\delta
\end{align}
where
\be
C=\frac{{1} }{\sqrt{2}\pi}\int_{c}^{\Psi^{-1}({\mathcal{E}_0} )}{\sqrt{{\mathcal{E}_0} -\Psi(x)}}~\eta(x)\;\mathrm{d}x
\ee
with $\displaystyle \eta=\frac{\Lambda'}{\Psi'}-\frac{\Lambda\Psi''}{(\Psi')^2}$
and $\displaystyle\Lambda=\frac{g'}{\Psi'}-g\frac{\Psi''}{(\Psi')^2}$. 
Eq.~(\ref{eq:F[E0,E0]}) yields
\be
\log_{10} f({\mathcal{E}})
\simeq\log_{10}(f({\mathcal{E}_0} ))
+\delta~
\frac{A-3B({\mathcal{E}_0} -\Psi(c))-2C\sqrt{{\mathcal{E}_0} -\Psi(c)}}{2\ln(10)f({\mathcal{E}_0} )\sqrt{{\mathcal{E}_0} -\Psi(c)}}, 
\ee
which notably enables to express the slope of the distribution function and to show that it primarily depends on $c_2$ and only weakly on $s_1$ for $\mathcal{E}_0/V_{\rm vir}^2\lesssim 5$.  
Expansions at higher order can be similarly obtained. 
Alternatively, \cite{Zhao1997} provides an analytic approximation to the distribution function.

\section{Analytical lensing properties with the Mellin transform method}
\label{appendix:mellin}

\subsection{Principle}

The Mellin transform method \citep{Marichev1985, Adamchick1996, Fikioris2007} enables to express definite integrals as Mellin-Barnes integrals. It was notably used by \cite{Mazure2002}, \cite{Baes2011a}, \cite{Baes2011b}, and \cite{Retana-Montenegro2012} to obtain analytic expressions for projected quantities relevent to gravitational lensing from the three-dimensional Einasto profile.

The Mellin transform $\mathfrak{M}_f(u)$ of a function $f(z)$ is defined as 
\be
\label{eq:Mellin}
\mathfrak{M}_f(u) = \phi(u)= \int_0^{+\infty} f(z)z^{u-1} dz
\ee
and its inverse
\be
\label{eq:Mellin_inv}
\mathfrak{M}^{-1}_{\phi}(z) = f(z) = \frac{1}{2\pi i} \int_{\mathcal{L}} \phi(u) z^{-u} du, 
\ee
where $\mathcal{L}$ is a vertical line in the complex plane. The Mellin convolution of two functions $f_1(z)$ and $f_2(z)$ is defined as
\be
\label{eq:Mellin_conv}
(f_1 \star f_2)(z) = \int_0^{+\infty} f_1(t) f_2\left(\frac{z}{t} \right) \frac{dt}{t}
\ee
and, as for the better-known Fourier transform, the Mellin transform of a Mellin convolution of two functions is equal to the product of their Mellin transforms, i.e., 
\be
\label{eq:Mellin_product}
\mathfrak{M}_{f_1 \star f_2}(u)=\mathfrak{M}_{f_1}(u) \times \mathfrak{M}_{f_2}(u).
\ee

It can be shown that any definite integral 
\be
f(z)= \int_0^{+\infty}g(t,z)dt
\ee
can be written as the Mellin convolution of two functions $f_1$ and $f_2$ and hence transformed into an inverse Mellin transform, 
\be
\label{eq:Mellin_integral}
f(z) = \frac{1}{2\pi i} \int_{\mathcal{L}} \mathfrak{M}_{f_1}(u) \mathfrak{M}_{f_2}(u) z^{-u} du. 
\ee
If $f_1$ and $f_2$ are hypergeometric functions, i.e., in a large number of cases, the integral of Eq.~(\ref{eq:Mellin_integral}) is a Mellin-Barnes integral that can be expressed as a Meijer $G$ or a Fox $H$ function \citep[e.g., ][]{Meijer1936, Fox1961, Mathai1978, Srivastava1982, Kilbas1999, Kilbas2004, Mathai2009}. Under certain conditions, these functions are analytical and the line integral can be evaluated using the residue theorem.

\subsection{Application to the surface density}

The integral entering the expression of the surface density $\widetilde{\Sigma}(X)$ (Eq.~(\ref{eq:Sigma_dekel_f})) can be expressed with $z=1$ as the Mellin convolution of 
\be
\label{eq:Mellin_f1}
f_1(t) = 2\rho_c r_c \frac{t^2}{t^a(1+t^{1/2})^{2(3.5-a)}}
\ee
and 
\be
\label{eq:Mellin_f2}
f_2(t) = 
\left\{
\begin{array}{ll}
	\displaystyle \frac{t}{\sqrt{1-X^2 t^2}} & {\rm if~} 0\leq t \leq X^{-1}\\
	\displaystyle 0 & {\rm if~} t > X^{-1}. 
\end{array}
\right.
\ee
The Mellin transform of $f_1$ is
\be
\mathfrak{M}_{f_1}(u)=4\rho_c r_c \mathcal{B}(4+2u-2a,3-2u)
\ee
with the variable change used for Eq.~(\ref{eq:U32}), while that of $f_2$ is
\be
\mathfrak{M}_{f_2}(u)= \frac{\sqrt{\pi} \Gamma(\frac{1+u}{2})}{\Gamma(\frac{u}{2})} \frac{1}{u X^{1+u}}
\ee
as in \cite{Baes2011a}, \cite{Baes2011b} and \cite{Retana-Montenegro2012}. Following Eq.~(\ref{eq:Mellin_integral}), the surface density can thus be expressed as 
\be
\label{eq:Sigma_Mellin}
\widetilde{\Sigma}(R)  = 4 \sqrt{\pi} \rho_c r_c \frac{1}{2\pi i} \int_{\mathcal{L}} \mathcal{B}(4+2u-2a,3-2u) \frac{\Gamma(\frac{1+u}{2})}{\Gamma(\frac{u}{2})uX^{1+u}} du,
\ee
which becomes 
\be
\label{eq:Sigma_int_app}
\widetilde{\Sigma}(R)  = 4 \sqrt{\pi} \rho_c r_c \frac{X}{2\pi i} \int_{\mathcal{L}} \frac{\Gamma(4y\!-\!2a) \Gamma(7\!-\!4y)}{\Gamma(7\!-\!2a)} \frac{\Gamma(y\!-\!\frac{1}{2})}{\Gamma(y)} \left[ X^2\right]^{-y} dy
\ee
with the variable change $y=1+u/2$ and $\mathcal{B}(a,b)=\Gamma(a)\Gamma(b)/\Gamma(a\!+\!b)$ (Eq.~(\ref{eq:Sigma_int})).
This integral is a Fox $H$ function, specified in Eq.~(\ref{eq:Sigma_H}).

As a sanity check, Eq.~(\ref{eq:Sigma_H}) can be used to retrieve the total mass $M_{\rm tot}=\mu M_{\rm vir}$ of an untruncated DZ halo by integrating the surface density $\widetilde{\Sigma}$ over the plane of the sky, 
\be
M_{\rm tot} = 2\pi \!\!\int_0^{+\infty} \!\!\!\!\!\!\!\!\widetilde{\Sigma}(R) R dR = \frac{4 \pi^{3/2} \rho_c r_c^3}{\Gamma(7-2a)} 
\!
\int_0^{+\infty} \!\!\!\!\!\!\!
t^{1/2}
H_{2,2}^{2,1} 
\left[
\left.
\!\!\!\!
\begin{array}{c}
	(-6,\!4), (0,\!1)\\
	(-\frac{1}{2},\!1), (-2a,\!4)
\end{array}
\!\!
\right|
t
\right]
\!
dt
\ee 
with the variable change $t=X^{1/2}$. This integral, which is the Mellin transform of a Fox $H$ function, can be calculated using Eq.~(2.8) of \cite{Mathai2009}. This yields
\be
M_{\rm tot} = 
\frac{4 \pi^{3/2} \rho_c r_c^3}{\Gamma(7-2a)} \frac{\Gamma(1)\Gamma(6-2a)\Gamma(1)}{\Gamma(\frac{3}{2})} = \frac{4\pi\rho_c r_c^3}{3-a} = \mu \Mvir
\ee 
since $\rho_c = (3-a)\mu \Mvir/4\pi r_c^3$.

\section{Series expansion of the lensing properties}
\label{appendix:lensing}

\subsection{Principle}

The Fox $H$ function 
\be
\label{eq:appendix_Fox}
H_{p,q}^{m,n} 
\left[
\left.
\!\!\!\!\!
\begin{array}{c}
	(\mathbf{a}, \!\mathbf{A})\\
	(\mathbf{b}, \!\mathbf{B})
\end{array}
\!\!\!
\right|
z
\right]
\!
= 
\!
\frac{1}{2\pi i} \int_{\mathcal{L}}
\!\!
\frac{\Pi_{j=1}^m \Gamma(b_j\!+\!B_j y) \Pi_{j=1}^n \Gamma(1\!-\!a_j\!-\!A_j y)}{\Pi_{j=m+1}^q \Gamma(1\!-\!b_j\!-\!B_j y) \Pi_{j=n+1}^p\Gamma(a_j\!+\!A_j y)} z^{-y} dy
\ee 
has analytical series expansions under certain conditions satisfied by the Fox $H$ functions considered in this paper \citep[cf.][]{Kilbas1999, Mathai2009, Baes2011a, Baes2011b}. As explicated in Appendix A of \cite{Baes2011a}, if all functions $\Gamma(b_i+B_i y)$ with $0\leq i \leq m$ have only single poles $\beta_{i,k}= -(b_i+k)/B_i$, 
\begin{multline}
\label{eq:appendix_expansion1}
H_{p,q}^{m,n} 
\left[
\left.
\!\!\!\!\!
\begin{array}{c}
	(\mathbf{a}, \!\mathbf{A})\\
	(\mathbf{b}, \!\mathbf{B})
\end{array}
\!\!\!
\right|
z
\right]
\!
= \\
\sum_{i=1}^{m}\! \sum_{k=0}^{\infty}\!\! \frac{(-1)^k}{k! B_i} 
\frac{
	\prod_{j=1,j\neq i}^m \!\Gamma\left(b_j\!-\!B_j\frac{b_i+k}{B_i}\right)
	\prod_{j=1}^n \!\Gamma\left(1\!-\!a_j+A_j\frac{b_i+k}{B_i}\right)
}{
	\prod_{j=m+1}^q \!\Gamma\left(1\!-\!b_j\!+\!B_j\frac{b_i+k}{B_i}\right)
	\prod_{j=n+1}^p \!\Gamma\left(a_j\!-\!A_j\frac{b_i+k}{B_i}\right)
}
z^{(b_i+k)/B_i}, 
\end{multline}
while if several gamma functions share the same pole, the Fox $H$ function can be expressed as a logarithmic-power series rather than a power series \citep{Kilbas1999}. 
In the case where two gamma functions $\Gamma(b_i+B_i y)$ with $1\leq i\leq m$ share at least one pole, 
\begin{multline}
\label{eq:appendix_expansion2}
H_{p,q}^{m,n} 
\left[
\left.
\!\!\!\!\!
\begin{array}{c}
(\mathbf{a}, \!\mathbf{A})\\
(\mathbf{b}, \!\mathbf{B})
\end{array}
\!\!\!
\right|
z
\right]
\!
= \\
{\sum_{i,k}} ^{\prime}  \frac{(-1)^k}{k! B_i} 
\frac{
	\prod_{j=1,j\neq i}^m \Gamma\left(b_j\!-\!B_j\frac{b_i+k}{B_i}\right)
	\prod_{j=1}^n \Gamma\left(1\!-\!a_j+A_j\frac{b_i+k}{B_i}\right)
}{
	\prod_{j=m+1}^q \Gamma\left(1\!-\!b_j\!+\!B_j\frac{b_i+k}{B_i}\right)
	\prod_{j=n+1}^p \Gamma\left(a_j\!-\!A_j\frac{b_i+k}{B_i}\right)
}
z^{(b_i+k)/B_i}
\\
+{\sum_{k_1}} ^{\prime\prime} \!\! \frac{(-1)^{k_1+k_2}}{k_1! k_2! B_1 B_2}
\frac{
	\prod_{j=3}^m \!\Gamma\!\left(b_j\!-\!B_j\frac{b_i+k_1}{B_i}\right)
	\prod_{j=1}^n \!\Gamma\!\left(1\!-\!a_j+A_j\frac{b_i+k_1}{B_i}\right)
}{
	\prod_{j=m+1}^q \!\Gamma\!\left(1\!-\!b_j\!+\!B_j\frac{b_i+k_1}{B_i}\right)
	\prod_{j=n+1}^p \!\Gamma\!\left(a_j\!-\!A_j\frac{b_i+k_1}{B_i}\right)
}
\\
\times z^{(b_i+k_1)/B_i} \left(C_{k_1}\!-\!\ln z \right)\\
\end{multline}
where the two gamma functions sharing poles are described by the first two indices $j=1$ and $j=2$, the first sum (with a prime) covers the single poles, the second sum (with a double prime) covers the second-order poles, $k_2=B_2(b_1+k_1)/B_1-b_2$, 
\begin{multline}
C_{k_1} = B_1 \psi (k_1+1) +B_2 \psi (k_2+1)\\
+\sum_{j=3}^m B_j \psi(b_j-B_j\frac{b_1+k_1}{B_1}) 
-\sum_{j=1}^{n} A_j \psi(1-a_j+A_j\frac{b_1+k_1}{B_1})\\
+\sum_{j=m+1}^q B_j \psi(1-b_j+B_j\frac{b_1+k_1}{B_1})
-\sum_{j=n+1}^{p} A_j \psi(a_j-A_j\frac{b_1+k_1}{B_1}), 
\end{multline}
and $\psi$ is the digamma function \cite[][Appendix A]{Baes2011a}. 
Poles of the Fox $H$ functions considered in this paper are of the form $a/2-k/4$ and $n/2-k$ where $a$ is the DZ slope parameter and the $n$ are odd integers. In most realistic cases, $a$ is either a non-rational number or a rational number with an odd denominator such that the two types of poles do not overlap. 
In this case, it is Eq. (\ref{eq:appendix_expansion1}) that enables to express the different Fox $H$ functions encountered in this article as series expansions. The following series expansions assume that $a$ is either a non-rational number or a rational number with an odd denominator, but other series expressions could be deduced from Eq.~(\ref{eq:appendix_expansion2}) if $a$ were rational with an even denominator.

\subsection{Surface density}

Assuming $a$ to be either non-rational or rational with an even denominator, Eq.~(\ref{eq:Sigma_H}) and the series expansion (\ref{eq:appendix_expansion1}) yields for the surface density of an untruncated DZ profile 
\begin{multline}
\label{eq:appendix_sigma_expansion}
\widetilde{\Sigma}(X) = \frac{4 \sqrt{\pi} \rho_c r_c}{\Gamma(7-2a)} 
\left[
\sum_{k=0}^{\infty} \frac{(-1)^k}{k!} \frac{\Gamma(-2a+2-4k)\Gamma(5+4k)}{\Gamma(\frac{1}{2}-k)} X^{2k}\right.\\
\left.+\sum_{k=0}^{\infty} \frac{1}{4}\frac{(-1)^k}{k!} \frac{\Gamma(-\frac{1}{2}+\frac{a}{2}-\frac{k}{4})\Gamma(7-2a+k)}{\Gamma(\frac{a}{2}-\frac{k}{4})} X^{1-a+k/2}
\right], 
\end{multline}
where $X=r/r_c$ the two-dimensional radius scaled by the DZ characteristic radius (cf. Eq.~(\ref{eq:rho32})). 
If $a>1$, this surface density is divergent in $X=0$, while if $a<1$ -- which is the most frequent case --, one can retrieve the central density expressed in Eq.~(\ref{eq:Sigma_0}) as a consistency check. 
We recall that the surface density of a DZ profile truncated at the virial radius is $\Sigma(X)=\widetilde{\Sigma}(X)-\widetilde{\Sigma}(c)$.

\subsection{Cumulative mass and deflection angle}

The series expansion of $\widetilde{\mathcal{M}}(X)$ can be obtained either by applying Eq.~(\ref{eq:appendix_expansion1}) to the Fox $H$ function of Eq.~(\ref{eq:M2D_H}) or by directly integrating Eq.~(\ref{eq:appendix_sigma_expansion}): 
\begin{multline}
\label{eq:appendix_M2D_expansion}
\widetilde{\mathcal{M}}(X)=\frac{4\pi^{3/2}\rho_c r_c^3}{\Gamma(7-2a)} 
\left[
\sum_{k=0}^{\infty} \frac{(-1)^k}{k!}
\frac{\Gamma(-2a+2-4k)\Gamma(5+4k)}{\Gamma\left(\frac{1}{2}-k\right)(k+1)} X^{2k+2}
\right.\\
\left.+\sum_{k=0}^{\infty} \frac{1}{4}\frac{(-1)^k}{k!}
\frac{\Gamma\left(-\frac{1}{2}+\frac{a}{2}-\frac{k}{4}\right)\Gamma(7-2a+k)}{\Gamma\left(\frac{a}{2}-\frac{k}{4}\right)\left(\frac{3}{2}-\frac{a}{2}+\frac{k}{4}\right)} X^{3-a+k/2}
\right]. 
\end{multline}
The series expansions of
$\mathcal{M}(X)=\widetilde{\mathcal{M}}(X)-\widetilde{\mathcal{M}}(c)$, 
the  deflection angle $\widetilde{\alpha}(X) = \widetilde{\mathcal{M}}(X)/\pi r_c^2 \Sigma_{\rm crit} X$, 
and $\alpha(X)=\widetilde{\alpha}(X)-\widetilde{\mathcal{M}}(c)/\pi r_c^2 \Sigma_{\rm crit} X$ 
can be deduced from Eq.~(\ref{eq:appendix_M2D_expansion}).

\subsection{Deflection potential}

For a thin axially-symmetric lens, the deflection potential such that the scaled deflexion angle $\alpha=\nabla \psi$ is
\be
\label{eq:lensing_potential_int}
\psi (X) = 2\int_0^X x \kappa(x) \ln\left( \frac{X}{x}\right) dx
\ee
\citep[][Eq.~(8.8)]{Schneider1992}. Injecting Eq.~(\ref{eq:Sigma_int}) and following similar steps as for Eq.~(\ref{eq:M2D_H}) yields 
\be
\label{eq:lensing_potential}
\widetilde{\psi} (X) = \frac{2 \sqrt{\pi} \rho_c r_c}{\Gamma(7-2a)\Sigma_{\rm crit}} X^3\! ~
H^{2,3}_{4, 4} \!
\left[
\left.
\!\!\!\!\!
\begin{array}{c}
	(-6,4), (-\frac{1}{2},1), (-\frac{1}{2},1), (0,1)\\
	(-\frac{1}{2},1), (-2a,4), (-\frac{3}{2},1), (-\frac{3}{2},1)
\end{array}
\!\!\!
\right|
X^2
\right]
\ee
for an untruncated DZ profile.
As for the cumulative mass, the series expansion of $\widetilde{\psi}(X)$ can be obtained either by applying Eq.~(\ref{eq:appendix_expansion1}) to the Fox $H$ function of Eq.~(\ref{eq:lensing_potential}) or by directly injecting Eq.~(\ref{eq:appendix_sigma_expansion}) into Eq.~(\ref{eq:lensing_potential_int}), yielding
\begin{multline}
\label{eq:appendix_psi_expansion}
\widetilde{\mathcal{\psi}}(X)=\frac{2\sqrt{\pi}\rho_c r_c}{\Gamma(7-2a) \Sigma_{\rm crit}} 
\left[
\sum_{k=0}^{\infty} \frac{(-1)^k}{k!}
\frac{\Gamma(-2a+2-4k)\Gamma(5+4k)}{\Gamma\left(\frac{1}{2}-k\right)(1+k)^2} X^{2k+2}
\right.\\
\left.+\sum_{k=0}^{\infty} \frac{1}{4}\frac{(-1)^k}{k!}
\frac{\Gamma\left(-\frac{1}{2}+\frac{a}{2}-\frac{k}{4}\right)\Gamma\left(7-2a+k\right)}{\Gamma\left(\frac{a}{2}-\frac{k}{4}\right)\left(\frac{3}{2}-\frac{a}{2}+\frac{k}{4}\right)^2} X^{3-a+k/2}
\right]. 
\end{multline}
For a DZ profile truncated at the virial radius, the deflection potential is $\psi(X)=\widetilde{\psi}(X)-X^2 \widetilde{\Sigma}(c)/\Sigma_{\rm crit}$, whose series expansion can be deduced from Eqs.~(\ref{eq:appendix_sigma_expansion}) and (\ref{eq:appendix_psi_expansion}). The associated Fermat potential $\phi(x,y)=(x-y)^2/2-\psi(x)$ and its series expansion can be similarly deduced.

\subsection{Average surface density}

As previously, the average surface density for an untruncated DZ profile yields
\begin{multline}
\label{eq:appendix_Sigmab_expansion}
\widetilde{\overline{\Sigma}}(X)=\frac{4\sqrt{\pi}\rho_c r_c}{\Gamma(7-2a)} 
\left[
\sum_{k=0}^{\infty} \frac{(-1)^k}{k!}
\frac{\Gamma(-2a+2-4k)\Gamma(5+4k)}{\Gamma\left(\frac{1}{2}-k\right)(1+k)} X^{2k}
\right.\\
\left.+\sum_{k=0}^{\infty} \frac{1}{4}\frac{(-1)^k}{k!}
\frac{\Gamma\left(-\frac{1}{2}+\frac{a}{2}-\frac{k}{4}\right)\Gamma\left(7-2a+k\right)}{\Gamma\left(\frac{a}{2}-\frac{k}{4}\right)\left(\frac{3}{2}-\frac{a}{2}+\frac{k}{4}\right)} X^{1-a+k/2}
\right], 
\end{multline}
while it is $\overline{\Sigma}(X)=\widetilde{\overline{\Sigma}}(X)-X\widetilde{\Sigma}(c)$ for a DZ profile truncated at the virial radius. Expressions for the lensing shear $\gamma(X)=[\overline{\Sigma}(X)-\Sigma(X)]/\Sigma_{\rm crit}$ and the magnification $\mu(X)=[(1-\kappa(X))^2-\gamma^2(X)]^{-1}$ can be deduced from Eqs.~(\ref{eq:appendix_sigma_expansion}) and (\ref{eq:appendix_Sigmab_expansion}).

\setcounter{section}{7}
\setcounter{figure}{0}

\begin{figure*}
	\includegraphics[width=\textwidth]{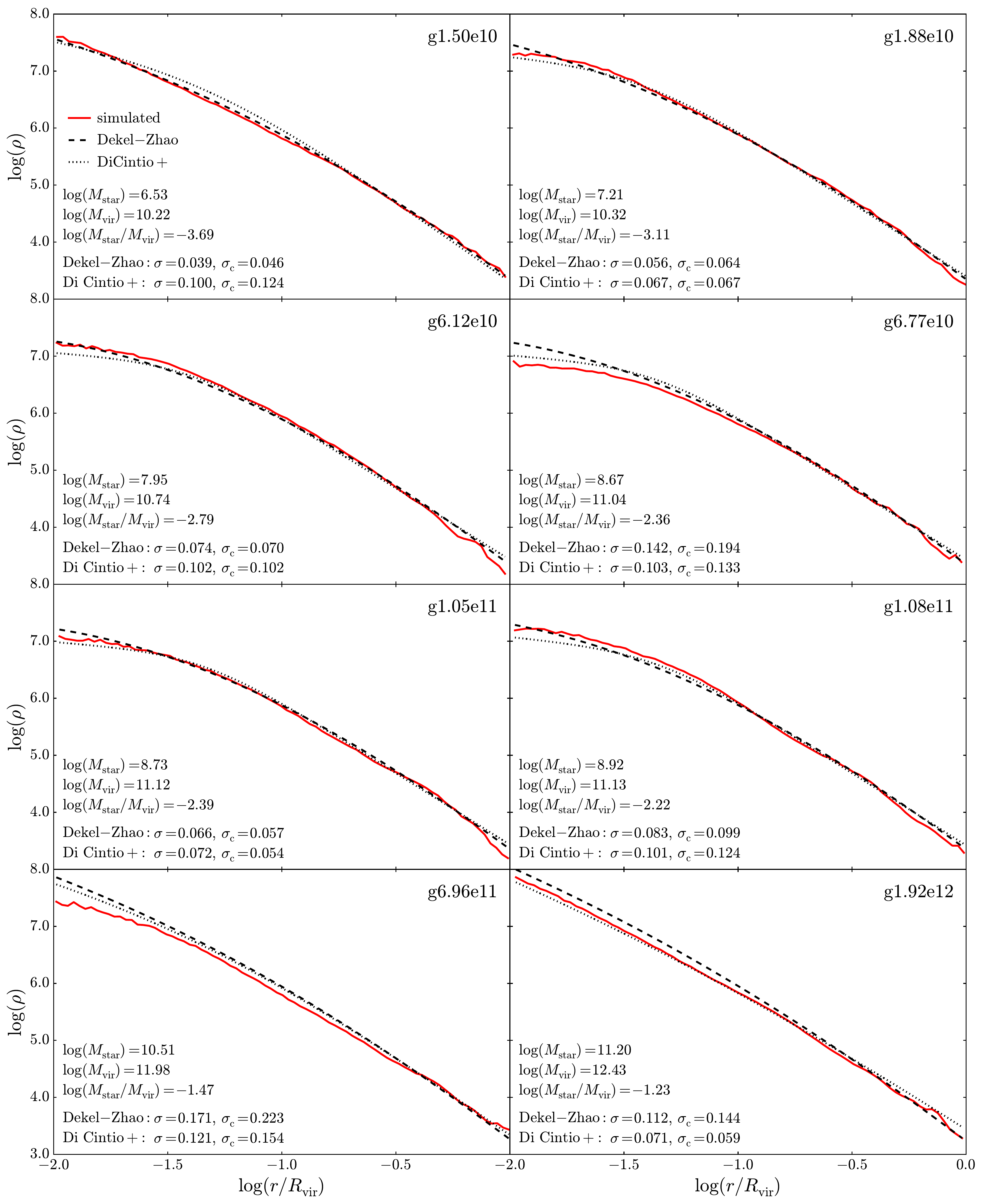}
	\vspace{-0.4cm}
	\caption{The dark matter density profiles at $z=0$ of the 8 arbitrary NIHAO galaxies shown in Fig.~\ref{fig:fits} (plain red line) with their current DZ (dashed) and \protect\citetalias{DiCintio2014b} (dotted) mass-dependent prescriptions. The masses $M_{\rm star}$, $M_{\rm vir}$, $M_{\rm star}/M_{\rm vir}$ and the rms errors $\sigma$ and $\sigma_{\rm c}$ are indicated.}
	\label{fig:prescriptions}
\end{figure*}

\begin{figure*}
	\includegraphics[width=\textwidth]{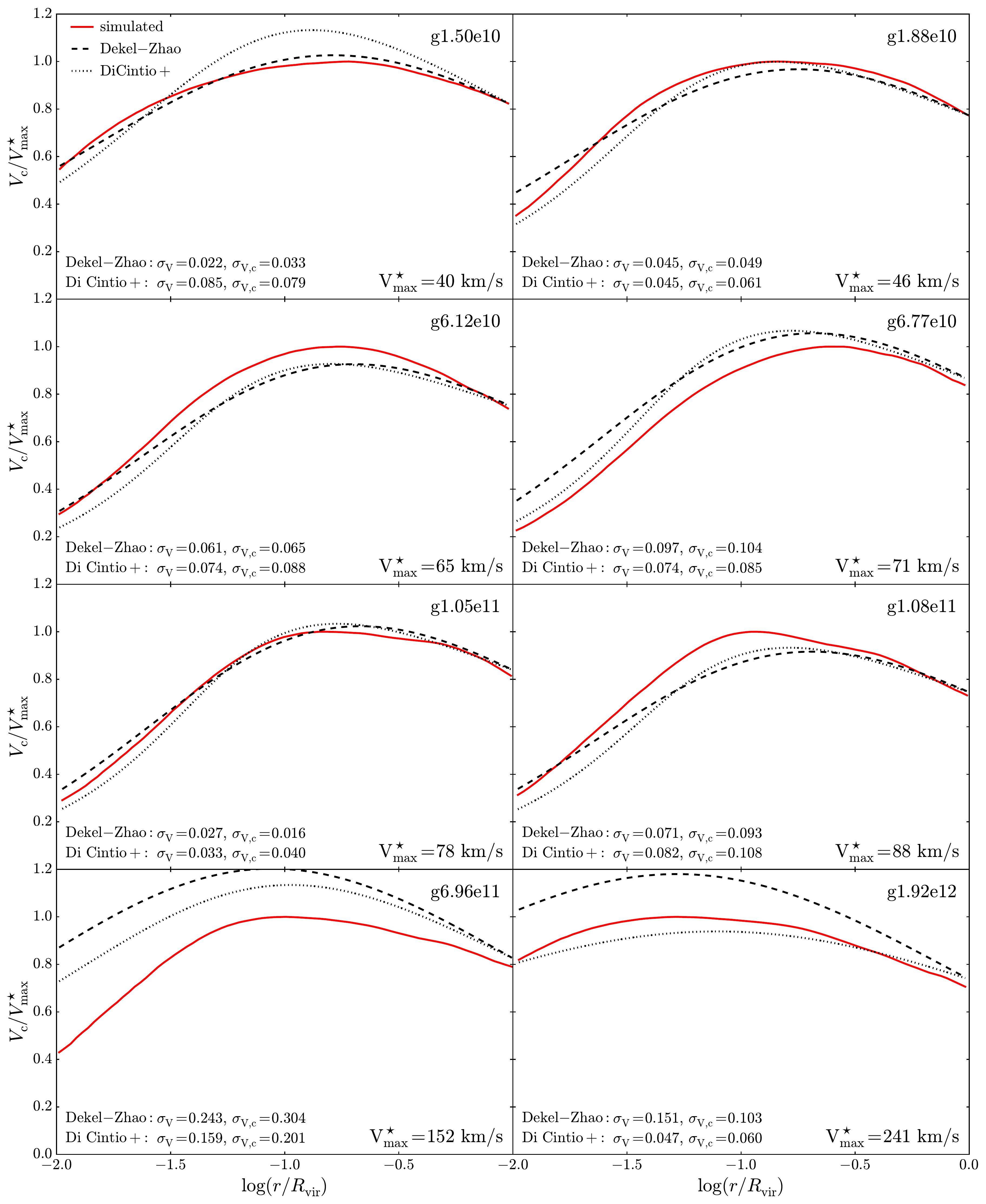}
	\vspace{-0.4cm}
	\caption{
		Dark matter circular velocity profiles, $V_{\rm c}(r) = \sqrt{GM(r)/r}$, of the eight $z=0$ NIHAO galaxies shown in Fig.~\ref{fig:fits} (plain red line) together with those inferred from the current DZ and \protect\citetalias{DiCintio2014b} mass-dependent prescriptions (dashed and dotted lines, respectively). The velocity of each galaxy is normalized to its maximum value $V_{\rm max}^\star$, which is an increasing function of mass. 
	}
	\label{fig:prescriptions_Vc}
\end{figure*}

\begin{figure*}
	\includegraphics[height=0.23\textwidth,trim={0.cm 0 1.5cm 1.cm},clip]{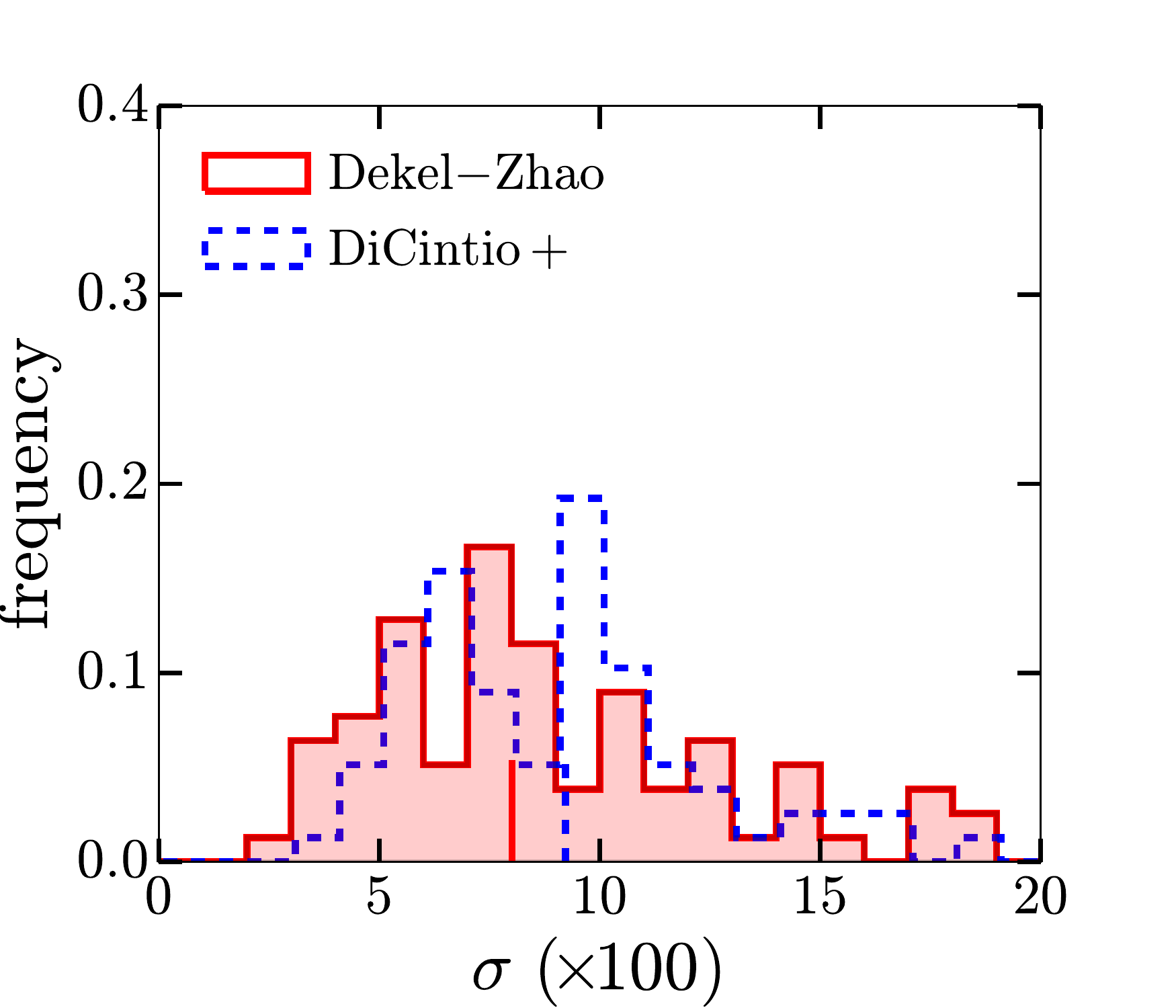}
	\includegraphics[height=0.23\textwidth,trim={1.5cm 0 1.5cm 1.cm},clip]{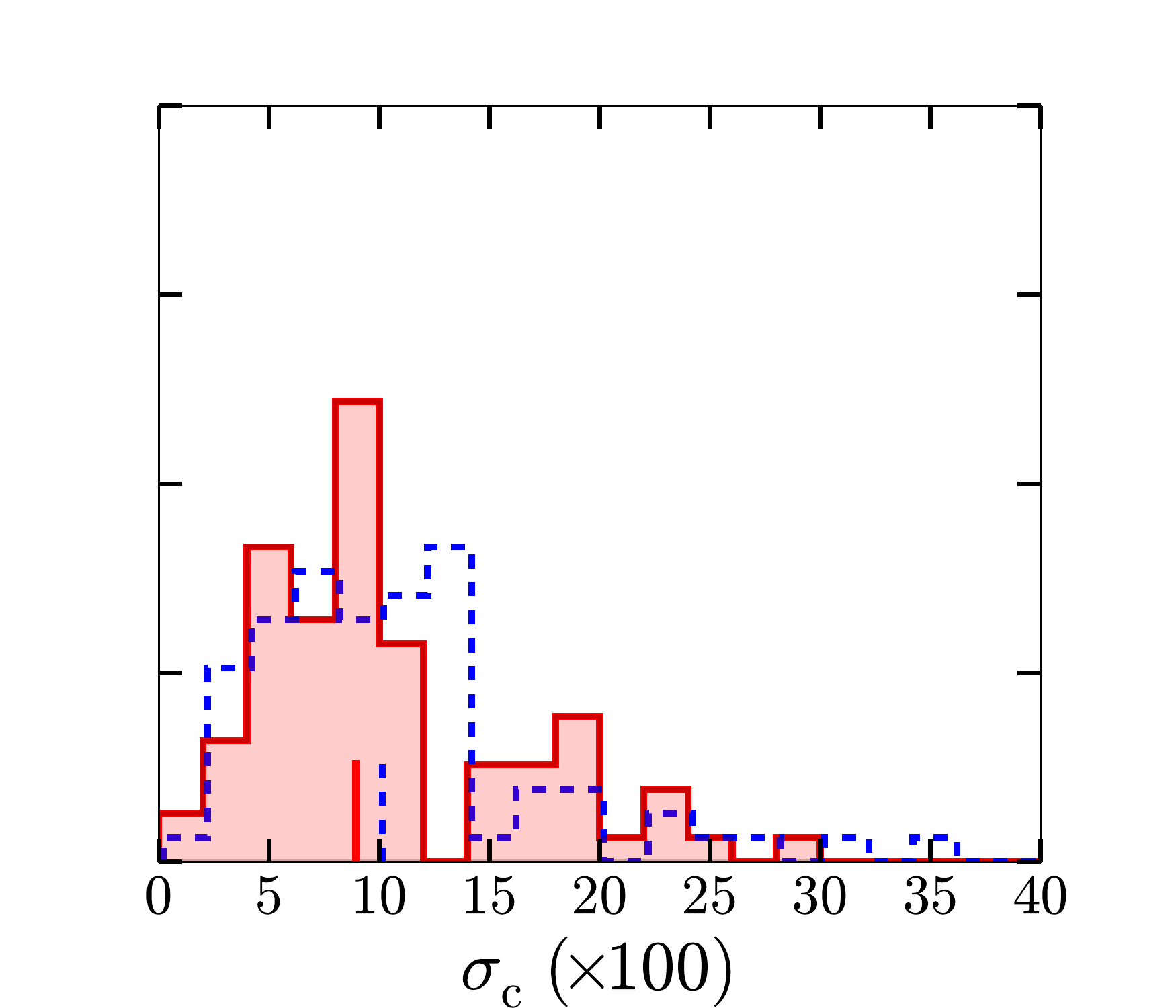}
	\includegraphics[height=0.23\textwidth,trim={1.5cm 0 1.5cm 1.cm},clip]{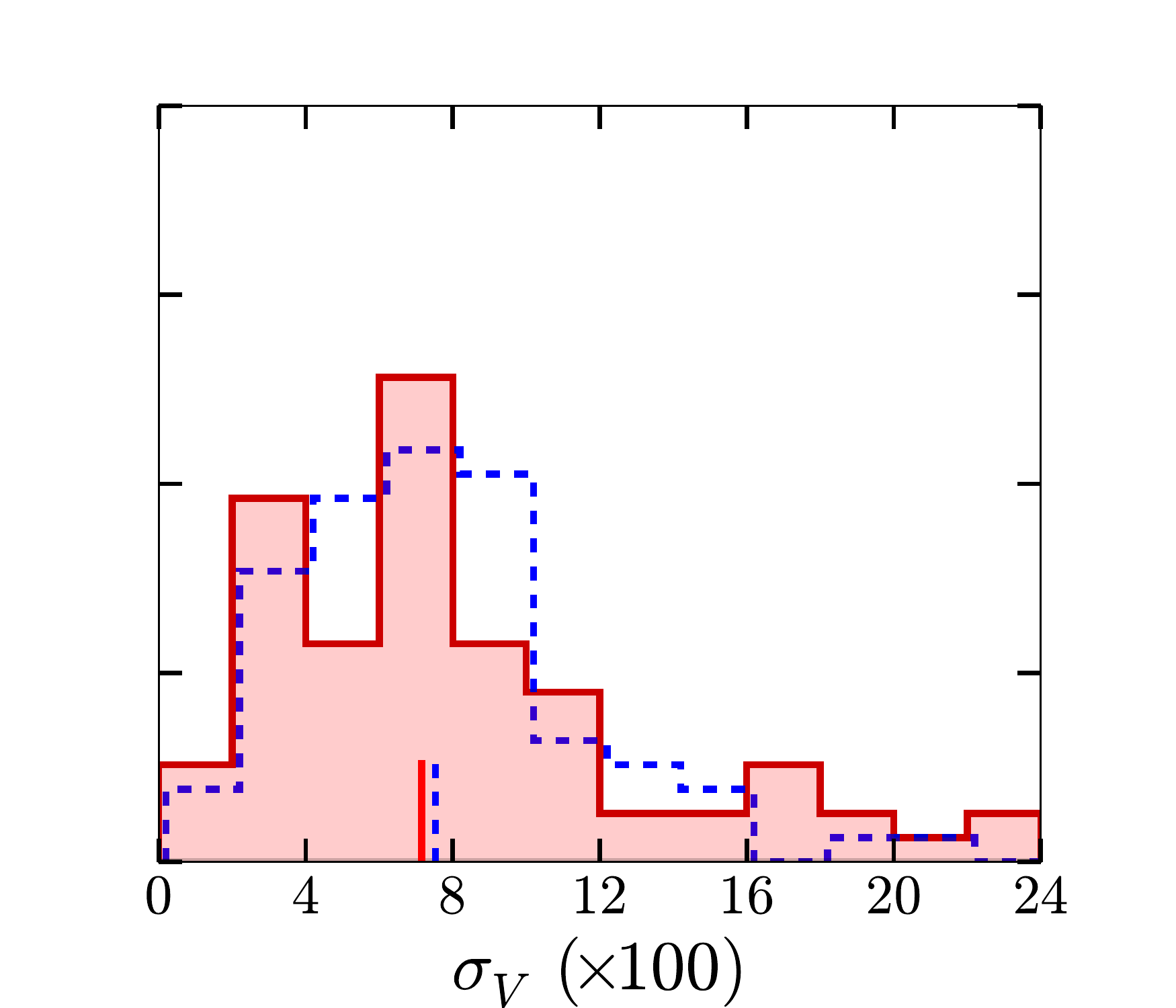}
	\includegraphics[height=0.23\textwidth,trim={1.5cm 0 1.5cm 1.cm},clip]{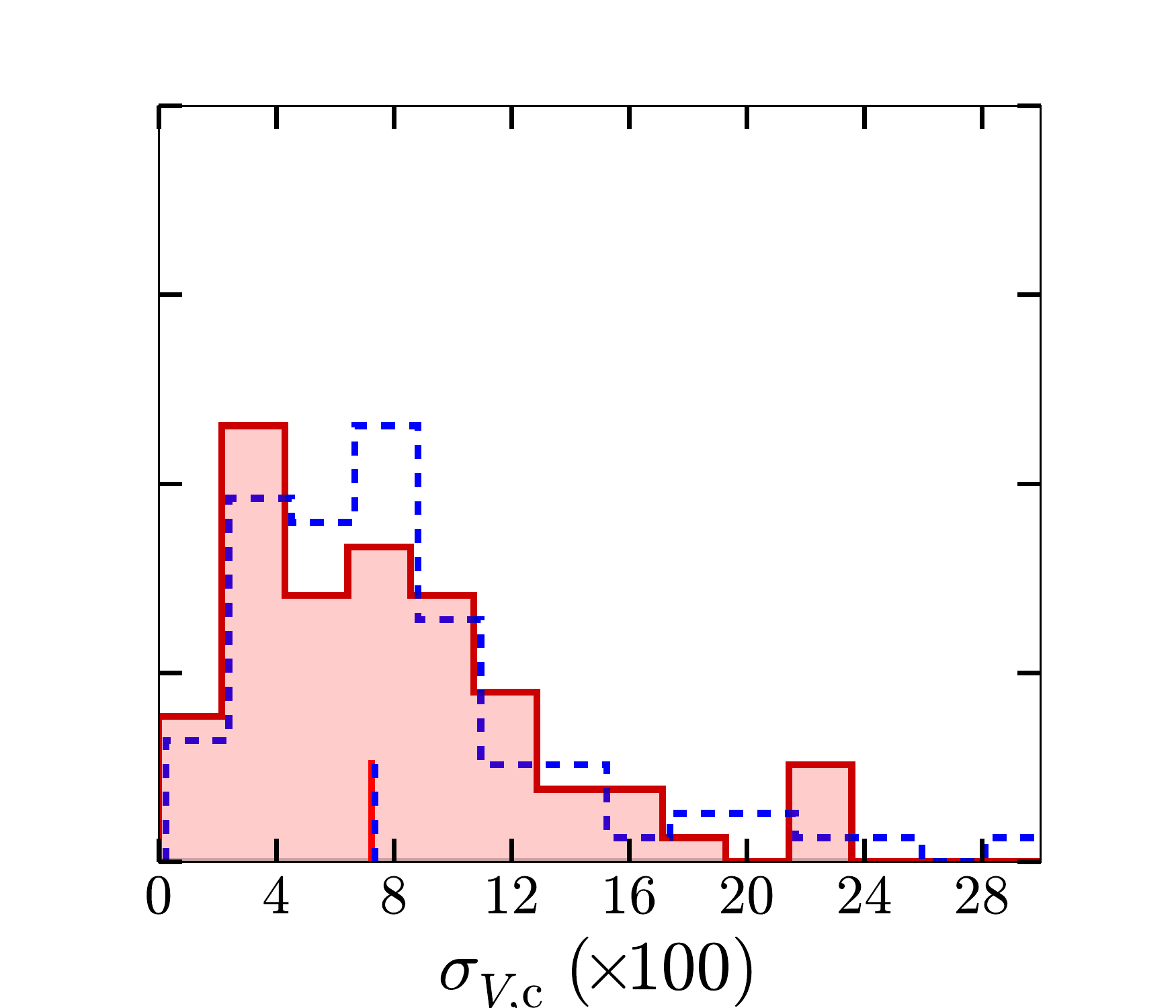}	
	\caption{Comparison between the current DZ (plain red line) and \protect\citetalias{DiCintio2014b} (blue dashed line) mass-dependent prescriptions in terms of their rms errors in $\log\rho$ and $V_c/V_{\rm max}$ over the ranges $0.01\Rvir-\Rvir$ and $0.01\Rvir-0.1\Rvir$ for all NIHAO galaxies at $z=0$. 
	The median values for the two prescriptions, which are highlighted by vertical lines above the x-axis, respectively yield 
		$0.080$ \& $0.092$ for $\sigma$,
		$0.089$ \& $0.101$ for $\sigma_{\rm c}$,
		$0.072$ \& $0.075$ for $\sigma_{\rm V}$, 
		$0.072$ \& $0.074$ for $\sigma_{\rm V, c}$. 
	The standard deviations respectively yield 
		$0.041$ \& $0.043$ for $\sigma$,
		$0.060$ \& $0.064$ for $\sigma_{\rm c}$, 
		$0.054$ \& $0.043$ for $\sigma_{\rm V}$, 
		$0.058$ \& $0.055$ for $\sigma_{\rm V, c}$. 
	The current prescriptions provides equivalent (or marginally better fits) to the DM density profile and the circular velocity profile than the \protect\citetalias{DiCintio2014b} prescriptions. 
	}
	\label{fig:prescriptions_sigma_hist}
\end{figure*}

\begin{figure*}
	\includegraphics[height=0.23\textwidth,trim={0.cm 0 1.5cm 1.cm},clip]{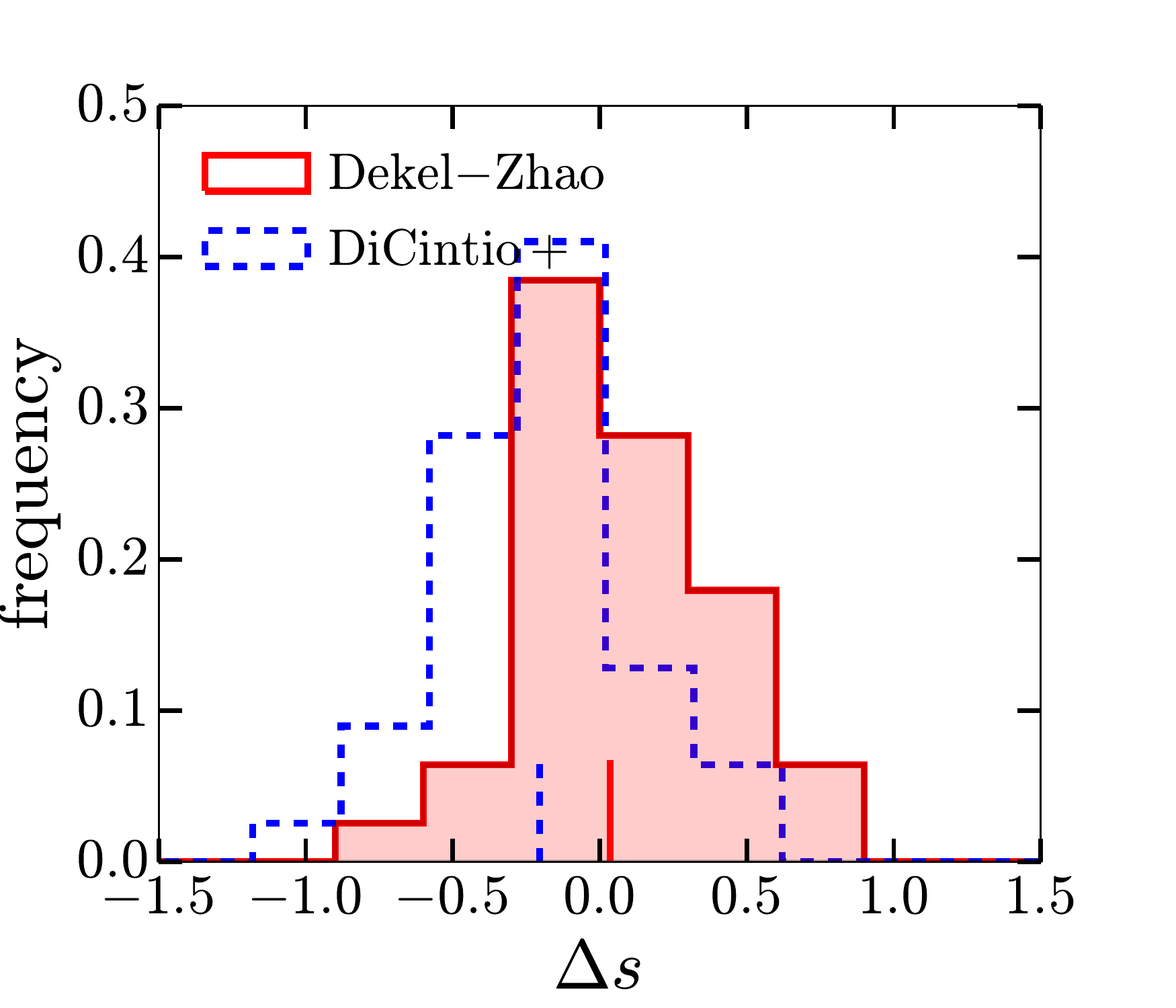}
	\includegraphics[height=0.23\textwidth,trim={1.5cm 0 1.5cm 1.cm},clip]{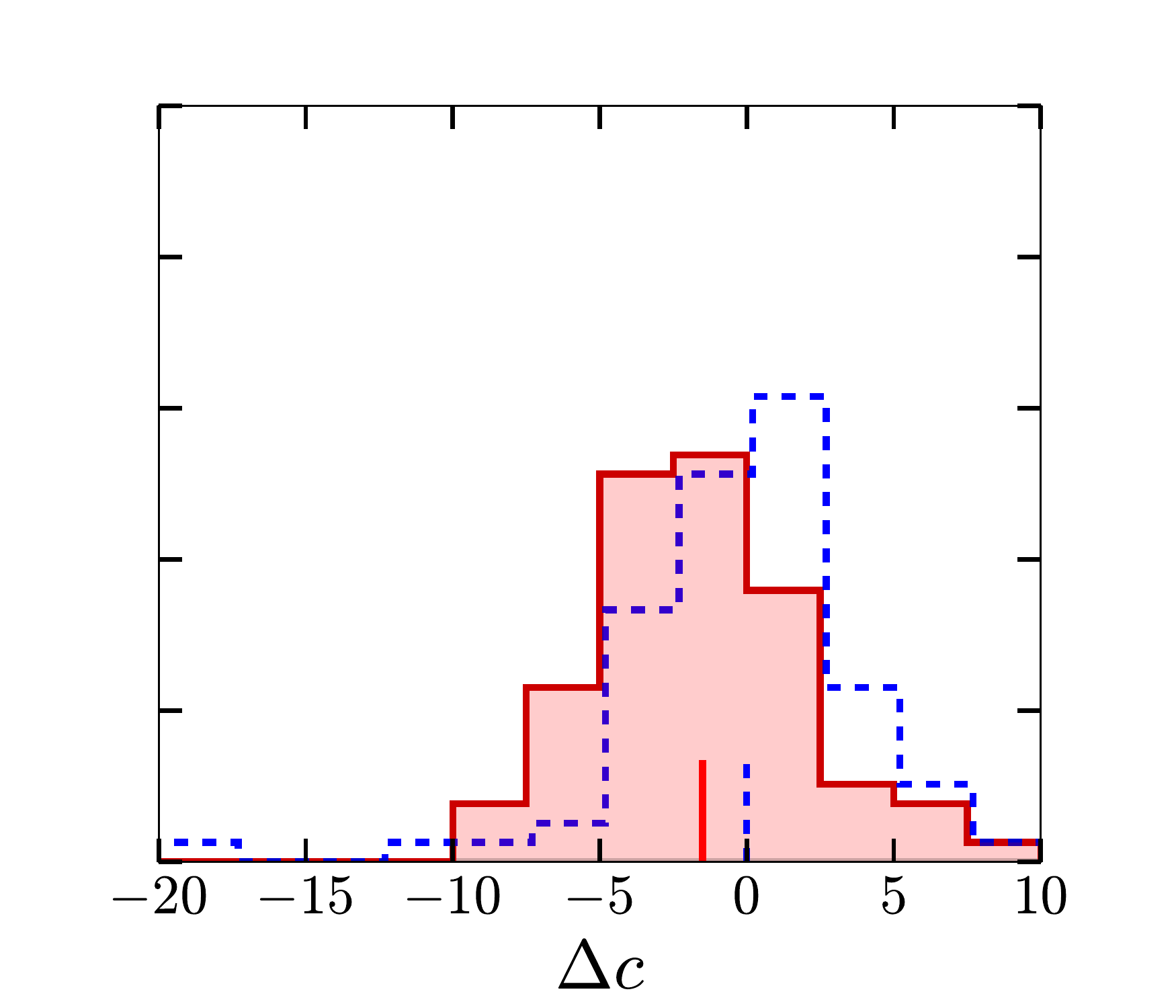}
	\includegraphics[height=0.23\textwidth,trim={1.5cm 0 1.5cm 1.cm},clip]{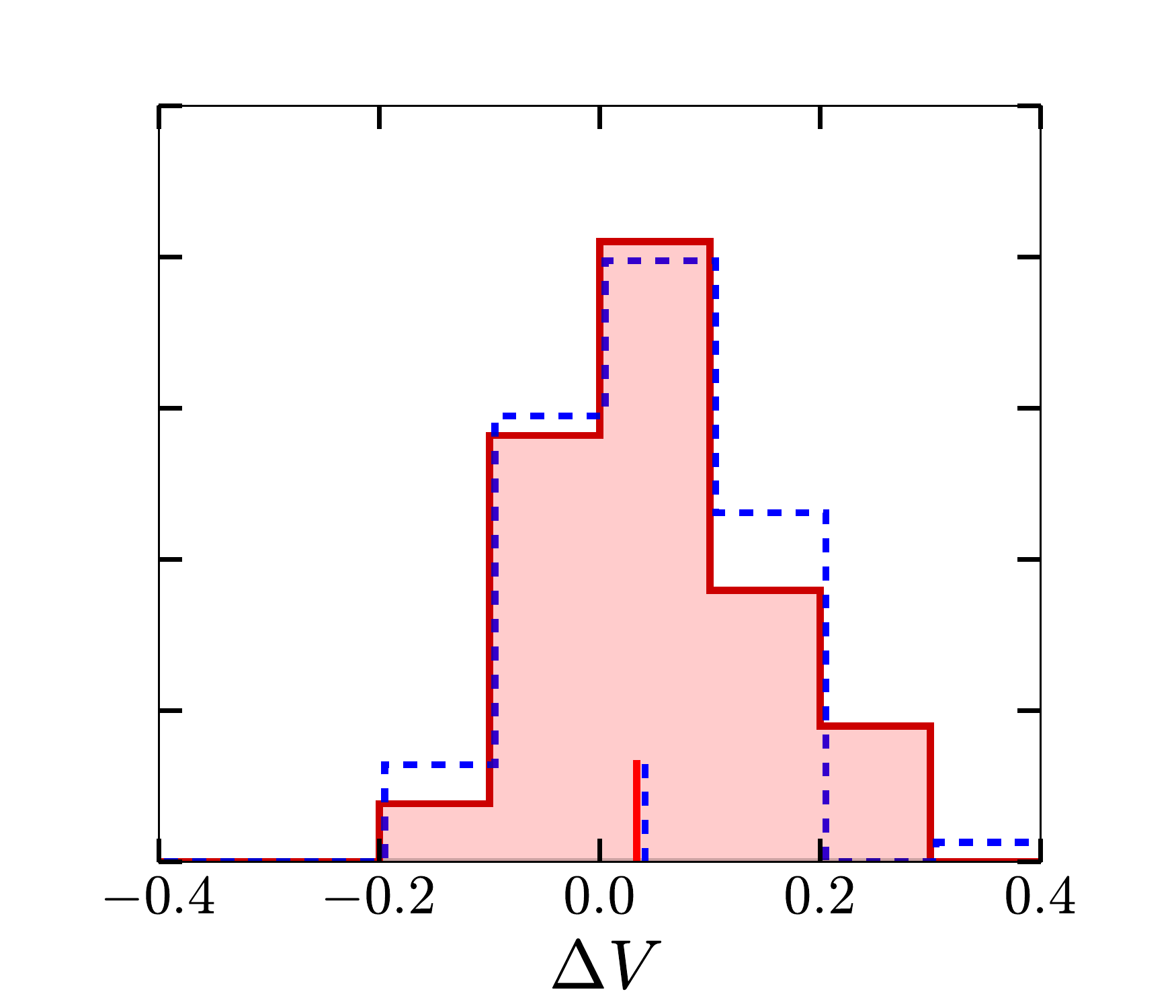}
	\includegraphics[height=0.23\textwidth,trim={1.5cm 0 1.5cm 1.cm},clip]{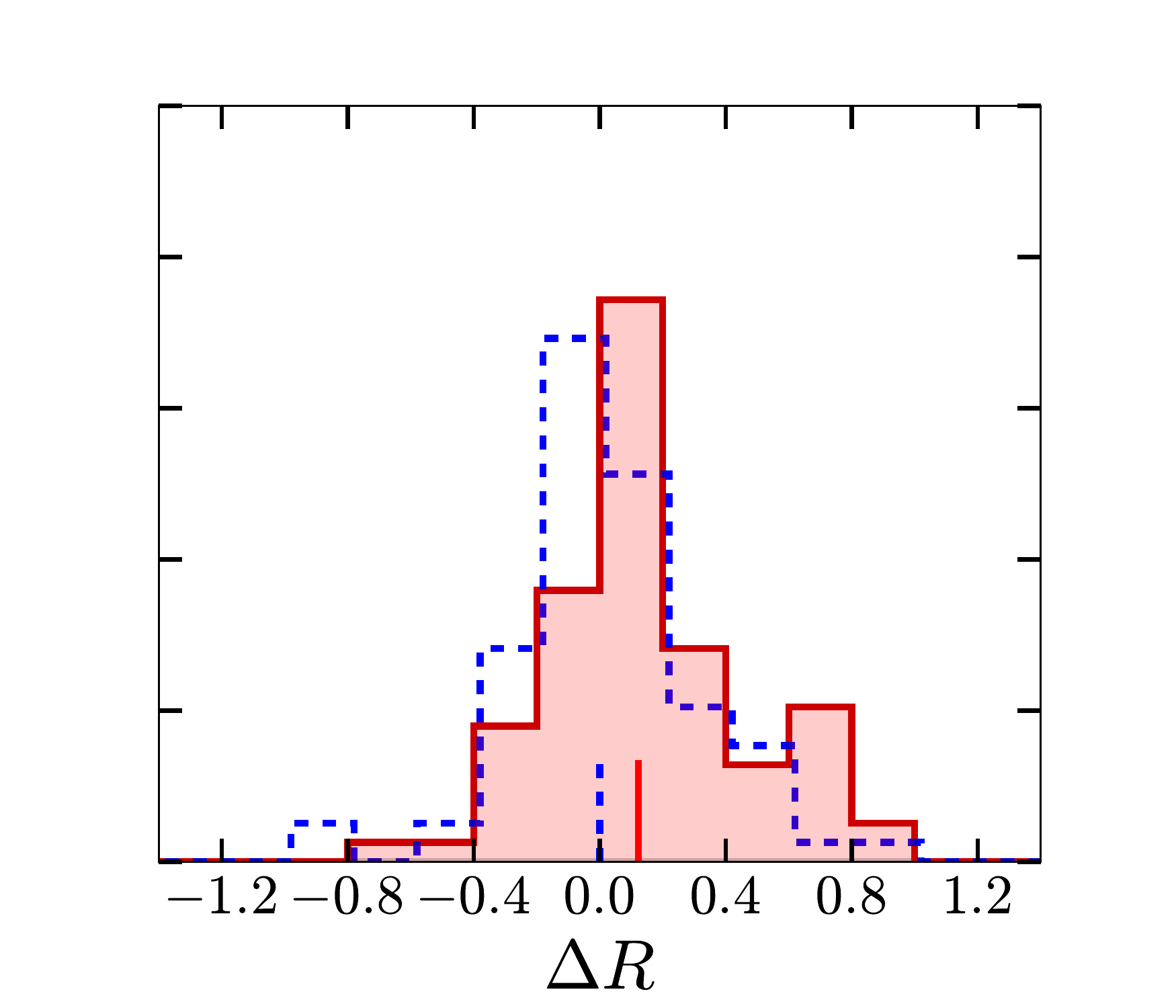}
	\caption{Comparing the current DZ and the \protect\citetalias{DiCintio2014b} mass-dependent prescriptions: inner slope and concentration differences, $\Delta s = s_{1,\rm model}-s_{1}^\star$ and $\Delta c = c_{\rm 2, model}-c_{\rm 2}^\star$, as well as the maximum velocity and radius relative differences, $\Delta V= (V_{\rm max, model}-V_{\rm max}^\star)/V_{\rm max}^\star$ and $\Delta R= (R_{\rm max, model}-R_{\rm max}^\star)/R_{\rm max}^\star$, between  the 
	current DZ (plain red line) and the \protect\citetalias{DiCintio2014b} (blue dashed line) mass-dependent prescriptions and the simulated profiles for all $z=0$ NIHAO galaxies simulated with baryons. 
	The median values for the two prescriptions,
	which are highlighted by vertical lines above the x-axis, respectively yield 
	$0.03$ \& $-0.20$ for $\Delta s$,  
	$-1.5$ \&  $-0.0$  for $\Delta c$,  
	$0.034$ \&  $0.041$ for $\Delta V$, 
	$0.123$ \&  $0.00$ for $\Delta R$. 
	The standard deviations respectively yield 
	$0.31$ \&  $0.31$ for $\Delta s$, 
	$8.7$ \&  $9.3$ for $\Delta c$, 
	$0.089$ \&  $0.086$  for $\Delta V$, 
	$0.312$ \&  $0.298$ for $\Delta R$.
	The current prescription provides fits whose accuracy is comparable to the \protect\citetalias{DiCintio2014b} prescription despite having two shape parameters instead of four (concentration included). 
	}
	\label{fig:prescriptions_comparison_histograms}
\end{figure*}

\setcounter{section}{6}
\section{Mass-dependent prescriptions}
\label{appendix:prescriptions}

Figs.~\ref{fig:prescriptions} and \ref{fig:prescriptions_Vc} complement Section~\ref{section:prescription_accuracy} by comparing the DM density and circular velocity profiles resulting directly from the mass-dependent prescriptions of Section \ref{section:prescriptions} (without leaving the concentration free as in Section \ref{section:rotation}) for the eight fiducial NIHAO haloes at different masses shown in Fig.~\ref{fig:fits} with their simulated profiles. 
The two shape parameters $s_1$ and $c_2$ of the DZ profiles shown in Figs.~\ref{fig:prescriptions} and \ref{fig:prescriptions_Vc} are set by the fitting functions shown in Fig.~\ref{fig:s1_c2_ratios}, expressed in Eqs.~(\ref{eq:s1(x)}) and (\ref{eq:exp_function}), and whose best-fit parameters are given in Table~\ref{table:fit_rho}. 
The figures also show the \citetalias{DiCintio2014b} mass-dependent prescriptions, for which the four shape parameters of the double power-law profile ($a$, $b$, $g$ in Eq.~(\ref{eq:rho_abc}) and $c_2$) are obtained from the stellar and halo masses as indicated in the Appendix of \cite{DiCintio2014b}. 
Fig.~\ref{fig:prescriptions_sigma_hist} systematically compares the rms of the residuals between the simulated density and circular velocity profiles and those stemming from the current DZ and \citetalias{DiCintio2014b} prescriptions. 
Figs.~\ref{fig:prescriptions_comparison_histograms} further shows the distributions of $\Delta s$, $\Delta c$, $\Delta V$ and $\Delta R$ resulting from the two mass-dependent prescriptions.

\section{Additional figures}
\label{appendix:figures}

\subsection{Measuring the inner slope and concentration}

\begin{figure}
	\hfill\includegraphics[width=1\linewidth,trim={0cm 0cm 0cm 0.2cm},clip]{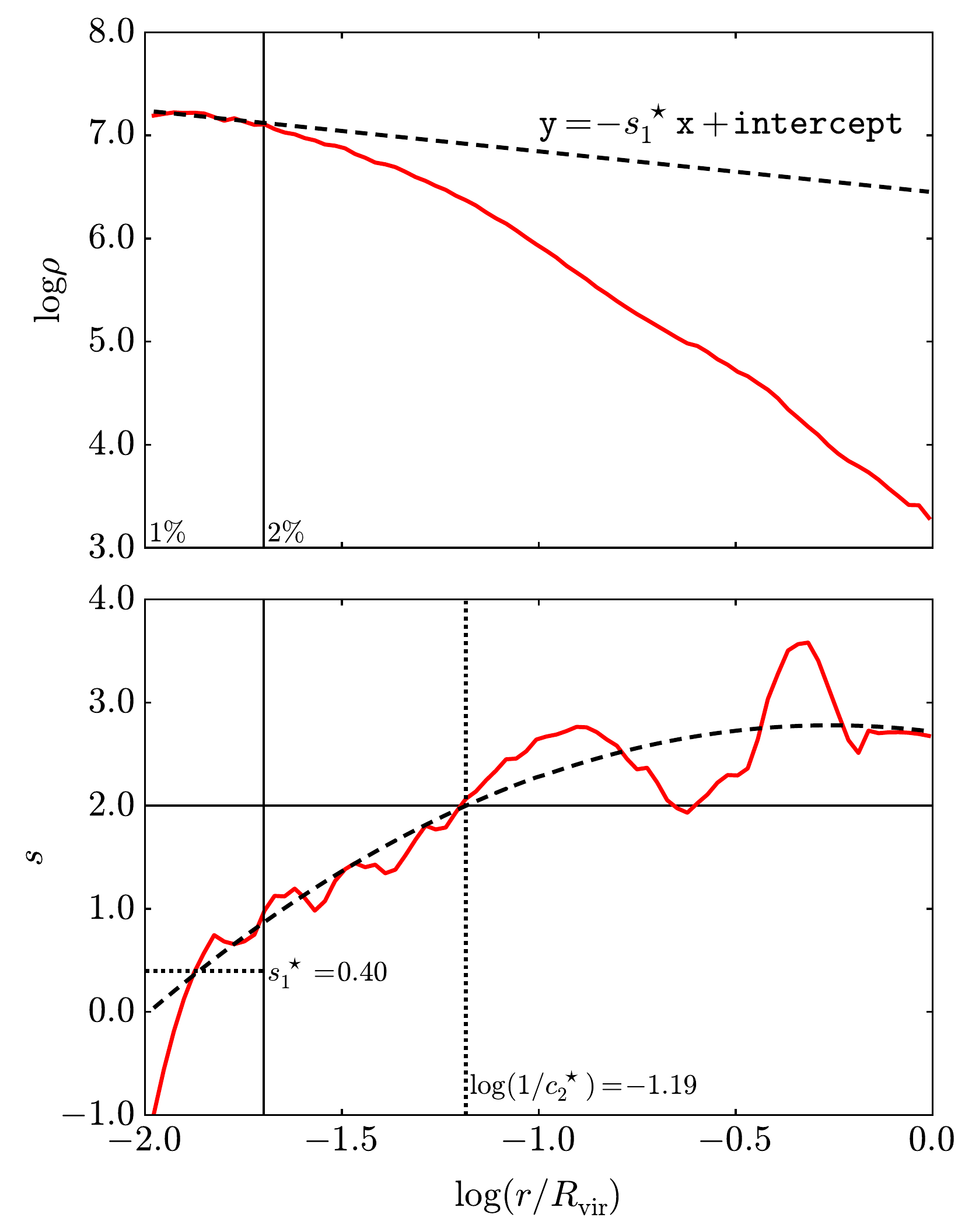}
	\vspace{-0.5cm}
	\caption{
		Measurement of the inner slope ${s_{1}^\star}$ and concentration ${c_2^\star}$ of NIHAO halo \texttt{g1.08e11} at $z=0$ from its density and logarithmic slope profiles. 
		The upper panel shows the density profile (plain red line). A linear least-square fit to $\log\rho$ between $0.01\Rvir$ and $0.02\Rvir$ (dashed line) enables to determine ${s_{1}^\star}$ the average inner slope in this radius range. 
		The lower panel shows the corresponding logarithmic slope $s$, obtained with a Savitsky-Golay smoothing filter \protect\citep[][implemented as \texttt{scipy.signal.savgol\_filter} in \texttt{scipy}]{SavitzkyGolay1964} with a polynomial order $n=3$. The plain red line correspond to a Savitsky-Golay window size $w=11$, the black dashed line to a smoother curve obtained with a maximal window size ($w=77$ here). 
		The concentration ${c_2^\star}$ corresponds to the radius ${r_2^\star}=\Rvir/{c_2^\star}$ where the smooth curve intersects the line $\mathtt{y}=2$. 
	}
	\label{fig:definition_s12_c2}
\end{figure}

Fig.~\ref{fig:definition_s12_c2} illustrates how the inner slope $s_1^\star$ and the concentration $c_2^\star$ are measured from the simulated density and logarithmic slope profiles. As explained in Section~\ref{subsection:params}, $s_1^\star$ is the average slope between $0.01 R_{\rm vir}$ and $0.02 R_{\rm vir}$, $c_2^\star$ corresponds to the radius where the logarithmic slope equals $2$. To define $c_2^\star$, the logarithmic slope profile is smoothed with a Savitsky-Golay filter with maximum window size. This window size choice maximizes the smoothing in order to have a measure of ${c_2^\star}$ that is not affected by the slope fluctuations, as examplified in the figure.

\subsection{Density residuals}

\begin{figure*}
	\includegraphics[width=\textwidth]{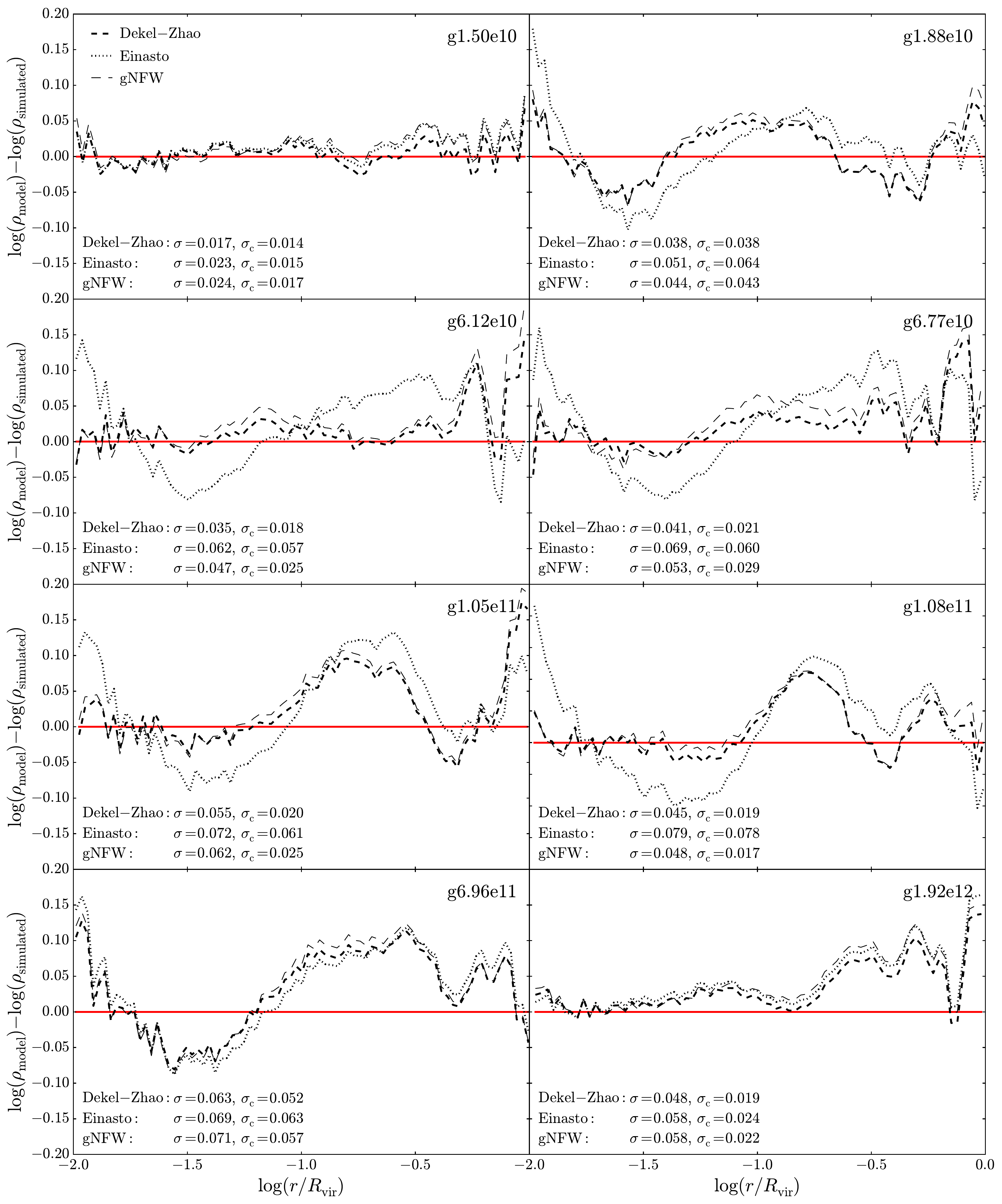}
	\vspace{-0.4cm}
	\caption{Residuals for the dark matter density profile DZ, Einasto and gNFW least-square fits shown in Fig.~\ref{fig:fits}, respectively traced as dashed, dotted, and thin dashed black lines. The rms errors $\sigma$ and $\sigma_{\rm center}$ are indicated.   
	}
	\label{fig:fits_rho_residuals}
\end{figure*}

\begin{figure*}
	\includegraphics[width=\textwidth]{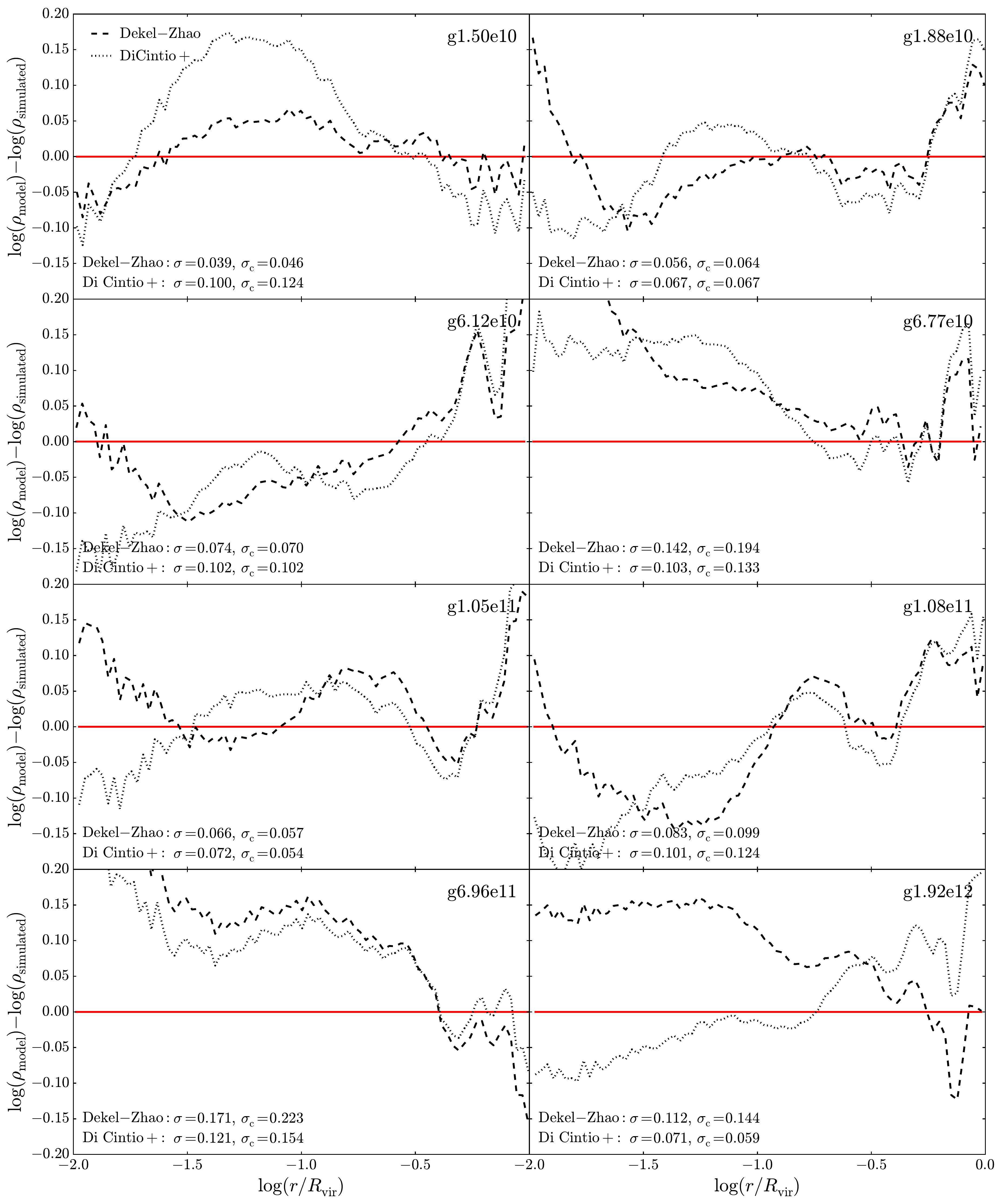}
	\vspace{-0.4cm}
	\caption{Residuals between the simulated density profiles shown in Fig.~\ref{fig:prescriptions} and the current DZ (dashed) and \protect\citetalias{DiCintio2014b} (dotted) mass-dependent prescriptions. The rms errors $\sigma$ and $\sigma_{\rm center}$ are indicated.   
	}
	\label{fig:prescription_rho_residuals}
\end{figure*}

\begin{figure*}
	\includegraphics[width=\textwidth]{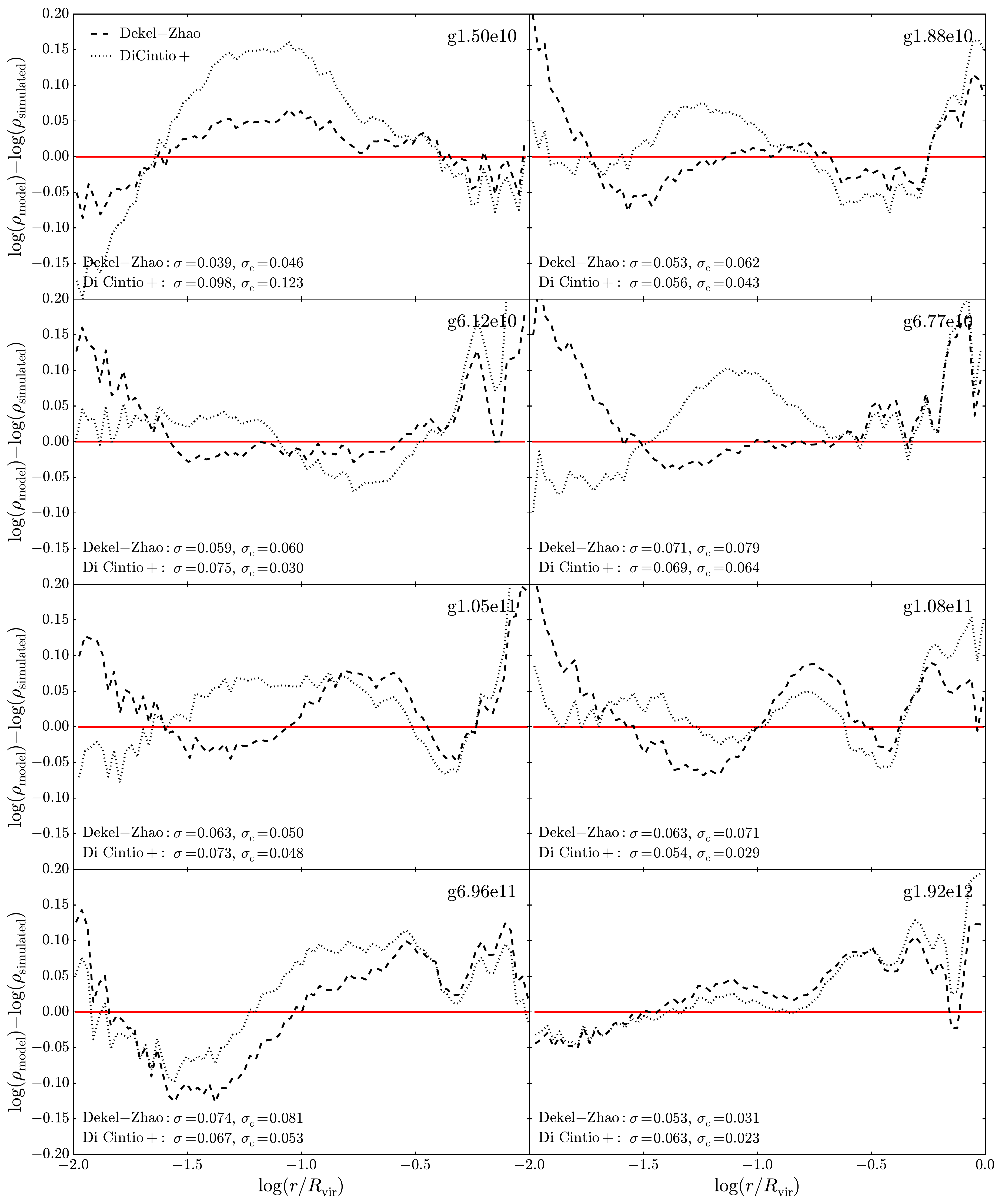}
	\vspace{-0.4cm}
	\caption{Residuals between the simulated density profiles shown in Fig.~\ref{fig:Vfit} and those stemming from one-parameter DZ (dashed) and \protect\citetalias{DiCintio2014b} (dotted) fits to the rotation curves. The rms errors $\sigma$ and $\sigma_{\rm center}$ are indicated.   
	}
	\label{fig:Vfit_rho_residuals}
\end{figure*}

Figs.~\ref{fig:fits_rho_residuals}, \ref{fig:prescription_rho_residuals}, and \ref{fig:Vfit_rho_residuals} respectively show the residuals (in $\log \rho$) between the simulated dark matter density profiles and (i) the DZ, Einasto and gNFW fits (Fig.~\ref{fig:fits}), (ii) the current DZ and \protect\citetalias{DiCintio2014b} mass-dependent prescriptions (Fig.~\ref{fig:prescriptions}), and (iii) the density profiles inferred from one-parameter DZ and \protect\citetalias{DiCintio2014b} fits to the rotation curves (Fig.~\ref{fig:Vfit}).
As stated in Sections~\ref{section:fitting}, \ref{section:prescription_accuracy}, and \ref{section:rotation}, the DZ profile provides significantly better fits to simulated density profiles than the Einasto profile and marginally better fits than the gNFW profile; its mass-dependent prescriptions are as successfull or even marginally better than the \protect\citetalias{DiCintio2014b} ones. We note a certain degree of stochasticity in the radial dependence of the residuals in $\log \rho$.

\subsection{Velocity residuals}

\begin{figure*}
	\includegraphics[width=\textwidth,trim={0cm 0cm 0cm 0cm},clip]{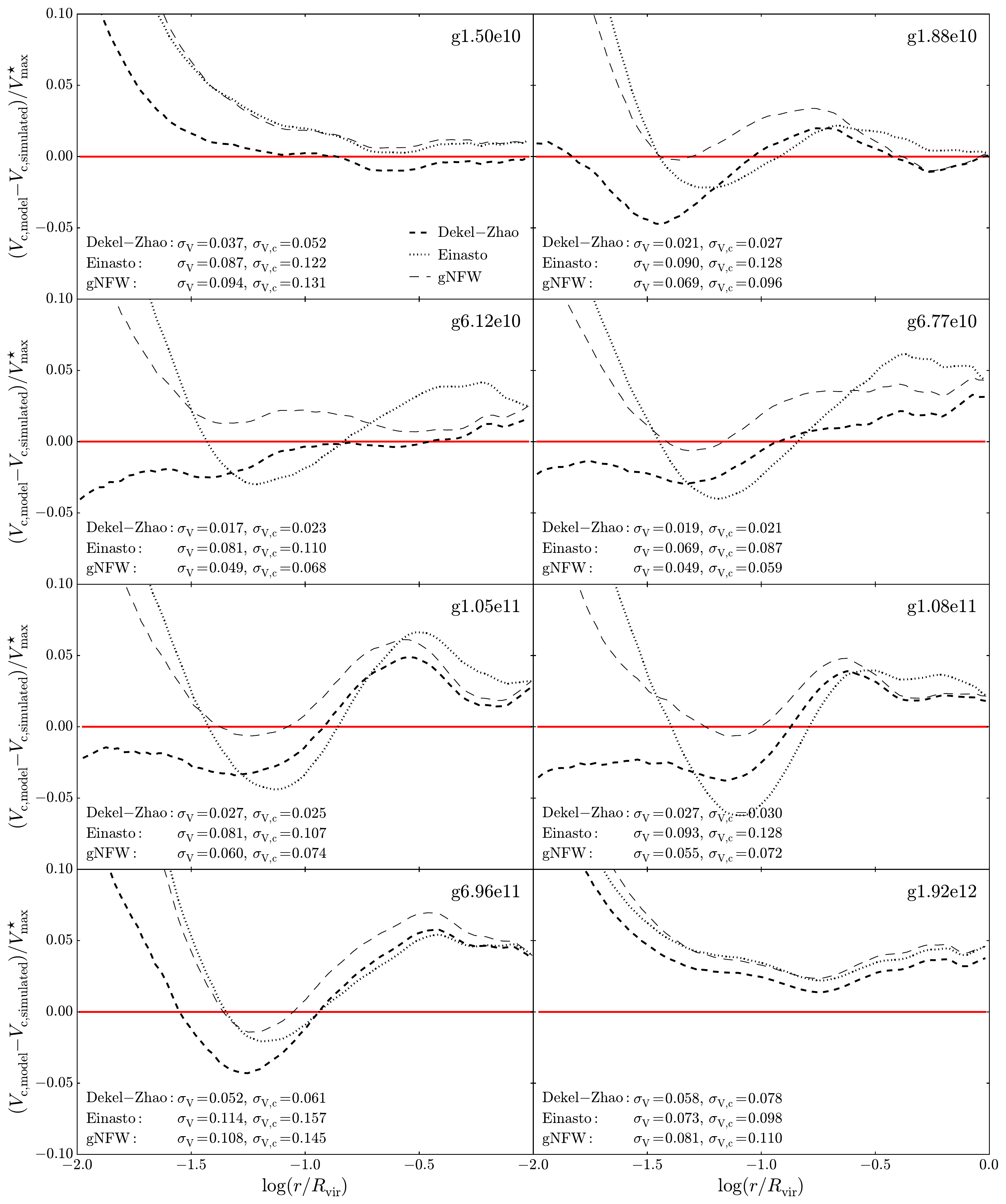}
	\vspace{-0.4cm}
	\caption{Residuals between the simulated rotation curves and those inferred from DZ (dashed), Einasto (dotted) and gNFW (thin dashed) fits to the density profiles shown in Fig.~\ref{fig:fits_Vc}. The rms errors $\sigma$ and $\sigma_{\rm center}$ are indicated.   
	}
	\label{fig:fits_Vc_residuals}
\end{figure*}

\begin{figure*}
	\includegraphics[width=\textwidth]{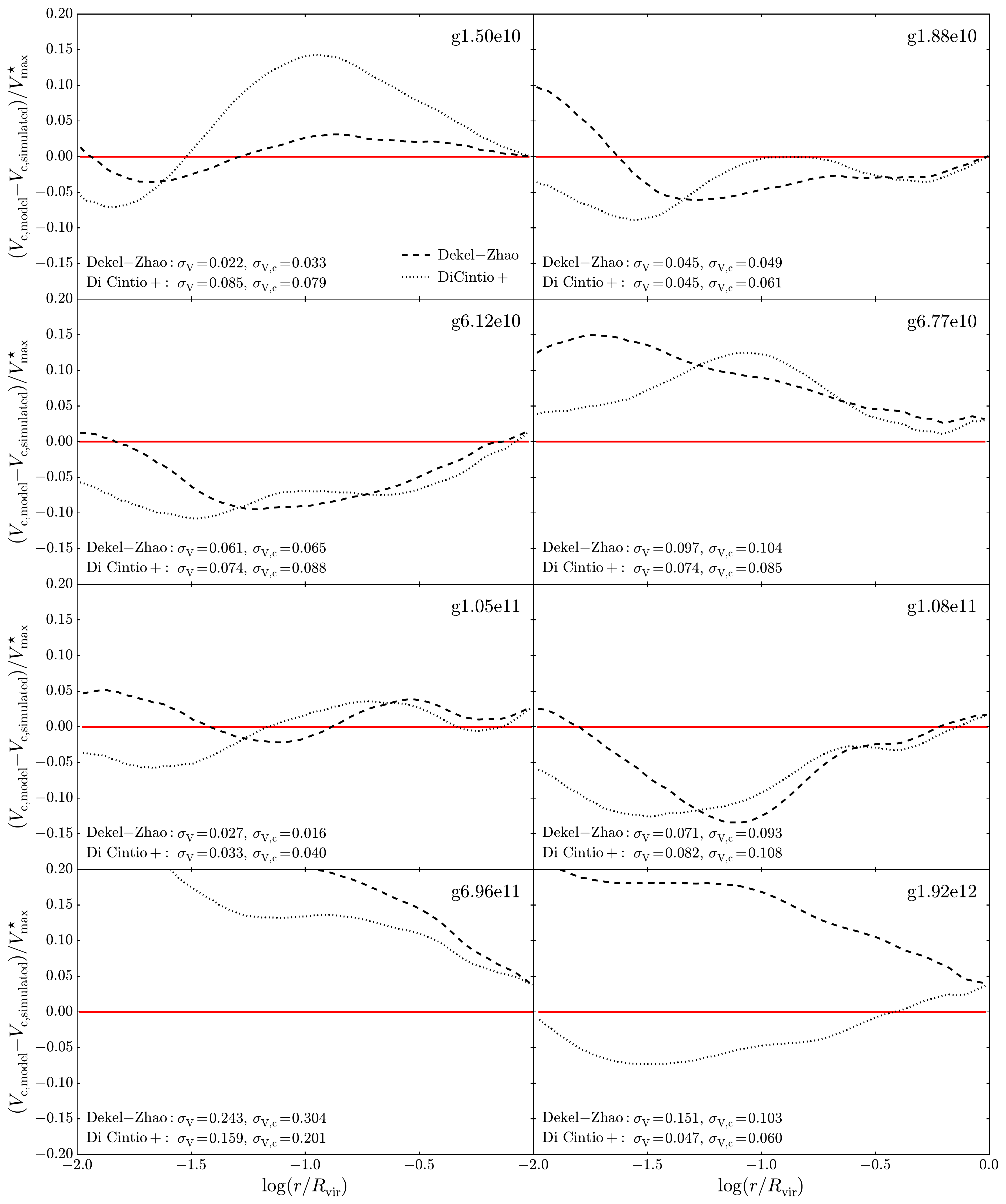}
	\vspace{-0.4cm}
	\caption{Residuals between the simulated rotation curves and the current DZ (dashed) and \protect\citetalias{DiCintio2014b} (dotted) mass-prescriptions shown in Fig.~\ref{fig:prescriptions_Vc}.  The rms errors $\sigma$ and $\sigma_{\rm center}$ are indicated.   
	}
	\label{fig:prescriptions_Vc_residuals}
\end{figure*}

\begin{figure*}
	\includegraphics[width=\textwidth]{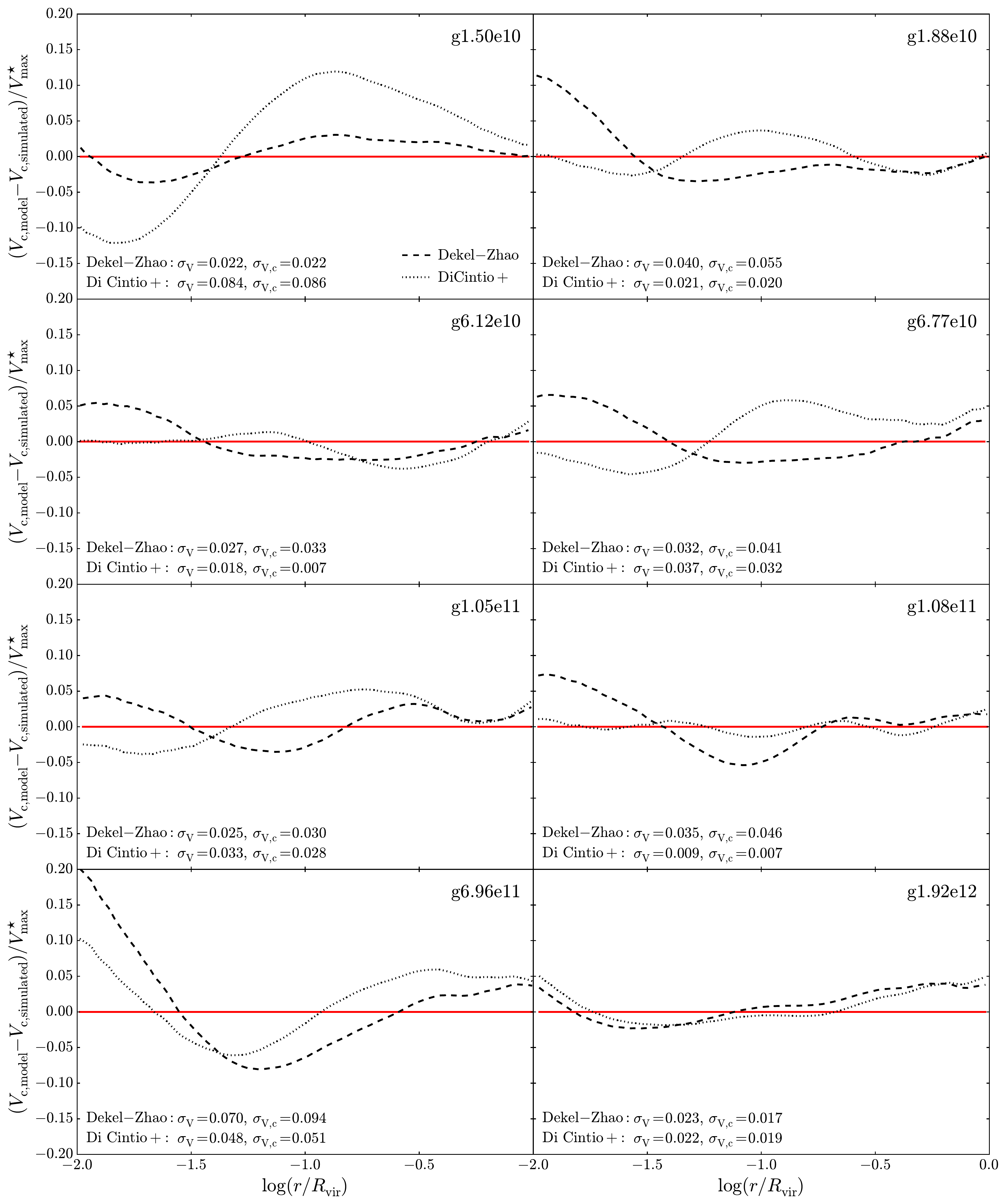}
	\vspace{-0.4cm}
	\caption{Residuals between the simulated rotation curves and the one-parameter DZ (dashed) and \protect\citetalias{DiCintio2014b} (dotted) fits to them shown in Fig.~\ref{fig:Vfit_Vc}. The rms errors $\sigma$ and $\sigma_{\rm center}$ are indicated.   
	}
	\label{fig:Vfit_Vc_residuals}
\end{figure*}

Figs.~\ref{fig:fits_Vc_residuals}, \ref{fig:prescriptions_Vc_residuals}, and \ref{fig:Vfit_Vc_residuals} respectively show the residuals  between the simulated rotation curves and (i) those stemming from DZ, Einasto and gNFW fits to the density profiles (Fig.~\ref{fig:fits_Vc}), (ii) the current DZ and \protect\citetalias{DiCintio2014b} mass-dependent prescriptions (Fig.~\ref{fig:prescriptions_Vc}), and (iii) one-parameter DZ and \protect\citetalias{DiCintio2014b} fits to the rotation curves (Fig.~\ref{fig:Vfit_Vc}).
As stated in Sections~\ref{section:fitting}, \ref{section:prescription_accuracy}, and \ref{section:rotation}, the DZ profile provides significantly better fits to simulated density profiles than  both the Einasto and the gNFW profiles; its mass-dependent prescriptions are as successfull or even marginally better than the \protect\citetalias{DiCintio2014b} ones.

\bsp	
\label{lastpage}
\end{document}